\documentclass[11pt, letterpaper]{article}

\usepackage{epstopdf}
\usepackage{doublespace}
\usepackage{caption}
\usepackage{url}
\usepackage{amsmath}
\usepackage{booktabs}       
\usepackage{graphicx}
\usepackage[breaklinks=true]{hyperref}
\usepackage{breakurl} 
\usepackage{enumitem}
\usepackage{subfigure}
\usepackage[margin=1in]{geometry}
\usepackage{fixltx2e}

\begin{document}
\begin{titlepage}
\begin{spacing}{1}
\begin{center}
\begin{flushleft}
Manuscript accepted for publication in \emph{Computational Intelligence and Neuroscience}\\ Special Issue on Ergonomic Issues in Brain-Computer Interface Technologies: Current Status, Challenges, and Future Direction
\end{flushleft}

\Large \vspace{2.5in}\textbf{\textsc{Evaluating a Semi-Autonomous Brain-Computer Interface Based on Conformal Geometric Algebra and Artificial Vision\vspace{0.75in}}}

\large M.\ A.\ Ramirez-Moreno and D.\ Guti\'{e}rrez\vspace{.75in}
	\\ \normalsize Centro de Investigaci\'{o}n y de Estudios Avanzados (CINVESTAV),\\
	 Unidad Monterrey, Apodaca, N.\ L.\ , 66600, M\'{e}xico.\\ \vspace{1in}
\end{center}

\noindent Corresponding author:\vspace{.25in}\\
D.\ Guti\'{e}rrez, Ph.D.\\
Centro de Investigaci\'{o}n y de Estudios Avanzados (CINVESTAV)\\
V\'{\i}a del Conocimiento 201, Parque de Investigaci\'{o}n e Innovaci\'{o}n Tecnol\'{o}gica (PIIT)\\
Autopista al Aeropuerto Km. 9.5, Lote 1, Manzana 29\\
Apodaca, N.\ L.\ , 66600, M\'{e}xico\\
Tel: (+52-81) 1156-1740 x 4513\\
Fax: (+52-81) 1156-1741\\
E-mail: dgtz@ieee.org
\end{spacing}
\end{titlepage}

\renewcommand{\thesection}{\arabic{section}}
\renewcommand{\thetable}{\Roman{table}}
\newcommand{\itb}[1]{\mbox{\boldmath $\mathit{#1}$}}

\settowidth{\baselineskip}{\em large}
\newcounter{subeqn}
\begin{spacing}{1.5}
\renewcommand{\abstractname}{\normalsize Abstract}
\begin{abstract}
  \normalsize In this paper, we evaluate a semi-autonomous brain-computer interface (BCI) for manipulation tasks. In such system, the user controls a robotic arm through motor imagery commands. In traditional process-control BCI systems, the user has to provide those commands continuously in order manipulate the effector of the robot step-by-step, which results in a tiresome process for simple tasks such as pick and replace an item from a surface. Here, we take a semi-autonomous approach based on a conformal geometric algebra model that solves the inverse kinematics of the robot on the fly, then the user only has to decide on the start of the movement and the final position of the effector (goal-selection approach). Under these conditions, we implemented pick-and-place tasks with a disk as an item and two target areas placed on the table at arbitrary positions. An artificial vision (AV) algorithm was used to obtain the positions of the items expressed in the robot frame through images captured with a webcam. Then, the AV algorithm is integrated to the inverse kinematics model to perform the manipulation tasks. As proof-of-concept, different users were trained to control the pick-and-place tasks through the process-control and semi-autonomous goal-selection approaches, so that the performance of both schemes could be compared. Our results show the superiority in performance of the semi-autonomous approach, as well as evidence of less mental fatigue with it.
\end{abstract}
\end{spacing}
\newpage

\section{Introduction}
A brain-computer interface (BCI) is a system that enables a real-time user-device communication pathway through brain activity. Through the years, development and research on BCI has mainly been oriented to the creation of rehabilitation systems, as well as systems that help disabled patients regain to some extent their lost or diminished capabilities~\cite{Bi13}. Some reported devices that have been successfully controlled using BCIs are spellers, electric wheelchairs, robotic arms, electric prostheses and humanoid robots~\cite{Cecotti10,Wang11,Murphy17,Bell08}. In BCI studies, the most common technique used to acquire brain non-invasively is electroencephalography (EEG). 

In order to manipulate the device through brain activity, the design of the BCI  must include the following stages: signal acquisition, filtering, feature extraction, classification, device modeling and control~\cite{Wolpaw02}. During the filtering stage, unwanted noise and artifacts are removed from the signals using temporal and spatial filters. Then, temporal or spatial features of interest are extracted from the signals to build feature vectors. These vectors are formed by characteristic components of the signals, which are then used in the classification stage to decipher user intention. Lastly, the device is manipulated based on the result of the classification algorithm. Depending on the device and the complexity of the system,  a model of the system is needed to perform with precision the desired tasks. BCIs can be divided into two groups based on their control strategy: \emph{process-control} and \emph{goal-selection}. In the process-control strategy, users are continuously controlling each part of the process. This is done by performing low-level commands continuously through the BCI, with no additional assistance. On the other hand, in the goal-selection strategy, users are responsible for selecting their desired goal and the system provides assistance to successfully perform the tasks with minimum effort~\cite{Royer12}.  In this case, the user performs high-level tasks by sending simple commands through the BCI. 

Common paradigms used as control commands in BCI include steady state visual evoked potentials (SSVEP), P$300$ waveform,  and motor imagery (MI). SSVEP is a resonance phenomenon occurring at occipital and parietal lobes as result of oscillatory visual stimulus presented to a user at a constant frequency~\cite{Iscan18}.
 The P$300$ is an EEG signal component that appears $300$-ms after an event of voluntary attention, and it is usually observed during visual or auditory stimulus presentation~\cite{Rezai12}. MI presents as an event-related desynchronization (ERD) found at the sensorimotor areas, which generates a contra-lateral power decrease in a frequency range from $8$-$13$~Hz (also known as the $\mu$ band)~\cite{McFarland00}. Controlling a BCI with SSVEP and P$300$ require less training in comparison to MI, as the first represent an unvoluntary response to a stimulus. However, its use in BCI is limited due to its requirement of a stimulus presentation device. The training process to control MI-based BCIs (MI-BCI) might involve stimulus presentation as well. However, it can be excluded for its final application on the BCI. Even though MI-BCIs require longer training periods, they are better suited for close-to-real-life environments and self-paced BCIs \cite{Leeb07}. 

Several studies present successfully implemented ERD based BCIs, most of them using a process-control strategy~\cite{Birbaumer99,Wolpaw04,Pfurtscheller03}. Some goal-selection BCIs have been reported as well~\cite{McFarland08,Friedrich09}. In~\cite{Royer12}, users were trained on process-control and goal-selection MI-BCIs to perform one-dimensional cursor movements on a screen. The results suggests that users performing on goal-selection strategy showed higher accuracy and faster learning in comparison to the process-control approach. However, the authors state that a direct comparison of goal-selection and process-control in a more complicated (real world) scenario has not yet been presented. In the proposed study, three-dimensional object manipulation tasks through a robotic arm are implemented in a MI-BCI. The complexity of the three-dimensional movements on real objects is higher than the one-dimensional movements  on virtual objects presented in~\cite{Royer12}. In \cite{Valbuena07}, a semi-autonomous BCI is implemented to manipulate a robotic arm to perform tasks such as pouring a beverage inside a glass in a tray, through SSVEP. In future research, similar tasks as in \cite{Valbuena07} could be implemented in our BCI using MI instead, allowing a more natural execution of daily-life context tasks without the need of a stimulus presentation screen. 

In a typical process-control MI-BCI, the user controls the direction of the final effector of a robotic arm through low-level commands, which means that 
the user has to maneuver the robot in a three-dimensional space to reach for a desired target. Clearly, the user remains in a high attention state during the maneuvers, as he/she is continuously aware of the final effector position during the whole task. This continuous awareness might lead to generation of mental fatigue or frustration, which is undesirable as it can directly affect user performance and learning~\cite{Grainmann08}. The analysis of P$300$ features, such as amplitude and latency, has been shown to be useful in identifying the depth of cognitive information processing~\cite{Cheng11}. The amplitude of P$300$ waveform tends to decrease when users encounter cognitive tasks with high difficulty~\cite{Isreal80}. On the other hand, P$300$ latency has shown to increase when the stimulus is cognitively difficult to process~\cite{Murata05}. Another study has reported correlation between changes in the P$300$ component and BCI performance~\cite{Mak12}. The evidence provided by these studies might suggest that the analysis of P$300$ could be implemented as a mental fatigue indicator during BCI training and control.

In order to diminish  mental fatiguing in BCI systems, a semi-autonomous BCI using a goal-selection strategy is here proposed. This system  assists the user to perform a specific task by calculating all the variables needed to successfully execute it. 
Some studies have previously presented BCI designs focusing on this semi-autonomous approach with successful results on performance, accuracy and comfort for the user~\cite{Perrin10,Valbuena07,Gohring13}. Therefore, this paper presents the implementation of a traditional low-level MI-BCI and a semi-autonomous MI-BCI designed to perform object manipulation tasks with a robotic arm. In the process-control strategy MI-BCI, the user commands the final effector of the robot to move in a three-dimensional space to reach for a target placed on a table. In the semi-autonomous MI-BCI, one small disk and two target areas are placed on a table. Here, the robot reaches for the disk and places it on a specific target, which is selected by the user. As proof-of-concept, two volunteers were trained on each BCI system and their performance was evaluated and compared. A statistical P300 analysis was performed on all users in order to observe mental fatigue differences induced by the operation of low-level and semi-autonomous BCIs. 

In order to model the robot used in this experiment, %
 a conformal geometric algebra (CGA) model was implemented in both the traditional and semi-autonomous BCIs to solve the inverse kinematics of the robotic arm, i.e., obtaining the joint angles needed for a specific position of the final effector. Additionally, an artificial vision (AV) algorithm was integrated in the semi-autonomous BCI in order to provide information about the positions of the items on the table referenced to robot frame. As the implementation of the semi-autonomous BCI implies a higher computational load, the CGA model was chosen for the solution of the inverse kinematics. CGA has shown to represent an operation reduction and in some cases, a decrease in computational load when compared to traditional inverse kinematics solution~\cite{Hildenbrand06}.

 This paper is organized as follows: the CGA model and AV algorithm are described in Section~\ref{sec:methods}, and the design of both BCIs is explained in Section~\ref{sec:bci_des}; evaluations on both algorithms and performance results of users controlling both BCIs are presented in Section~\ref{sec:Res}. Preliminary short reports of the system's implementation (but not its evaluation) have been presented at~\cite{Ramirez18} and~\cite{Ramirez19}.

\section{Robot modeling and artificial vision}\label{sec:methods}
In this section we describe each of the components required to compute the inverse kinematics of a robotic arm by using CGA. Furthermore, here we explain in detail the AV algorithm used to obtain the positions of the objects to be manipulated by the robot. 

\subsection{Conformal Geometric Algebra}

Traditional methods to solve the inverse kinematics of robots include several matrix operations, as well as many trigonometric expressions. All this can result in a quite complex solution depending on the modeled robot~\cite{Spong05}. In  this study, a conformal geometric algebra (CGA) model is proposed instead, as it is considered to be computationally lighter, easier to implement, and highly intuitive.  CGA has proved to be a powerful tool when solving the inverse kinematics of robotic arms~\cite{Carbajal15,Hildenbrand08}. It also offers an operation reduction when compared to traditional methods and provides runtime efficient solutions. More information on computational efficiency characteristics can be found in~\cite{Perwass09}. 

With this model, the joint angles of the robot are obtained for a specific position of the final effector.  In CGA, two new dimensions ($e_{0},e_{\infty}$) are defined, representing a point in the origin and a point in the infinity, respectively, in addition to the three-dimensional Euclidean space ($e_{1}, e_{2}, e_{3}$)~\cite{Carbajal15}. In this space, geometric entities (points, lines, circles, planes and spheres) and calculations involving them (distances and intersections) can be represented with simple algebraic equations.

Also, the geometric product between two vectors \emph{a} and \emph{b} is defined as a combination of the inner product and the outer product:
\begin{equation}
ab=a\cdot b + a\wedge b.
\label{eq:geo}
\end{equation}

 The inner product is used to calculate distances between elements, and the outer product generates a bivector, which is an element occupying the space spanned by both vectors. It is also used find the intersection between two elements.  The intersection $M$ of two geometric objects $A$ and $B$ represented in CGA is given by $M^{*}=A^{*} \wedge B^{*}$, or $M^{*}=A^{*} \cdot B$. The element $A^{*}$ is the dual of $A$ and is expressed as
\begin{equation}
A^{*}= AI_{c}^{-1},
\label{eq:dual}
\end{equation}
where $I_{c}^{-1}=e_{0}e_{3}e_{2}e_{1}e_{\infty}$, which allows for a change in representation of the same element. Standard and dual representations of commonly used geometrical objects in CGA are shown in Table~\ref{tab:tab1}. There, $x$ and $n$ are points represented as a linear combination of the $3$D base vectors:
\begin{equation}
x = x_{1}e_{1} + x_{2}e_{2} + x_{3}e_{3}. 
\label{eq:pointinspace}
\end{equation}

There are two possible representations of the same element, as shown in Table~\ref{tab:tab1}. A circle can be represented as the space spanned by three points in space, as well as the intersection of two spheres. Also, a line can be expressed as the intersection of two planes, as well as the space spanned by two points expanded to the infinity.

\begin{figure}
\centerline{\includegraphics[width=0.75\textwidth]{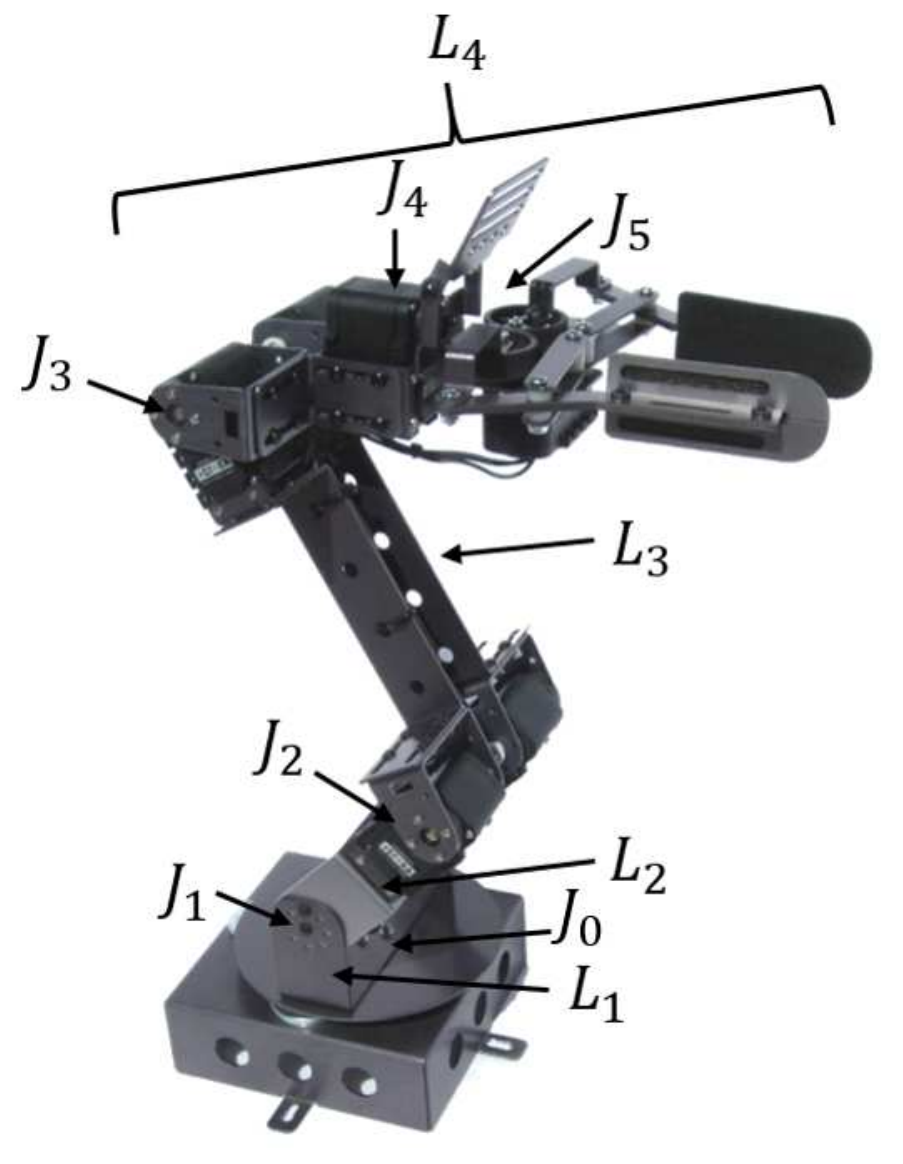}}
\caption{Joints and links of our $5$-DOF Dynamixel AX-$18$A Smart Robotic Arm.}
\label{fig:robot}
\end{figure}

Making use of the previous equations and relationships, a CGA model to solve the inverse kinematics of a manipulator robot was obtained following the proposed method in~\cite{Kleppe16}. The modeled robot was the Dynamixel AX-$18$A Smart robotic arm, which is a five degrees-of-freedom (DOF) manipulator robot. Figure~\ref{fig:robot} shows the modeled robot as well as its joints and links. The DOF of this robot correspond to its shoulder rotation, elbow flexion-extension, wrist flexion-extension, wrist rotation, and hand open-close function~\cite{Griggs14}. The inverse kinematics solution was obtained for joints $J_{0}$, $J_{2}$, and $J_{3}$. For the particularities of the manipulation tasks, joints $J_{4}$ and $J_{5}$ were not considered for simplicity.  

\subsection{Our CGA model}
Next, we describe the required CGA model that we implemented specifically for our system.

\subsubsection{Fixed joints and planes}
\begin{figure}
\centerline{\includegraphics[width=0.75\textwidth]{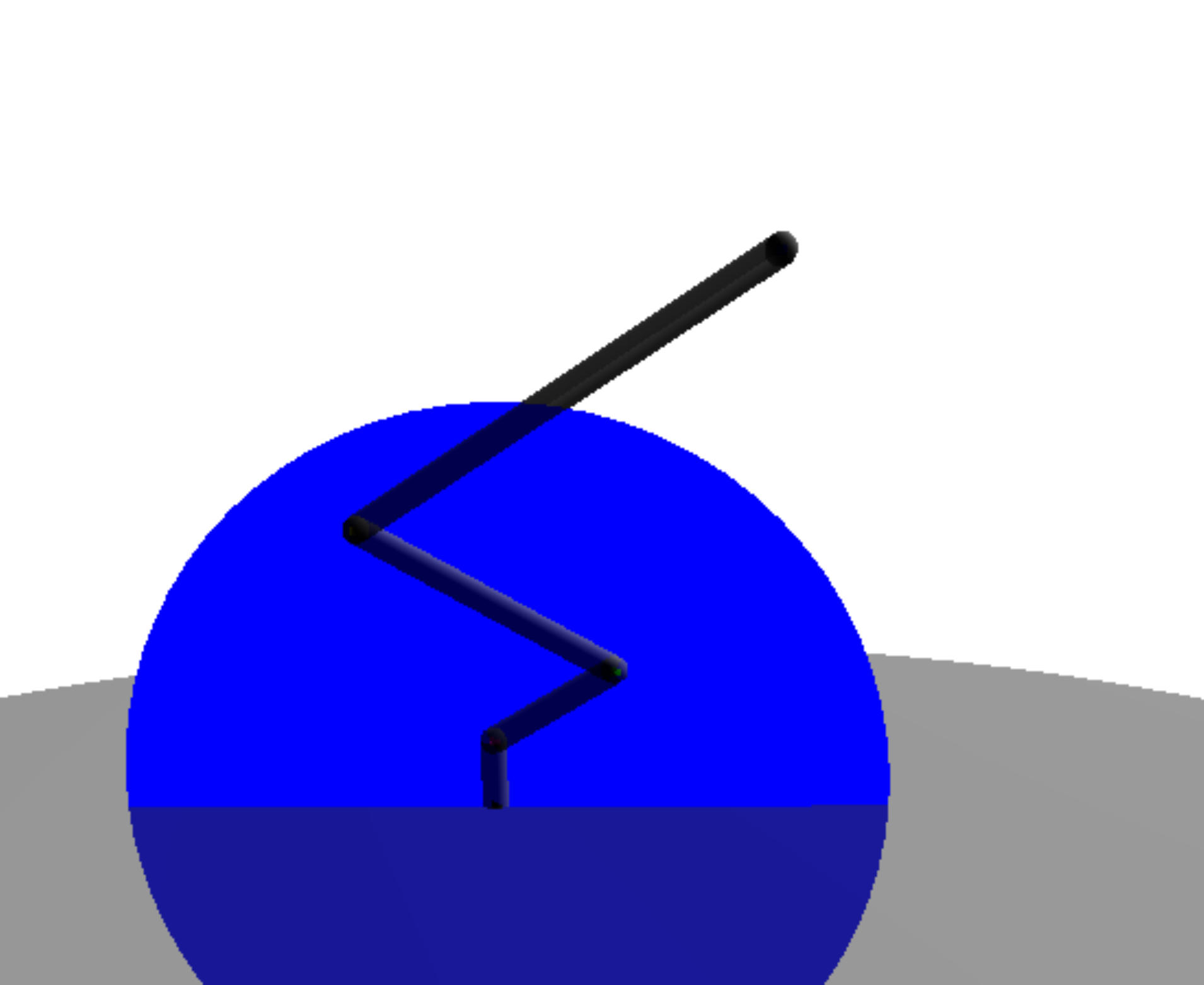}}
\caption{Planes $\pi_{e}$  and $\pi_{b}$  representing the orientation of the final effector and the robot base, respectively.}
\label{fig:planos}
\end{figure}

The origin of the CGA model was located at joint $J_{0}$, located at the center of the rotational base of the robot, therefore $J_{0}=e_{0}$. Joint $J_{1}$ is also a fixed joint with constant position, found directly above joint $J_{0}$. The position for joint $J_{1}$ was defined as  $x_{1}=[0,0,0.036]$. Now, let us consider the desired final effector position as a point in space $x_{e}$. Then, a vertical plane $\pi_{e}$ representing the direction of the final effector is described as
\begin{equation}
\pi_{e} = e_{0} \wedge e_{3} \wedge x_{e} \wedge e_{\infty}, 
\label{eq:ple}
\end{equation}
where $e_{0}$ represents the origin in robot frame and $e_{3}$ the Euclidian \emph{z} axis. As the position of the final effector is used to define $\pi_{e}$, the direction of plane changes consistently with $x_{e}$. A plane $\pi_{b}$, representing the rotational base of the robot is defined as
\begin{equation}
\pi_{b}=e_{0} \wedge e_{1} \wedge e_{2} \wedge e_{\infty},
\label{eq:plb}
\end{equation}
where $e_{1}$ and $e_{2}$ represent the Euclidean \emph{x} and \emph{y} axes. Planes $\pi_{e}$ and $\pi_{b}$ are shown in Figure~\ref{fig:planos}. 

\subsubsection{Calculation of joint's position}
In a kinematic chain model of a robotic arm using CGA, the implemented method to find joint $J_{n}$  is based on the intersection of two spheres centered at joints $J_{n-1}$ and $J_{n+1}$ with radii equal to the lengths of the links connecting $J_{n-1}$ with $J_{n}$, and $J_{n}$ with $J_{n+1}$, respectively. The intersection of both spheres results in a circle, which is then intersected to the plane of the final effector to obtain a point pair representing two possible configurations for joint $J_{n}$. One point is then selected as $J_{n}$, depending on the desired configuration. The process requires the following:
\begin{itemize}
\item Spheres centered at point $P$ and with radius $r$ are given by
  \begin{equation}
s=P-\frac{1}{2}r^{2}e_{\infty};
\label{eq:spheres}
  \end{equation}
\item There are two methods for creating a circle. We can either intersect two spheres $s_{j}$ and $s_{k}$ by
  \begin{equation}
c=(s_{j}^{*} \wedge s_{k}^{*})^{*},
\label{eq:circles_a}
  \end{equation}
or we can intersect a plane $\pi$ and a sphere $s$ by
\begin{equation}
  c=(\pi^{*} \wedge s^{*})^{*};
\label{eq:circles_b}
\end{equation}
\item The intersection of a circle $c$ and a plane $\pi$ to create a point pair $Pp$ is given by
\begin{equation}
Pp=(c^{*}\wedge \pi^{*})^{*};
\label{eq:pointpair}
\end{equation}
\item Finally, to obtain a point $P$ from $Pp$, we have
\begin{equation}
P= \frac{Pp \pm \sqrt{Pp^{2}}} {-e_{\infty}\cdot Pp}.
\label{eq:point}
\end{equation}
\end{itemize}

Based on the previous expressions, and in order to find the position of joint $J_{2}$ in our modeled robot, two spheres must be constructed, and they have to be centered at $J_{1}$ and $J_{3}$. However, the position of joint $J_{3}$ is yet unknown in our model. A similar situation occurs if the desired position is instead, joint $J_{3}$. In this particular case, $x_{e}$ is known but not $J_{2}$. Given this situation, another approach was implemented in order to find joint $J_{2}$. 

\subsubsection{Position of joint $J_2$}
Using (\ref{eq:spheres}), sphere $s_{1}$ was centered at $x_{1}$ with radius equal to the length of link $L_{2}$. Hence, in order to find joint  $J_{2}$, another sphere $s_{h}$ must be intersected to $s_{1}$. In order to construct $s_{h}$, its center must be defined. This is achieved by first creating an auxiliary sphere $s_{0}$, centered at the origin with radius $L_{a}$ equal to the horizontal component of the distance from $J_{0}$ to $J_{2}$. This is valid as the distance from $J_{0}$ to $J_{2}$ is constant for any position of the final effector $x_{e}$. 

\begin{figure}
\centerline{\includegraphics[width=0.75\textwidth]{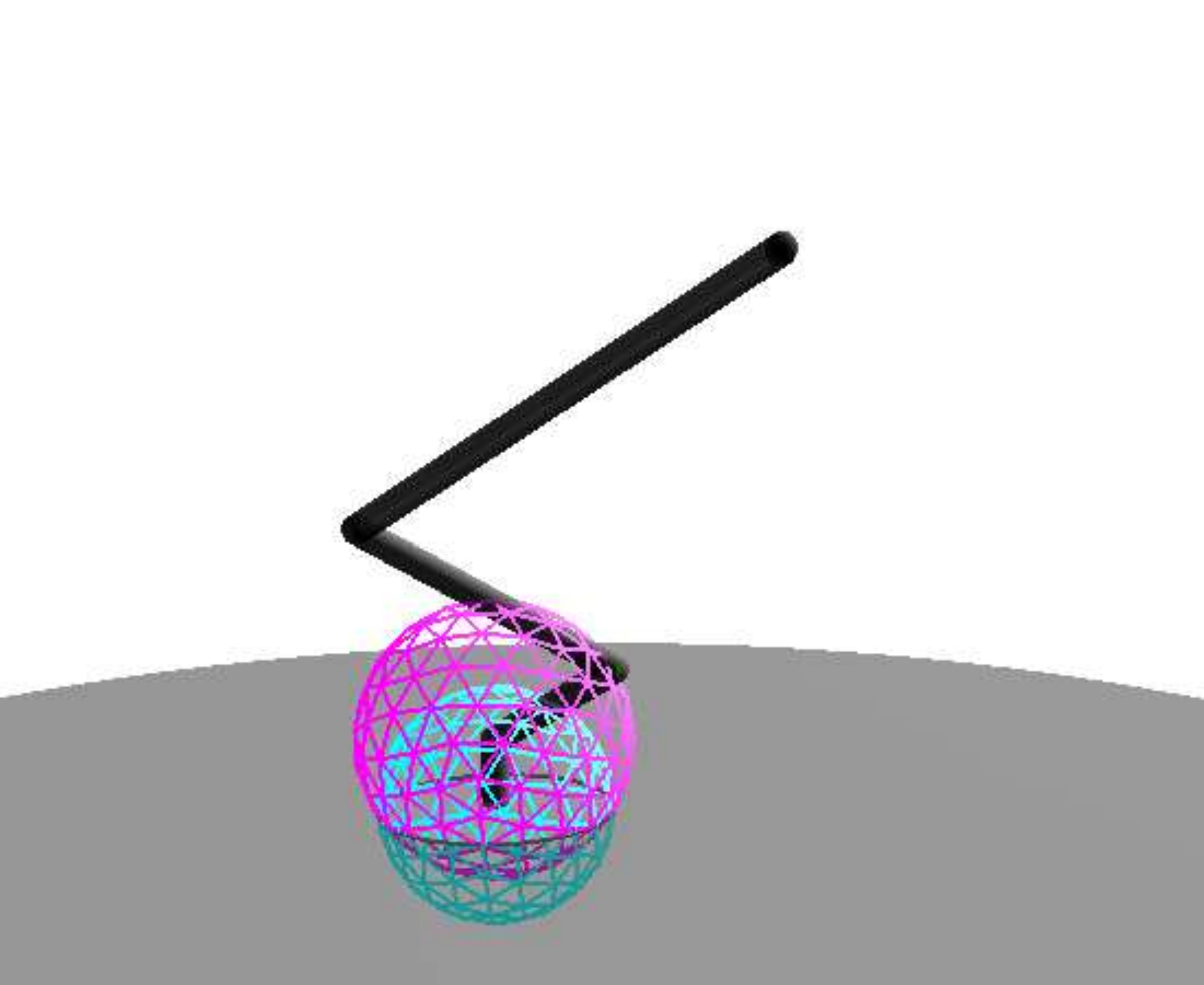}}
\caption{The intersection of spheres $s_{0}$ (bottom) and $s_{1}$ (top) define the circle $c_{0}$, from which we find the position of point $x_{h}$.}
\label{fig:xh}
\end{figure}

Then, using (\ref{eq:circles_b}), $s_{0}$ is intersected to plane $\pi_{e}$ to obtain circle $c_{0}$. Next, using (\ref{eq:pointpair}), $c_{0}$ is intersected to plane $\pi_{b}$ to produce point pair $Pp_{0}$, from which one point is selected as $x_{h}$ using (\ref{eq:point}). The procedure to find point $x_{h}$, which corresponds to the center of the desired sphere to be intersected with $s_{1}$, is shown in Figure~\ref{fig:xh}.

\begin{figure}
\centerline{\includegraphics[width=0.75\textwidth]{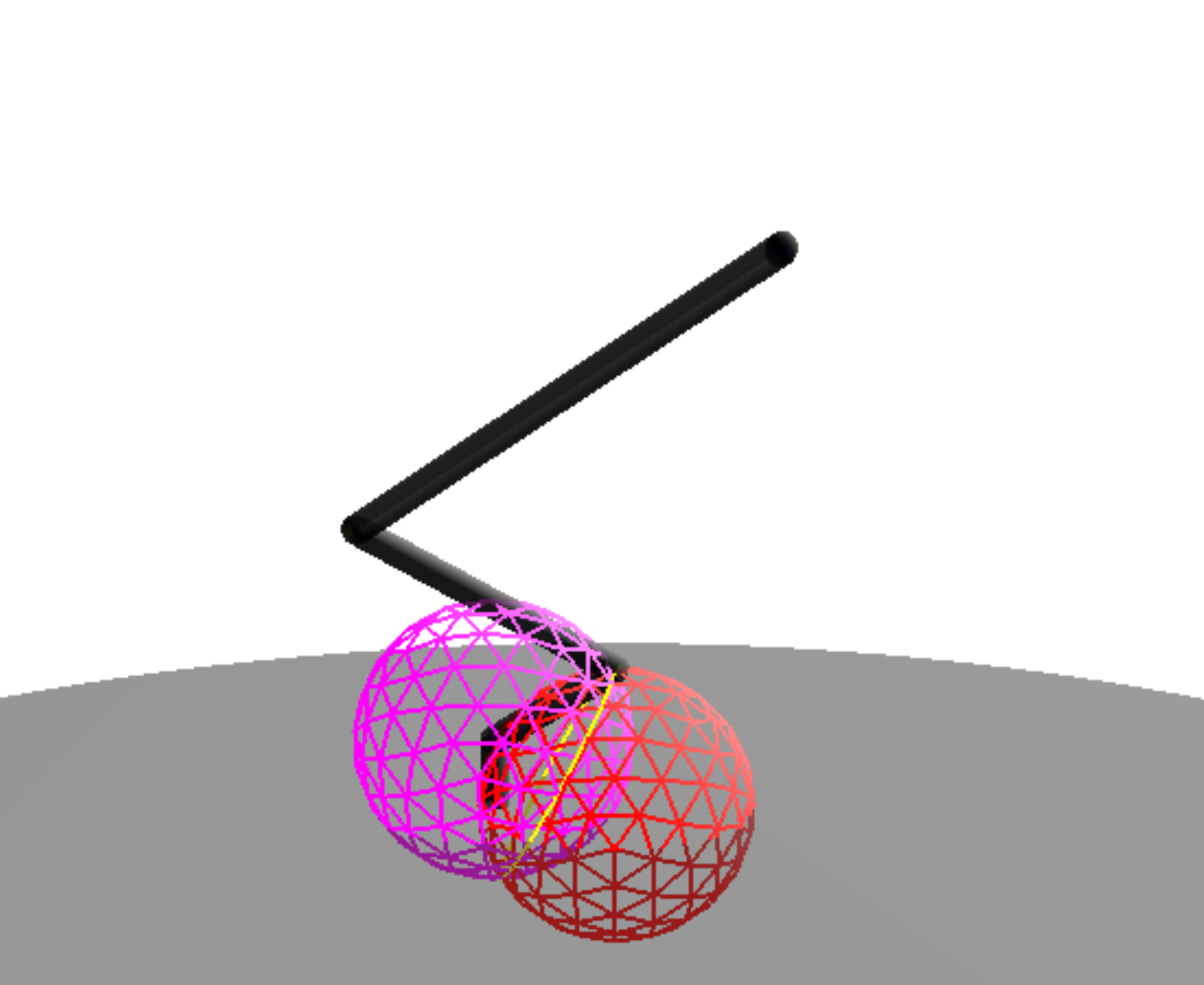}}
\caption{The intersection of spheres $s_{1}$ (left) and $s_{h}$ (right) define the circle $c_{2}$,  from which we find the position of joint $J_{2}$.}
\label{fig:j2}
\end{figure}

Using (\ref{eq:spheres}), sphere $s_{h}$ is centered at $x_{h}$ with radius $L_{b}$ equal to the vertical component of the distance from $J_{0}$ to $J_{2}$. Then, the intersection of spheres $s_{1}$ and $s_{h}$  is given by (\ref{eq:circles_a}), which results in circle $c_{2}$. Using (\ref{eq:pointpair}), the intersection of $c_{2}$ with plane $\pi_{e}$ renders point pair $Pp_{2}$. Finally, the position of $J_{2}$ is obtained from $Pp_{2}$ given by (\ref{eq:point}). The whole procedure previously detailed to obtain the position of joint $J_{2}$ is represented in Figure~\ref{fig:j2}.

\subsubsection{Position of joint $J_3$}

\begin{figure}
\centerline{\includegraphics[width=0.75\textwidth]{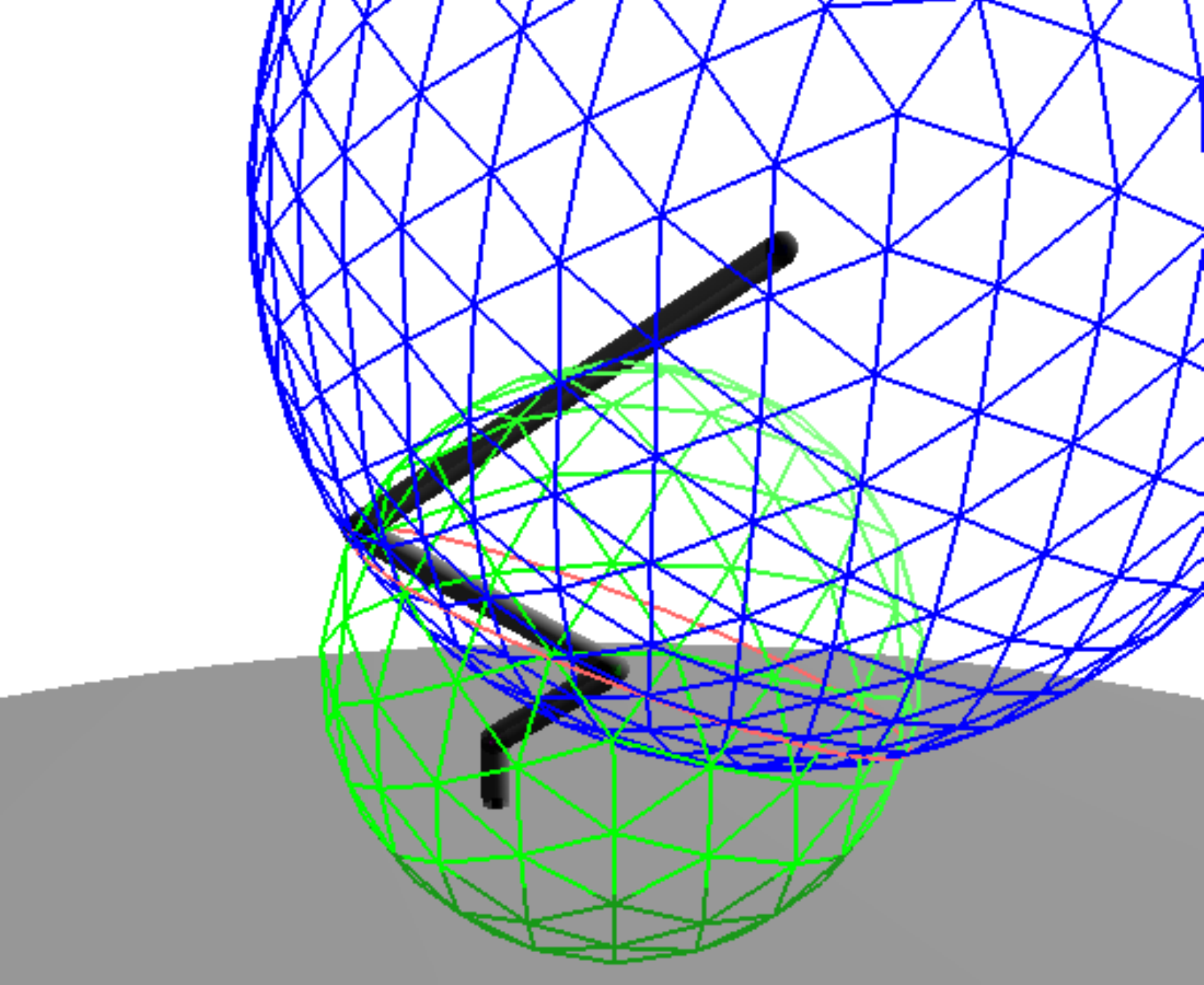}}
\caption{The intersection of spheres $s_{2}$ (bottom) and $s_{e}$ (top) define the circle $c_{3}$, from which it is possible to find the position of joint $J_{3}$.}
\label{fig:j3}
\end{figure}

The procedure to find the position of joint $J_{3}$ is straight-forward once the position of joint $J_{2}$ was calculated. For that, two spheres $s_{2}$ and $s_{e}$  are  defined using~(\ref{eq:spheres}), centered at $x_{2}$ and $x_{e}$, and with radii equal to the length of link $L_{3}$ and $L_{4}$, respectively. Both spheres are intersected to obtain circle $c_{3}$ using~(\ref{eq:circles_a}). With (\ref{eq:pointpair}), $c_{3}$ is then intersected to plane $\pi_{e}$ to obtain point pair $Pp_{3}$. From $Pp_{3}$, $J_{3}$ is easily obtained using~(\ref{eq:point}). A representation of the  procedure to find joint $J_{3}$ is shown in Figure~\ref{fig:j3}.
  
\subsubsection{Angle calculation}
In order to calculate the angles formed by two vectors $\alpha$ and $\beta$,  their corresponding unit vectors are defined as $\hat{\alpha}=\alpha/||\alpha||$ and $\hat{\beta}=\beta/||\beta||$. The normalized bivector spanning the space formed by those vectors is expressed as
\begin{equation}
\hat{N}=\pm\frac{\hat{\alpha}\wedge\hat{\beta}}{||\hat{\alpha}\wedge\hat{\beta}||}.
\label{eq:n}
\end{equation}
As explained in~\cite{Kleppe16}, the angle $\theta$ between $\alpha$ and $\beta$ is given by 
\begin{equation}
\theta=\mathrm{Atan2}[(\alpha\wedge \beta)/\hat{N},\alpha\cdot \beta],
\label{eq:angle}
\end{equation}
where $\mathrm{Atan2}$ corresponds to the \emph{four-quadrant} inverse tangent. This operator gathers information on the signs of its two arguments in order to return the appropriate quadrant of the computed angle~\cite{Organick66}. Such result is not possible to be obtained from the conventional single-argument $\arctan$ function. Also note that the plus sign in~(\ref{eq:n}) applies if the rotation from $\alpha$ to $\beta$ is counter-clockwise, while the minus sign applies in the opposite rotation. 

In order to find the joint angles using~(\ref{eq:angle}), vectors formed by the links of the robot need to be calculated. First, lines representing each link are defined: 
\begin{equation}
l_{01} = e_{0}\wedge J_{1}\wedge e_{\infty},
\label{eq:L01}
\end{equation}
\begin{equation}
l_{12} = J_{1}\wedge J_{2}\wedge e_{\infty},
\label{eq:l12}
\end{equation}
\begin{equation}
l_{23} = J_{2}\wedge J_{3}\wedge e_{\infty},
\label{eq:l23}
\end{equation}
and
\begin{equation}
l_{3e} = J_{3}\wedge J_{e}\wedge e_{\infty}.
\label{eq:l3e}
\end{equation}
The previous expressions define lines passing through links $L_{1}$, $L_{2}$, $L_{3}$, and $L_{4}$, respectively (see Figure~\ref{fig:robot}). $L_{4}$ was considered a straight line from joint $J_{3}$ to the final effector $x_{e}$, i.e., we ignored wrist-rotation and hand open-close joints.

In (\ref{eq:angle}), the parameters $\alpha$ and $\beta$ need to be \emph{directional} vectors for the purpose of computing our joint angles. Therefore, the directional vectors of plane $\pi_{e}$, as well as lines $l_{23}$ and $l_{3e}$, were calculated, which represent the base and links of the robot, respectively. From a given line $l$, its directional vector can be obtained as
\begin{equation}
(l \cdot e_{0}) \cdot e_{\infty}, 
\label{eq:vecline}
\end{equation}
and the directional vector normal to a plane $\pi$ is given by
\begin{equation}
(\pi^{*} \wedge e_{\infty}) \cdot e_{0}. 
\label{eq:vecplane}
\end{equation}

Based on all the previously defined elements, the vectors involved in the calculation of joint angles $\theta_{k}$, for $k=0,2,3$, are summarized in Table~\ref{tab:tab2}. There, $\alpha$ and $\beta$ in (\ref{eq:n}) and (\ref{eq:angle}) are replaced by $\alpha_{k}$ and $\beta_{k}$, respectively, to calculate $\theta_{k}$. Please note that, as joint $J_{1}$ is fixed, $\theta_{1}$ does not need to be calculated. 

\subsection{Artificial Vision Algorithm}
An AV algorithm was implemented to calculate the positions of items on a table, so the robotic arm could 
perform the desired manipulation tasks. 
An ATW-1200 Acteck web camera was used to record images at 30 fps with a resolution of $640\times 480$ pixels. The acquired images 
were processed and analyzed in real-time using the OpenCV library (\url{www.opencv.org}) from Python. 

\begin{figure}
\centerline{\includegraphics[width=0.75\textwidth]{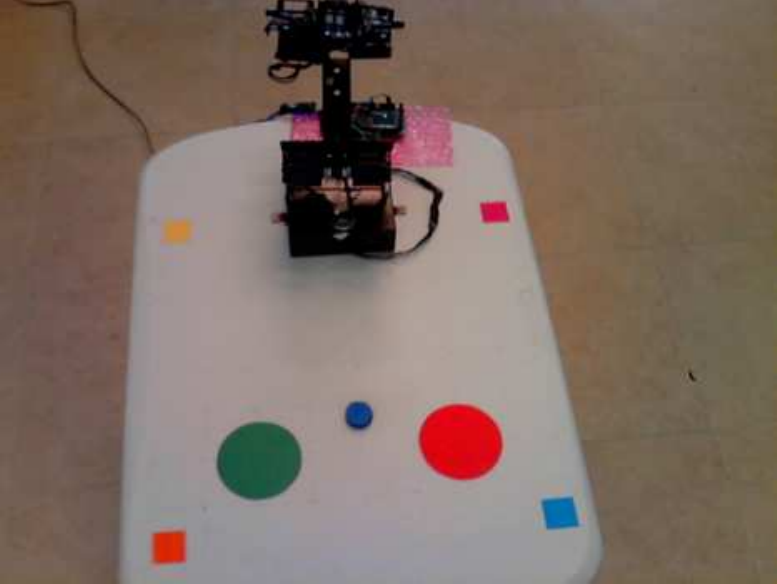}}
\caption{Robot and items on the table as seen by the camera during semi-autonomous BCI trials.}
\label{fig:view}
\end{figure}

The robotic arm was fixed on a white table, centered at one end of it. A plane was delimited on the table, defined as a $400\times 400$~mm square. Four $30\times 30$~mm square markers of different colors (cyan, orange, magenta and yellow) were placed inside the delimited square, one at each corner. A blue disk with height of $6$~mm and radius of $13$~mm was used as the item to be picked, while two stickers with radius of $42$~mm  (green and red) were used to indicate target areas. The camera was fixed in a high angle, so that all markers and items were inside its field of view. The setup of the robotic arm and items in the table are shown in Figure~\ref{fig:view}. 

In order to perform object manipulation tasks, the \emph{real-world} coordinates of the plane (in reference to robot frame) had to be obtained from the \emph{image} coordinates obtained by the camera. To achieve this, a homography transformation was performed on the acquired images. In general, a two-dimensional point $(u,v)$ in an image can be represented as a three-dimensional vector $(x,y,z)$ by letting $u=x/z$ and $v=y/z$. This is called the homogeneous representation of a point and it lies on the projective plane $P^{2}$~\cite{Dubrofsky09}. A homography is an invertible mapping of points and lines on the projective plane $P^{2}$, thus allowing to obtain the real-world coordinates of features in an image from its image coordinates. 

In our case, the desired transformation is such 
the image obtained from the camera is turned into a two-dimensional view of the same setup. In this transformation, the image shows a planar representation of the original view, as if the camera was placed directly above the delimited square. In order to obtain this representation, the following homography transformation was applied~\cite{Dubrofsky09}:
\begin{equation}
\begin{bmatrix}u \\ v \\ \end{bmatrix} = H \begin{bmatrix}x \\ y \\ \end{bmatrix},
\label{eq:homo}
\end{equation}
where vectors $[u \hspace{0.2cm} v]^{T}$ and $[x \hspace{0.2cm} y]^{T}$ represent the positions of selected points in the image and their corresponding positions in real-world coordinates, respectively, $H=K[R|t]$ is the homography matrix that defines the desired perspective change to be performed on the image, $K$ is the calibration matrix which contains the intrinsic parameters of the camera, while $R$ and $t$ are, respectively, the rotation matrix and translation vector applied on the camera in order to perform this transformation view. In (\ref{eq:homo}), $z$ is ignored as all items are considered to be at $z=0$.

\begin{figure}
	\centering
         \subfigure[Segmentation and binarization]{\includegraphics[width=0.5\textwidth]{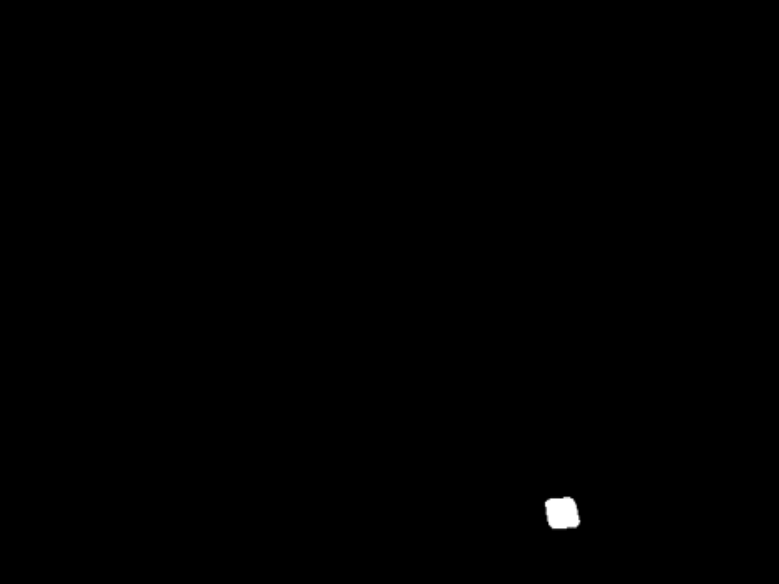}}
	 \subfigure[Centroid calculation]{\includegraphics[width=0.5\textwidth]{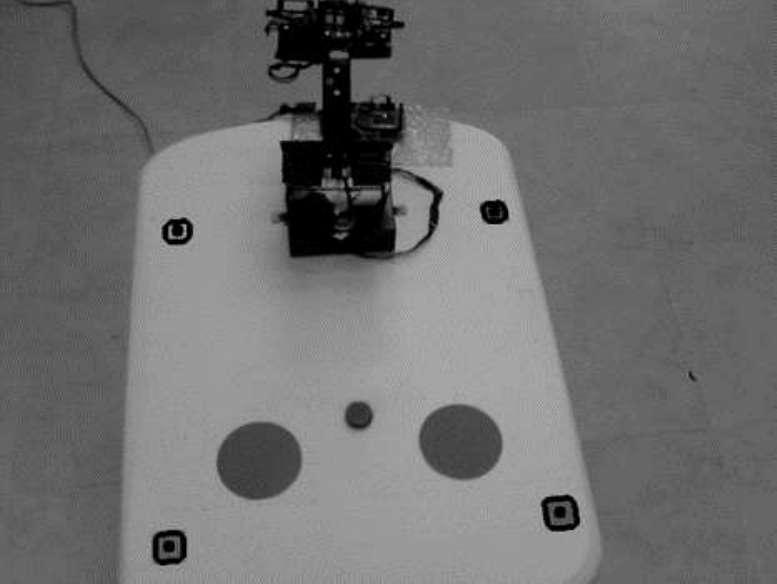}}
	 \subfigure[Homography transformation]{\includegraphics[width=0.5\textwidth]{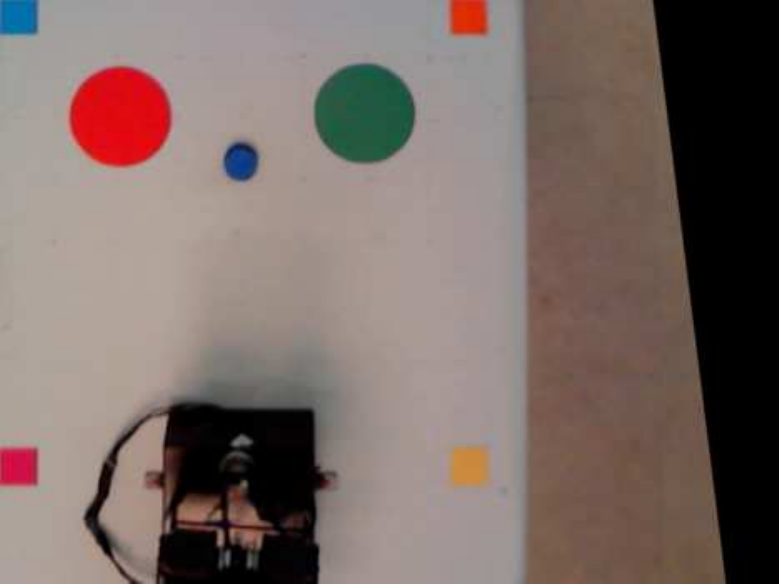}}
	\caption{Required steps of the AV algorithm.}
\label{fig:avprocess}
\end{figure}

In order to compute matrix $H$, both real-world and image coordinates of the centroids of the square markers were obtained. First, markers were detected through color segmentation and binarization, as shown in Figure~\ref{fig:avprocess}(a). This process was performed separately on each marker and their contours were detected. After that, the centroids of the markers in the image were calculated. The contours and centroids of each marker are shown in Figure~\ref{fig:avprocess}(b). 

 Since the markers have known dimensions ($30\times 30$~mm), the positions of their centroids in real-world coordinates relative to the plane are known as well. These positions were defined as: cyan at $[15, 15]$~mm, orange at $[385, 15]$~mm, magenta at $[15, 385]$~mm, and yellow at $[385, 385]$~mm, all inside the $400\times 400$~mm area available of the table. Then, both sets of coordinates are used to obtain $H$ with OpenCV's command \emph{findHomography}, and the resulting matrix is applied to transform the image, as shown in Figure~\ref{fig:avprocess}(c).

\begin{figure}
\centering
\includegraphics[width=0.75\textwidth]{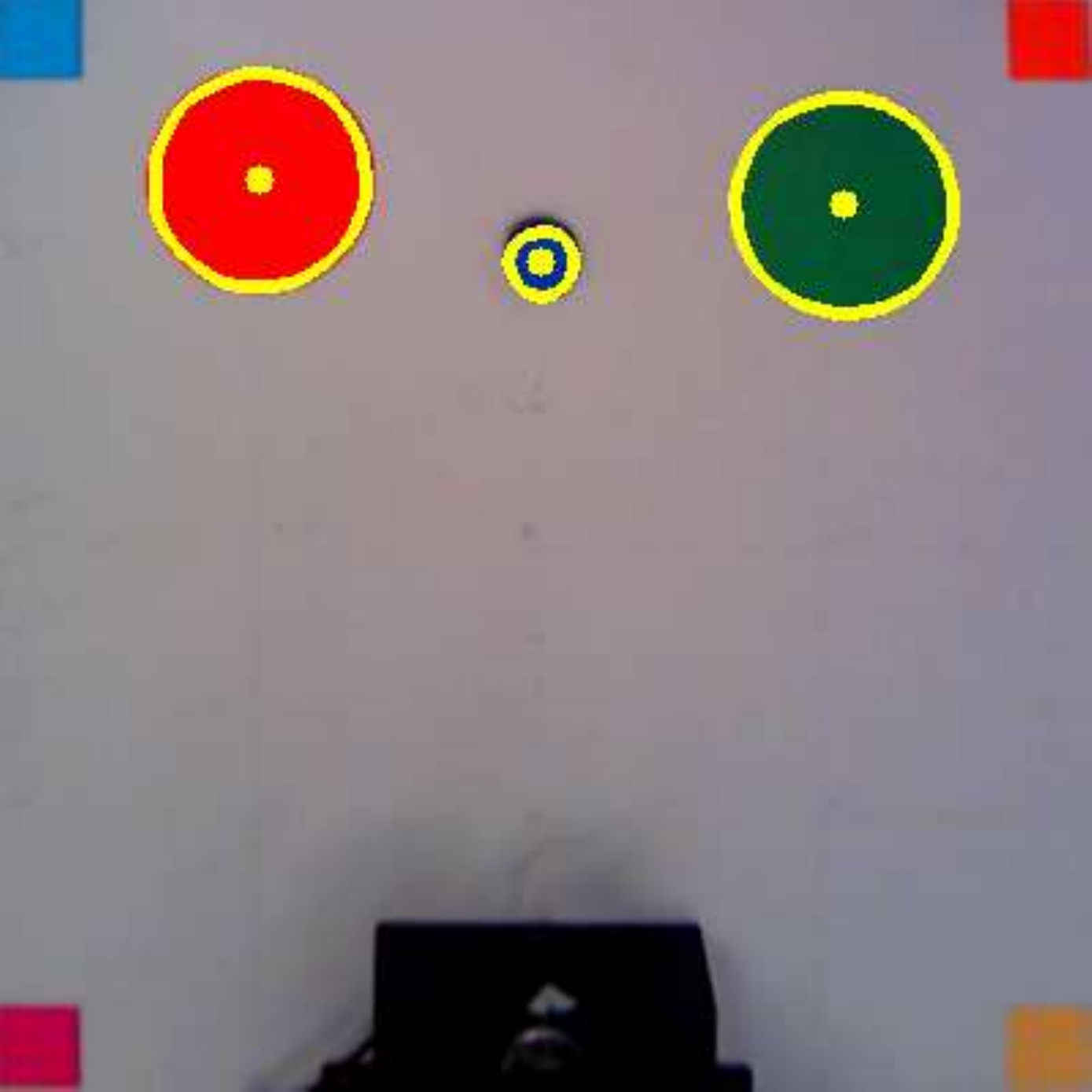}
 \caption{Visual representation of  contours and centroids of the items in the table, calculated by the AV algorithm in order to obtain their real-world coordinates.}
\label{fig:homo}
\end{figure}

Then, using the same procedure as with the markers, the centroids of the disk and targets in the new image were calculated. However, the reference frame from the image is different from the reference frame from the robot. Therefore, the first was transformed by applying the following rotation matrix:
\begin{equation}
R = \begin{bmatrix} \cos\pi & -\sin\pi \\ 
\sin\pi & \cos\pi \\ \end{bmatrix}.
\label{eq:rotation}
\end{equation}
Furthermore, a translation vector $[-200 \hspace{0.2cm} -400]^{T}$ was applied, as well as a sign switching of the \emph{x} axis to obtain the desired positions.  In robot frame, the \emph{x} axis of the delimited square goes from $-20$ to $20$, while the \emph{y} axis goes from $0$ to $40$, and the robot is located at the origin. After applying all those transformations, the centroids of all items are finally expressed in robot frame, and they can be detected by the AV system together with the contours of all items. This is shown in Figure~\ref{fig:homo}.

\section{Implementation of BCI systems}\label{sec:bci_des}
As proof-of-concept, four participants volunteered in this study (two female and two males, with average age of $22.25$ years, SD$=\pm 0.95$). The experimental protocol was divided in three stages for both the process-control and goal-selection BCIs: (i) training, (ii) cued manipulation, and (iii) uncued manipulation. 
Both BCIs were MI-based, therefore users were trained to control the corresponding $\mu$ band desynchronization at will.  In all trials, volunteers sat in front of a computer screen first showing a black screen (baseline) in which the user was meant to be in a resting state. Then, different types of stimulus were presented to the user, representing each a different command. The duration for the baseline ($15$~seconds) and stimulus presentation ($4$~seconds) was the same for all trials and stages. During stimulus presentation,  users were expected to react accordingly, either by imagining the movement of either left or right hand, or to remain in a resting state. In training trials, EEG signals were acquired and analyzed off-line to build and evaluate the performance of classifiers, which were then used on-line during the manipulation trials. In cued manipulation trials, the user was expected to manipulate the device as indicated by the stimuli. On the other hand, the user was encouraged to manipulate the device at will during uncued manipulation trials. 

\subsection{Training trials}\label{subsec:trainingtrials}
The training protocol was identical for both the process-control and goal-selection BCIs. Three types of stimuli were presented to the user: \emph{right hand imaginary movement} (RHIM), \emph{left hand imaginary movement} (LHIM),  and \emph{rest}. A total of $30$ stimuli ($10$ for each command) were randomly presented to the user.  Stimuli were represented in the computer screen with a red arrow pointing to the right (for RHIM), pointing to the left (for LHIM), and a black screen for rest. A $2$-seconds green cross appeared before all stimuli as a pre-stimulus, and there was a variable inter-stimulus resting period of $2$-$4$ seconds between stimuli. Users underwent three training sessions on different days, each comprised by five repetitions of the mentioned experimental protocol, while EEG recordings were obtained. 

\subsection{Signal acquisition}
EEG signals were recorded with the Mobita equipment from TMSi systems, using a measuring cap of $19$-channels: FP$1$, FP$2$, F$3$, F$4$, C$3$, C$4$, P$3$, P$4$, O$1$, O$2$, F$7$, F$8$, T$3$, T$4$, T$5$, T$6$, Cz, Fz, and Pz. Impedance of all electrodes was set below $5$~k$\Omega$ for all experiments. Signals were acquired with a sampling frequency of $1000$~Hz. Recordings were band-pass filtered with a fourth order $1$-$100$~Hz Butterworth, and a $60$~Hz Notch filter to eliminate power line interference. The OpenVibe software was used for the BCI design and implementation. More information about this software can be found in~\cite{Renard10}. 

\subsection{Classification algorithm}\label{methclas}
\begin{figure}
  \centering
  \includegraphics[width=0.75\textwidth]{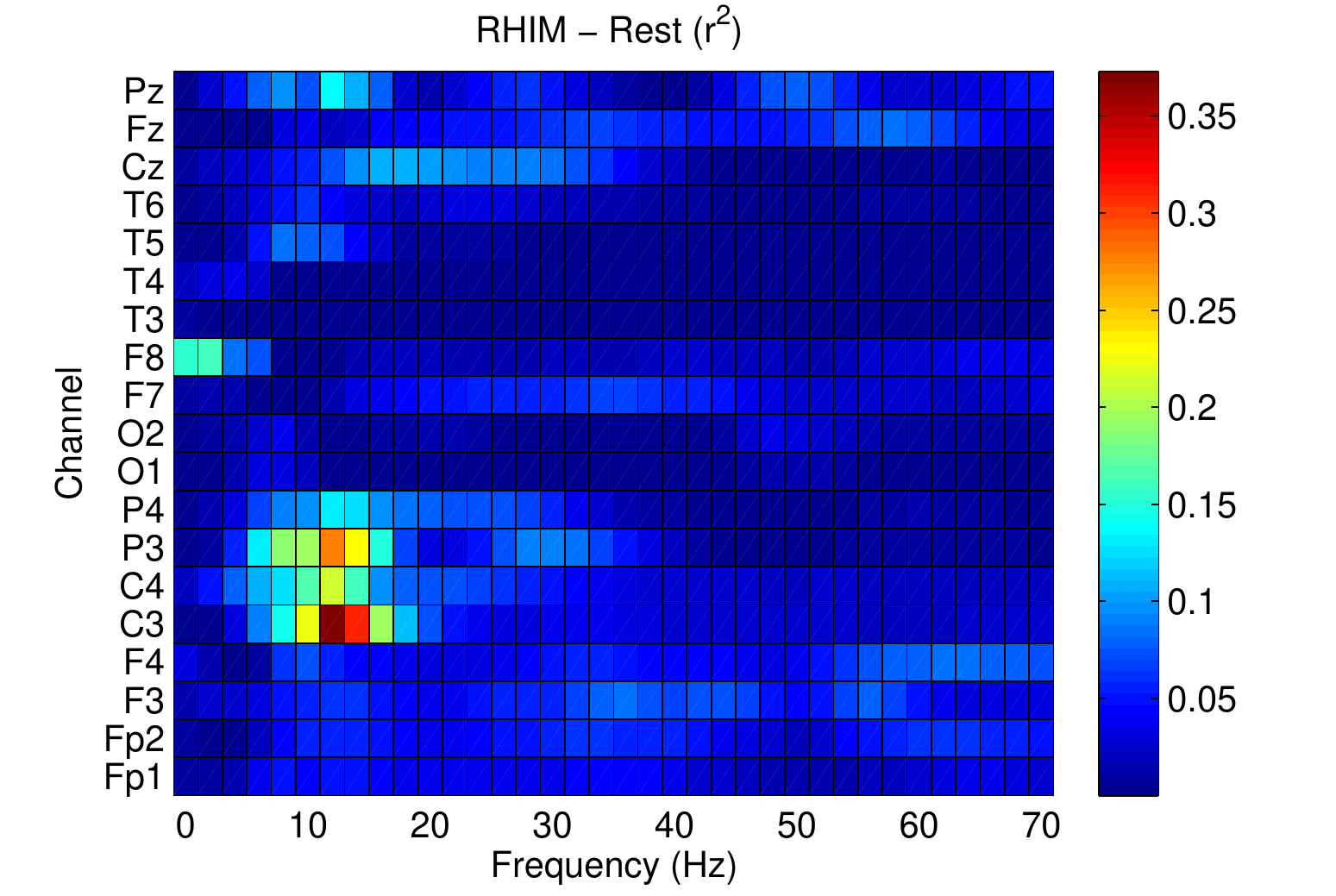}
\caption{Representative $r^{2}$ map obtained during one training session. $r^{2}$ values here shown were measured under conditions RHIM-Rest for all channels and frequencies.}
\label{fig:alma3}
\end{figure}

Feature extraction was performed using the BCI2000 offline analysis tool (\url{https://www.bci2000.org/mediawiki/index.php/User_Reference:BCI2000_Offline_Analysis}), where the $r^{2}$ value calculated.  A higher $r^{2}$ value is related to a higher discrimination of a signal under two stimulus conditions. More details about the statistic meaning of $r^{2}$ can be found at \url{https://www.bci2000.org/mediawiki/index.php/Glossary}. After each training session, signals from the five training trials were used to calculate $r^{2}$. Three $r^{2}$ maps (one per stimulus combination) were obtained per training session, showing the $r^{2}$ values in the $19$ available channels and frequencies ranging from $1$ to $70$~Hz. Each map, as the one shown in Figure~\ref{fig:alma3}, represents the channels and frequencies which, for a specific combination of conditions, showed higher discrimination. Through this procedure, the selected channels and frequencies were used as features for the classification algorithm. 

Signals were spatially filtered using a Laplacian filter on the selected channels, as well as through a fourth-order Butterworth band-pass filter tuned to the selected frequencies. Power values were then obtained from the filtered signals to build the feature vectors, which then became the input for a linear discriminant analysis (LDA) classifier, which separates data representing different classes by finding a hyperplane which maximizes the distances between the means of the classes, while minimizing the variance within classes~\cite{Lotte07}. 

In our case, three pair-wise classifiers per training session were obtained using this procedure: LHIM versus RHIM, LHIM versus rest, and RHIM versus rest. The three classifiers were tested online on the recorded signals to evaluate their performance as a percentage of correctly classified stimuli. The classification was performed on each four-seconds stimulus, divided into overlapped sub-epochs using a window function. Each four-seconds epoch was formed by $64$ sub-epochs of two seconds, separated by $0.0625$ seconds. One pair-wise classifier labeled each sub-block as one of two possible classes, and the four-seconds epoch was classified as the mode of the classification result for all its sub-blocks. Then, one general classifier  was built, based on the results of the three pair-wise classifiers. Here, the four-seconds epoch of each stimuli was labeled as class $I=1,2,$ or $3$ (LHIM, rest, or RHIM, respectively), if two out of the three pair-wise classifiers labeled the same epoch identically. The mean performance of the general classifiers across trials are shown in Table \ref{tab:tab4} for all subjects and training sessions, as well as their selected features. After training sessions, each user proceeded to perform the subsequent trials using the classifier with the highest performance obtained at the last training session. 

\subsection{Process-control BCI}\label{subsec:lowlev}
The process-control BCI was designed in such manner that users were able to perform three-dimensional movements to complete reaching tasks. In this system, the position of the final effector as well as the desired axis in which the effector moves, can be controlled through low-level MI-based commands.  To achieve this, the user has two choices: moving along a selected axis (\emph{y}-axis at the initial step) or change between axes.  In the design of this BCI, the classification of a LHIM results in a -$10$~mm displacement, while the classification of a RHIM results in a +$10$~mm  displacement on the selected axis. The classification of a rest event holds the position of the final effector with no displacement.  The consecutive classification of two rest events in a row allowed the user for a change of axis. This change of axis takes place in the following sequence: $y\rightarrow z$, $z\rightarrow x$, and $x\rightarrow y$.

\subsubsection{Cued manipulation}
\begin{figure}
\centering
\includegraphics[width=0.75\textwidth]{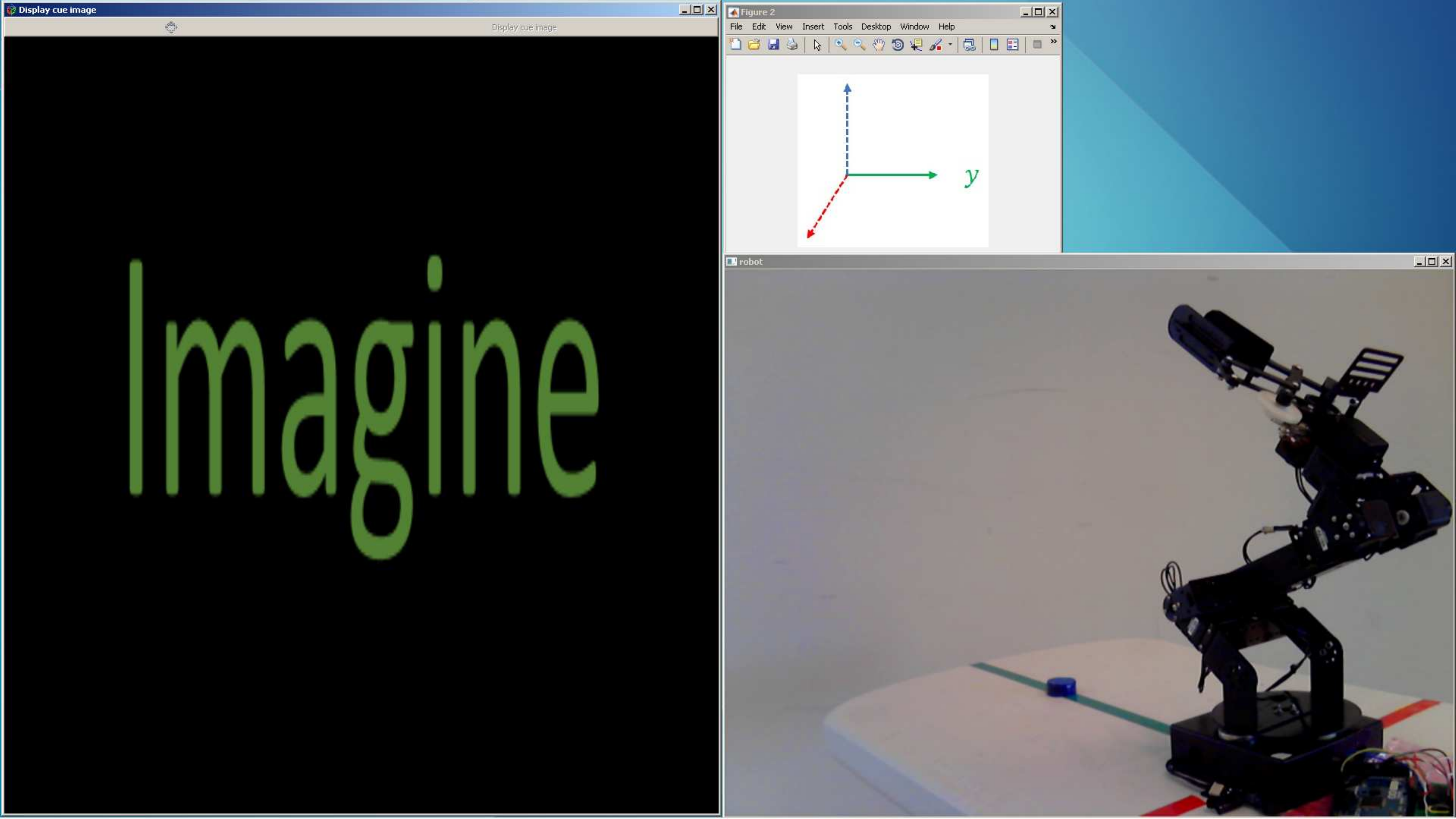}
 \caption{Setup of the process-control BCI. The windows shown in the screen are used for visualization of stimuli, indicating the current axis of the movement, and viewing of the robot performing the manipulation tasks.}
\label{fig:mont_convencional}
\end{figure}

In these trials, users sat in front of a computer showing three windows on the screen. The first window was used for stimulus presentation, the second was used to display in which axis the movement of the robot took place, and the third was used visualize the robot and its movements. The setup for these experiments is shown in Figure~\ref{fig:mont_convencional}. After the baseline period, $15$ random stimuli ($5$ for each type) were presented to the user. Pre-stimulus, stimulus, and inter-stimulus duration were the same as in training trials (see Section~\ref{subsec:trainingtrials}). After the stimulus was presented, the user was expected to emit the instructed command through the BCI. Then, the robot performed a specific movement based on the classification result. In these trials, performance was evaluated as the percentage of correctly classified stimuli. The intention of these trials was to get the users acquainted to the BCI, and they were performed immediately before the uncued manipulation trials. Users performed three sessions on different days, each formed by three repetitions of this protocol.  

\subsubsection{Uncued manipulation}
The same screen display was used as in cued trials, but here subjects were asked to complete reaching tasks on their own. At the start of each trial, the final effector was fixed at home position $[0, 155.5, 284.3]$ and a target was placed at $[0, 300, $-$49]$. At this initial step, the distance of the final effector to the target was $360$~mm. Note that the target is placed at $z=$-$49$, as the robot base is $49$~mm above the table. A baseline period was followed by the presentation of $20$ stimuli showing the word ``\verb|Imagine|'', in which the user was expected to emit MI commands through the BCI. The duration of pre-stimulus, stimulus, and inter-stimulus periods were the same as in training trials (see Section~\ref{subsec:trainingtrials}). The user was instructed to move the final effector as close as possible to the target within the $20$ stimuli, using the protocol described in Section~\ref{subsec:lowlev}. Performance was evaluated as the percentage of stimuli where: the user moved the final effector closer to the target, and changed successfully to the \emph{y}-axis. Users performed three sessions on different days, each formed by five repetitions of the described protocol. 

\subsection{Goal-selection BCI}\label{subsec:semiauto}
The goal-selection BCI was designed to perform in a semi-autonomous way pick-and-place tasks with the disk and two possible targets. Users were able to perform these tasks for any position of the items (randomly chosen before a trial), inside of the robot workspace. The centroids $C=[C_{x}, C_{y}]$ of the two target stickers were calculated in these trials by the AV algorithm. In this case, the classification of three types of events resulted in different manipulation tasks: 
\begin{itemize}
\item if an event was classified as RHIM, the robot reached for the disk, placed it on the target located to the right (greater $C_{x}$ component), and returned to home position;
\item if an event was classified as LHIM, the robot reached for the disk, placed it on the target located to the left (smaller $C_{x}$ component), and returned to home position;
\item if an event was classified as rest, the robot remained at home position. 
\end{itemize}
After the robot performed a manipulation task, all the items in the table were manually changed to random positions, in preparation for the next trial. 

\subsubsection{Cued manipulation trials}\label{subsubsec:cmt}
\begin{figure}
\centering
\includegraphics[width=0.75\textwidth]{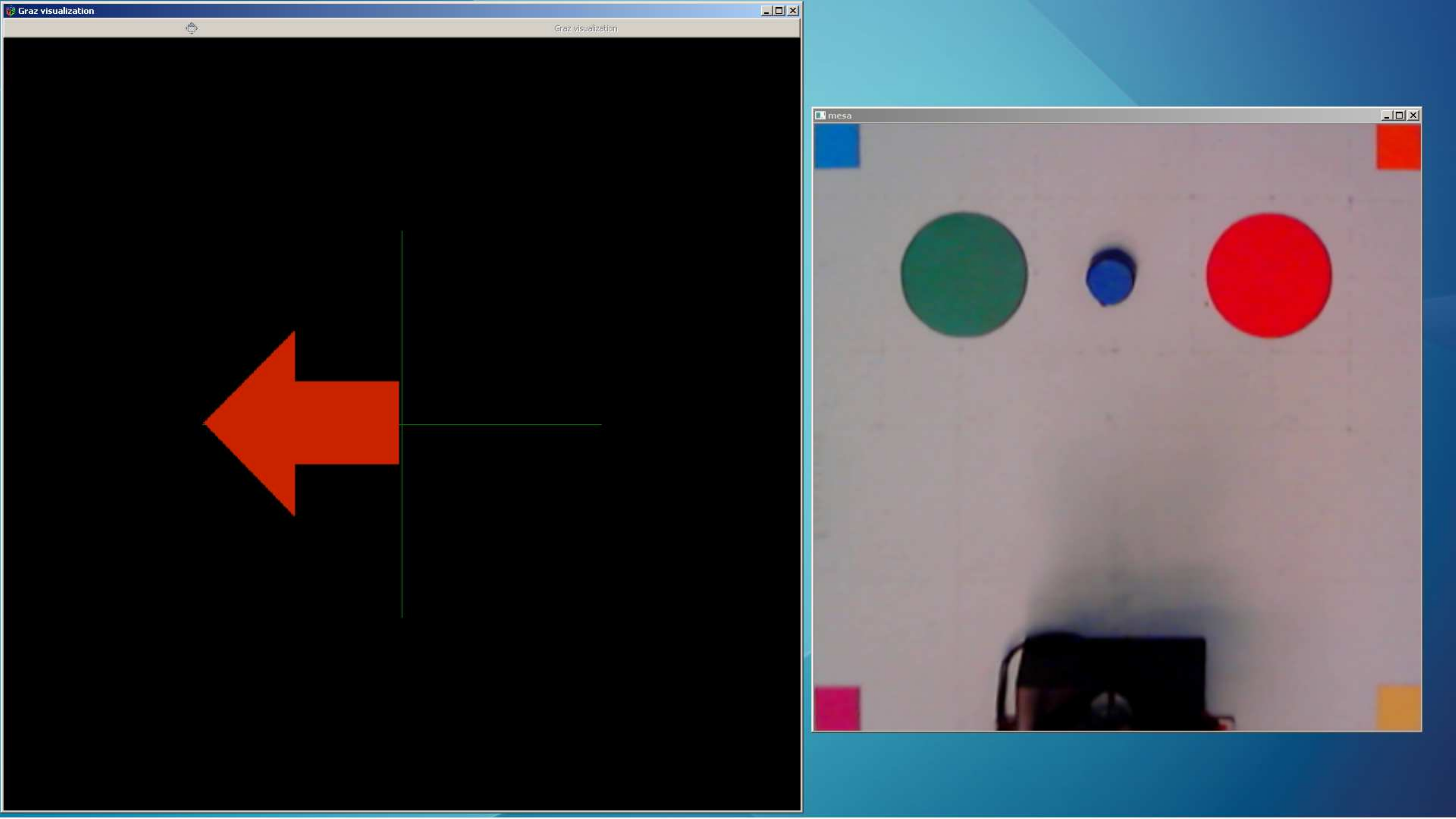}
 \caption{Setup of the semi-autonomous goal-selection BCI, as seen by the user. The windows are used for stimulus presentation and visualization of the manipulation tasks.}
\label{fig:mont_semiauto}
\end{figure}

In these trials, the subject sat in front of a computer screen which showed two screens. The first one was used for stimulus presentation, while the second was used to present the transformed image, as shown in Figure~\ref{fig:mont_semiauto}.  After the baseline period, a stimulus (RHIM, LHIM, or rest) was randomly presented. A total of $15$ stimuli ($5$ for each type) were presented in each trial. A one-second \emph{beep} sound followed a two-seconds green cross as pre-stimulus, with a $27$-$29$ seconds inter-stimulus period.  Manipulation tasks were performed according to the result of the classification and performance was evaluated as the percentage of correctly classified stimuli. The total duration of these trials was considerably longer than in the low-level BCI. This is mainly due to the longer inter-stimulus period, in which the manipulation tasks took place.  Users underwent three sessions on different days, performing  five trials in each session.

\subsubsection{Uncued manipulation trials}
For uncued manipulation trials, all stimuli were replaced with the word ``\verb|Imagine|'', and the user freely decided the task to perform, as explained in Section~\ref{subsec:semiauto}. A total of $15$ stimuli were presented in each trial. The stimulus, pre-stimulus, and inter-stimulus duration were the same as in the goal-selection BCI cued manipulation trials (see Section~\ref{subsubsec:cmt}). Immediately after the classification was performed, and before the robot executed the task, the user was asked the type of intended stimulus to emit. In these trials, performance was evaluated as the percentage of coincidences between the intended  and the classified stimulus type.   

\subsection{Analysis of data through P300 estimation}

\begin{figure}
\centering
\includegraphics[width=0.75\textwidth]{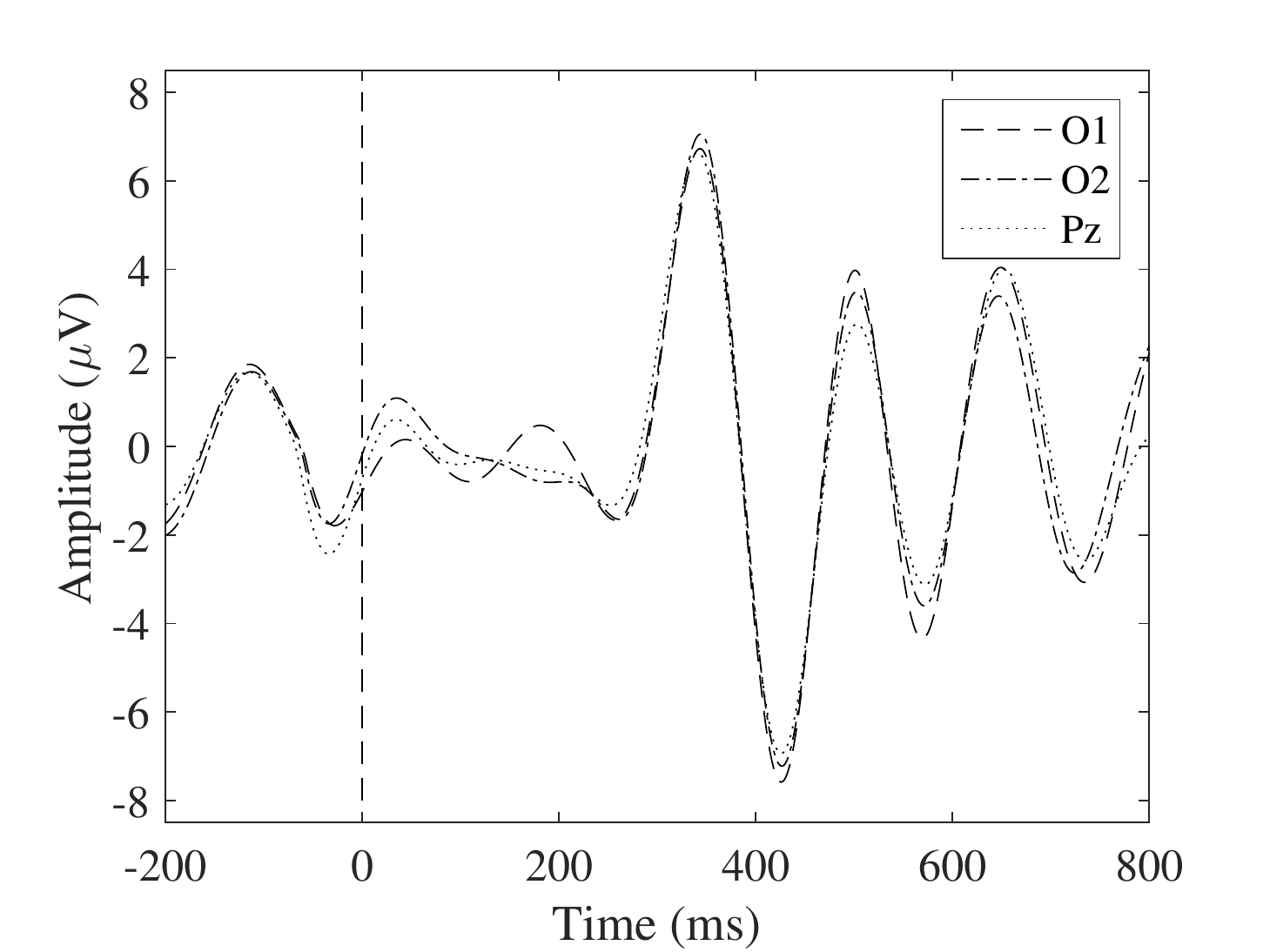}
 \caption{Representation of a P$300$ waveform calculated for channels O$1$, O$2$ and Pz.}
\label{fig:p300}
\end{figure}

Reported assessments of mental fatigue through P$300$ amplitude and latency can be found in~\cite{Uetake00}, and~\cite{Cheng11}. In~\cite{Cheng11}, mental fatigue was evaluated through EEG measurements. Participants' P$300$ were measured during a modified Eriksen flanker task, replacing word stimuli with arrows, before and after performing mental arithmetic tasks. A decreased P$300$ amplitude and an increased latency were observed after performing arithmetic tasks, when users were mentally fatigued. Statistical analysis revealed the most significant changes in amplitude and latency at channels  O$1$, O$2$ and Pz, probably as a reflection of visual processing during stimulus presentation of arrows.  Similar to the protocol used in~\cite{Cheng11} to assess mental fatigue,  signals were segmented into $1$-s stimulus-locked EEG epochs from $200$~ms before and $800$~ms after stimulus presentation. These epochs were obtained for the presentation of the word  ``\verb|Imagine|'' during uncued manipulation trials for both the process-control and goal-selection BCI. For each trial, a representative waveform was obtained by averaging the epochs from all stimuli. Then, the averaged waveforms were  band-pass filtered at $1$-$10$~Hz and were used to calculate P$300$ amplitude and latency. The amplitude was considered as the most positive peak within a $200$-$500$~ms window immediately after stimulus presentation. Latency was obtained as the time this peak appeared. Amplitude and latency values were obtained through this procedure for all trials, sessions and subjects, in channels O$1$, O$2$ and Pz.  A representation of an obtained P$300$ waveform is shown in Figure \ref{fig:p300} for these three channels.  

In order to examine the differences of mental fatigue within and between-users in relationship with the use of our two different BCI schemes, two-way ANOVA tests were performed on all users
: one for amplitude and one for latency. In this tests, influence of \emph{trial repetition} ($1$-$5$), \emph{channel location} (O$1$, O$2$ and Pz), and their interaction were anayzed on both P$300$ features. The number of replications was considered as three, representing the three uncued manipulation sessions performed by the users. To further analyze mental fatigue related to continuous BCI manipulation, one-way ANOVA ($p<0.05$) tests were performed on each subject. Six one-way ANOVA tests were performed per subject: three channels (O$1$, O$2$ and Pz) $\times$ two P$300$ features (amplitude and latency). These tests were performed in order to find which channel showed significant relationship to the trial repetition factor. Then, amplitude and latency values of all users were compared using the most significant channel from this analysis.

\section{Results}\label{sec:Res}
A preliminary validation of our CGA model and AV algorithm can be found in~\cite{Ramirez18} and~\cite{Ramirez19}, respectively, hence we omit those details here. Therefore, this section shows the results of evaluating the whole system in the context of our BCI implementations for four subjects (two on each BCI type). Performance values were obtained for all subjects in training, cued and uncued trials, according to the particularities of each experimental protocol. For training trials, performance values correspond to the classifier accuracies  shown in Table \ref{tab:tab4}. Performance for cued and uncued manipulation trials were obtained as explained in Sections~\ref{subsec:lowlev} and~\ref{subsec:semiauto}. Performance values included in these results represent the average across trials for each session.  

\subsection{Performance of process-control BCI}
\begin{figure}
	\centering
	\subfigure[$S_{1}$, Training.]{\includegraphics[width=0.3\textwidth]{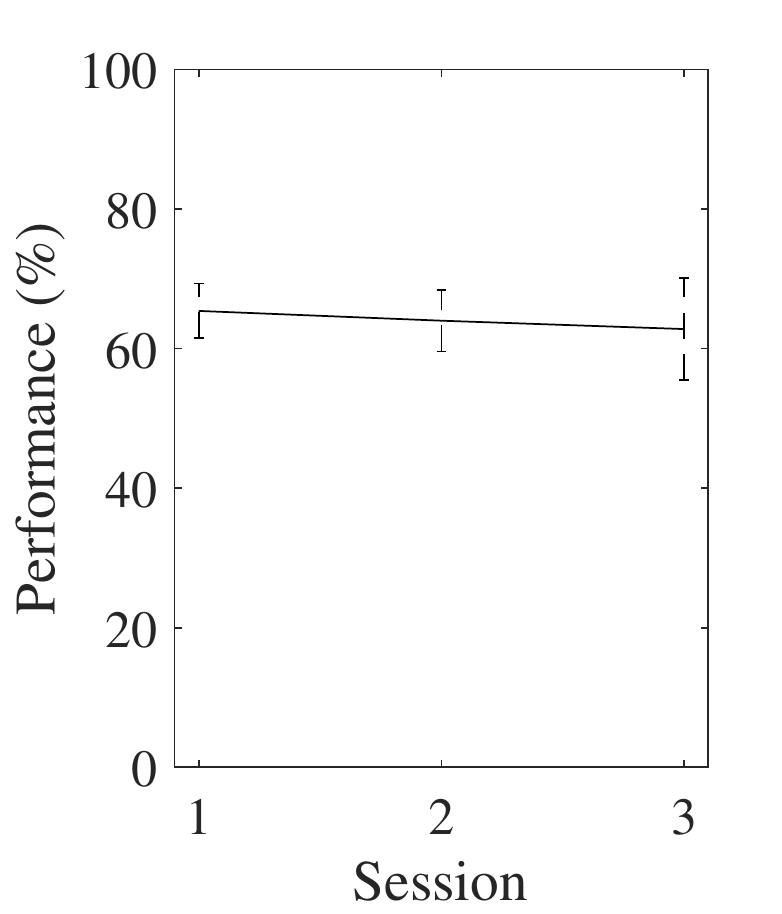}}
           \subfigure[$S_{1}$, Cued manipulation.]{\includegraphics[width=0.3\textwidth]{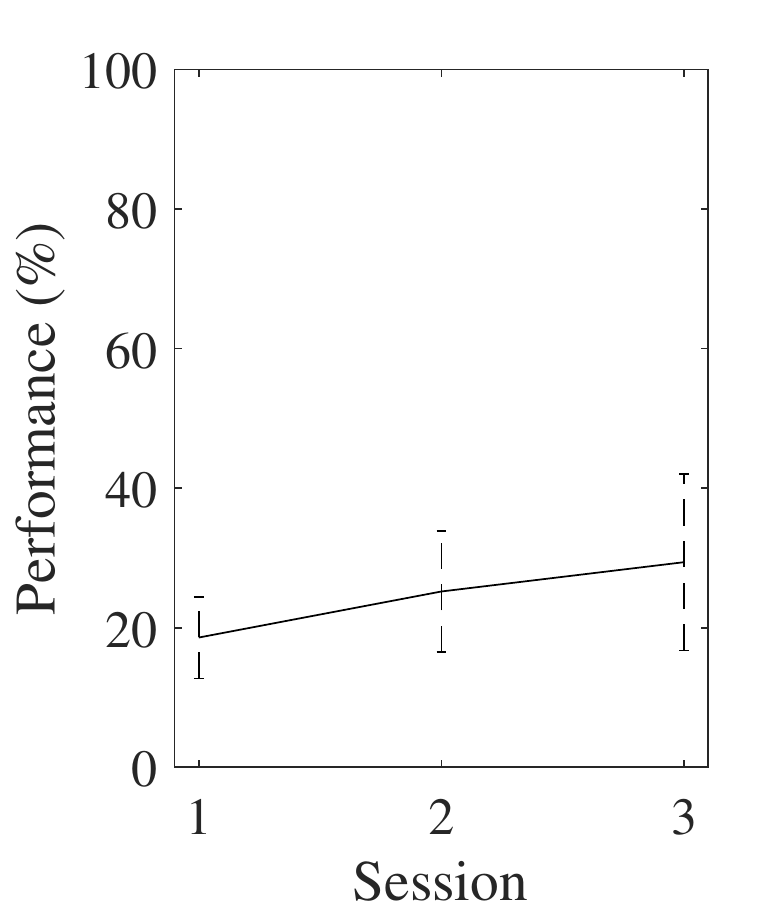}}
	\subfigure[$S_{1}$, Uncued manipulation.]{\includegraphics[width=0.3\textwidth]{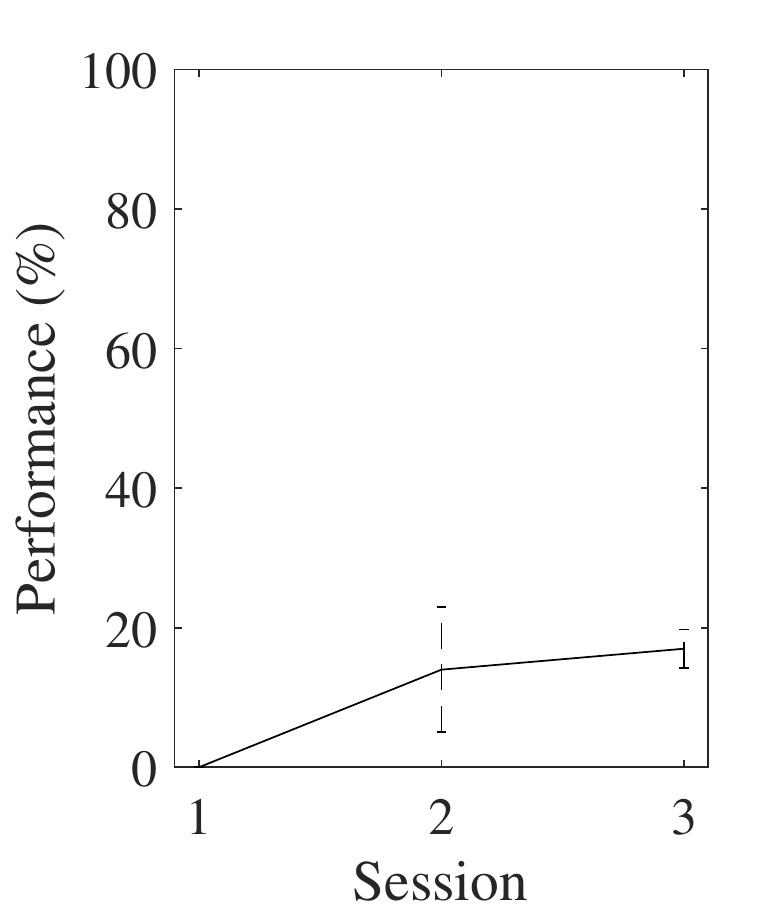}}
	 \\
	\subfigure[$S_{2}$, Training.]{\includegraphics[width=0.3\textwidth]{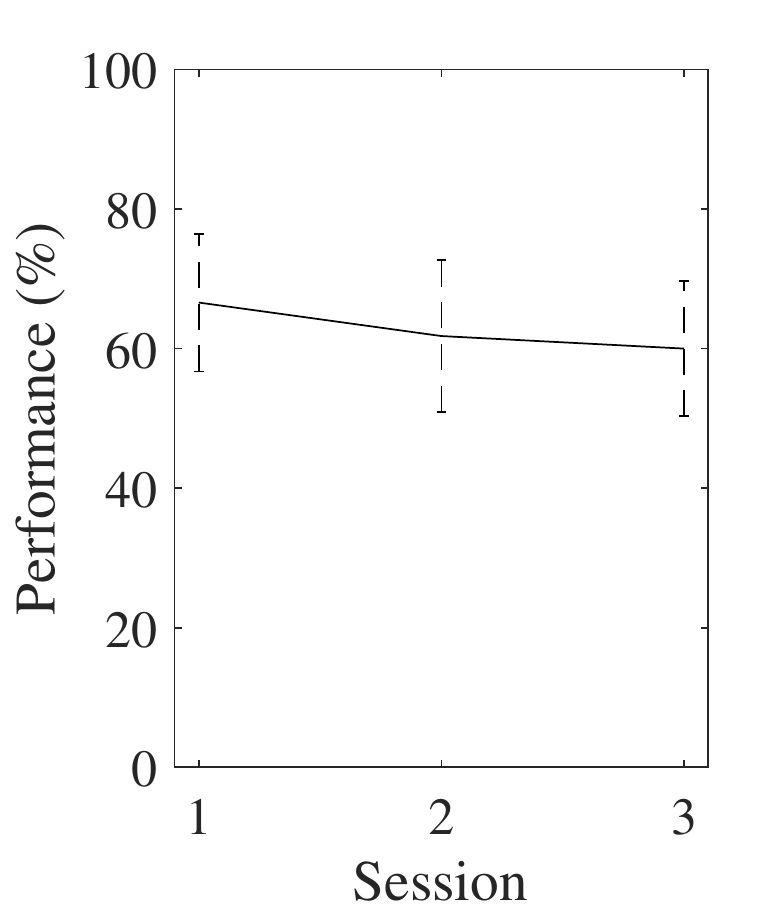}}
           \subfigure[$S_{2}$, Cued manipulation.]{\includegraphics[width=0.3\textwidth]{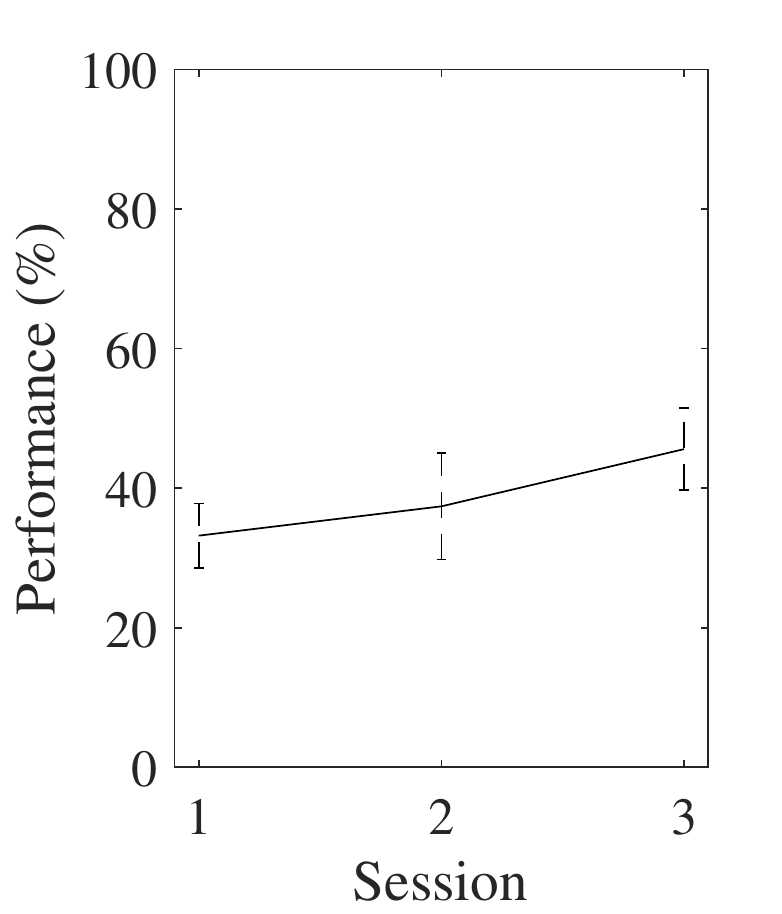}}
           \subfigure[$S_{2}$, Uncued manipulation.]{\includegraphics[width=0.3\textwidth]{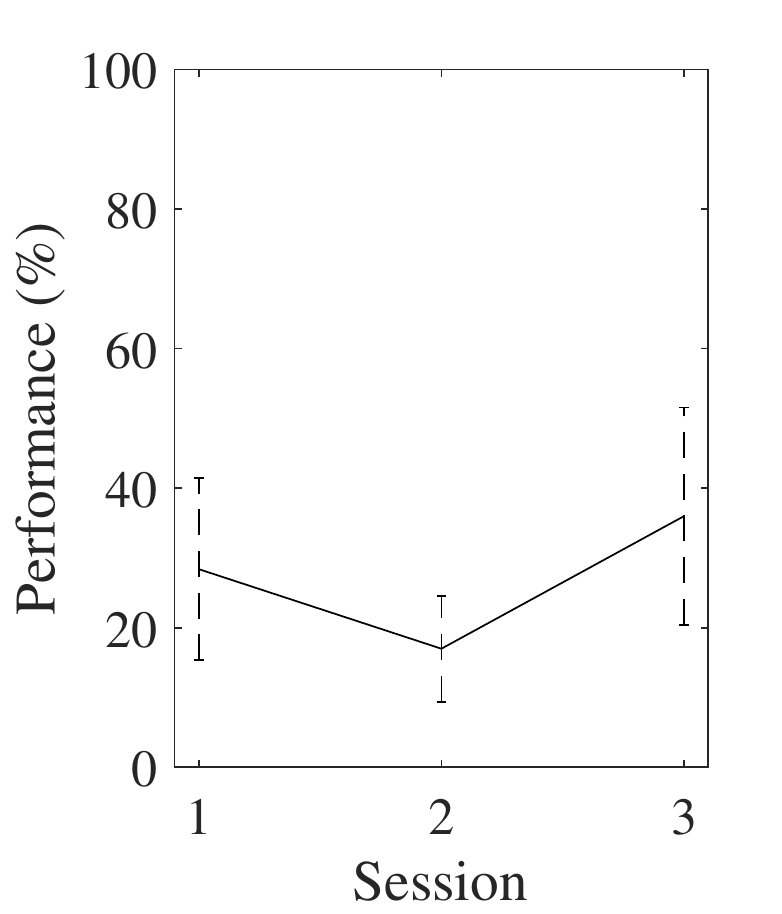}}
	\caption{Performance for Users~$S_{1}$ (top) and $S_{2}$ (bottom) in process-control BCI during training,  cued, and  uncued  manipulation trials (left, middle and right columns respectively). Bars indicate one standard deviation.}
\label{fig:S1S2_per}
\end{figure}

Subject~$S_{1}$ reached an accuracy level of $65 \%$ at its first training session, $64\%$ at the second, and  $63 \%$  by the third.  During cued manipulation trials, performance started at $18 \%$, then increased to $25\%$ and  $29\%$ by the second and thirds sessions respectively. For uncued manipulation trials, the user only moved far from the target at the first  session ($0\%$). For the second and third sessions, User~$S_{1}$ obtained performances of $14\%$ and $17\%$. Subject~$S_{2}$ showed a similar behavior to $S_{1}$ during training trials, starting at $65\%$, and decreasing to $62\%$ and $60\%$ by the second and third sessions. In cued manipulation trials, performance started at $33 \%$, then increased  to $37 \%$ by the second session, and $45 \%$ by the third. For  uncued manipulation trials, performance values started at $28\%$ for the first session, then decreased to $17\%$ at the second, and increased to $37\%$ by the third. Results for process-control BCI performance are shown in Figure~\ref{fig:S1S2_per} for Users~$S_{1}$ and $S_{2}$, respectively.\\
 
\subsection{Performance of goal-selection BCI}
\begin{figure}
	\centering
	\subfigure[$S_{3}$, Training.]{\includegraphics[width=0.3\textwidth]{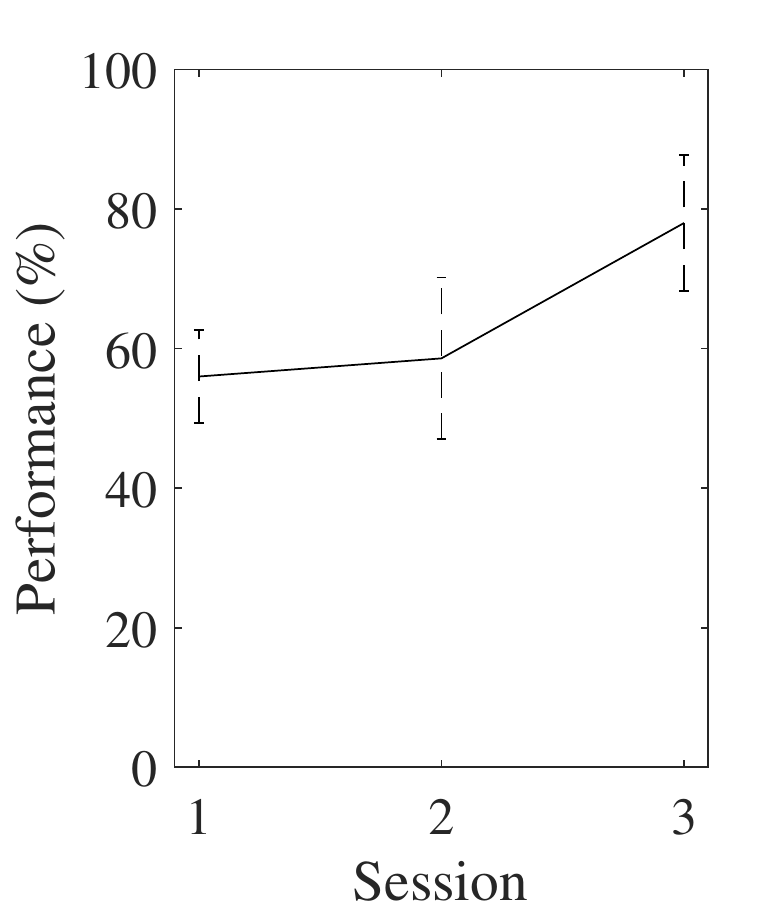}}
           \subfigure[$S_{3}$, Cued manipulation.]{\includegraphics[width=0.3\textwidth]{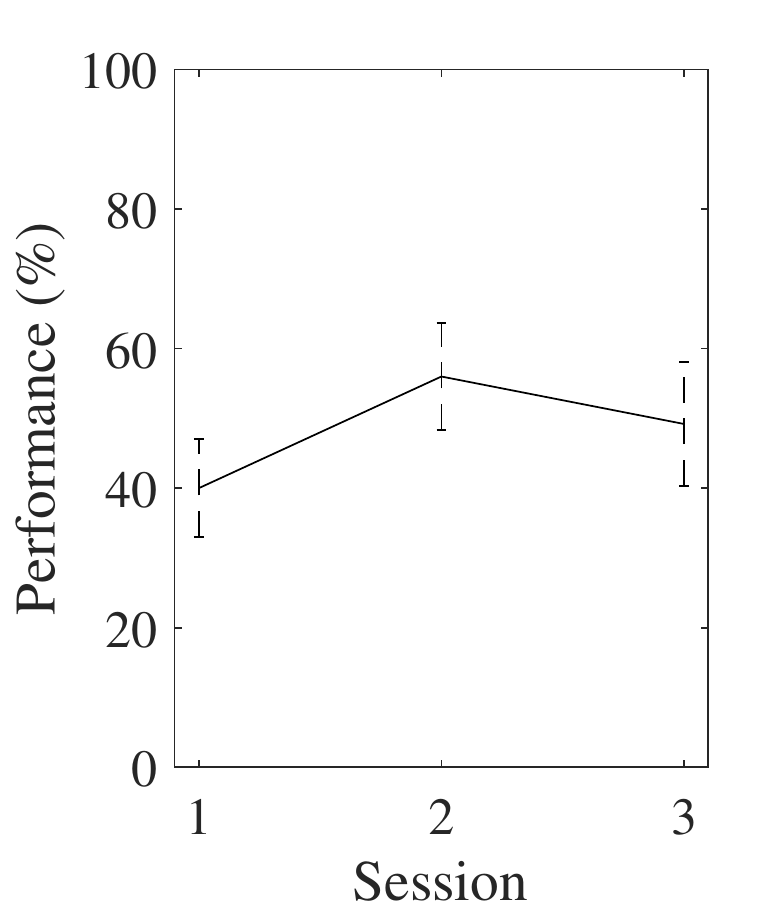}}
	\subfigure[$S_{3}$, Uncued manipulation.]{\includegraphics[width=0.3\textwidth]{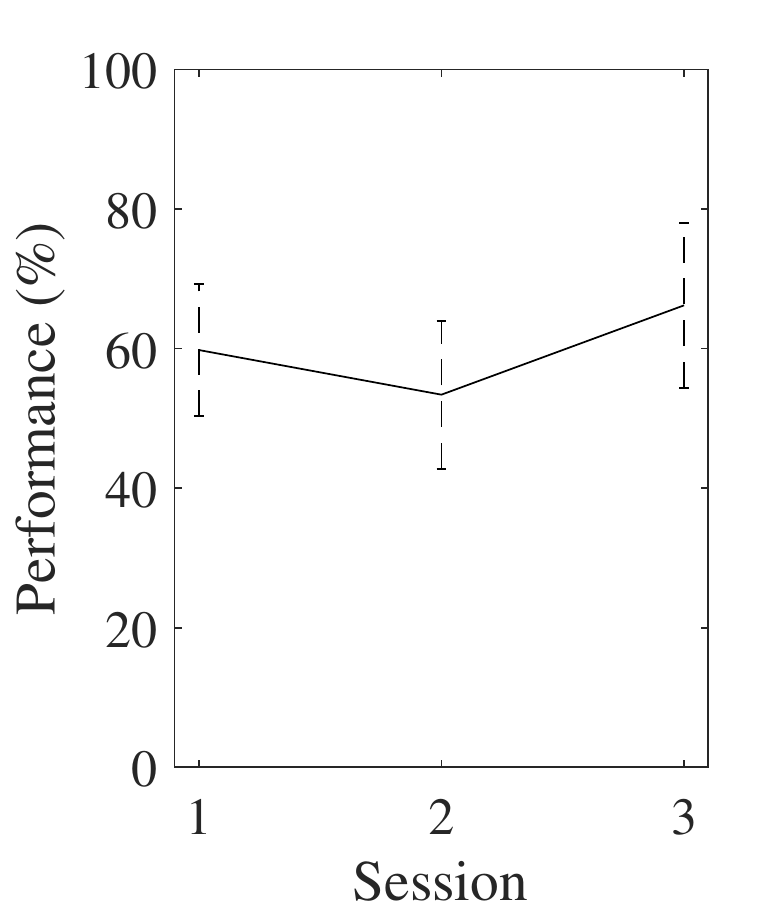}}
	 \\
	\subfigure[$S_{4}$, Training.]{\includegraphics[width=0.3\textwidth]{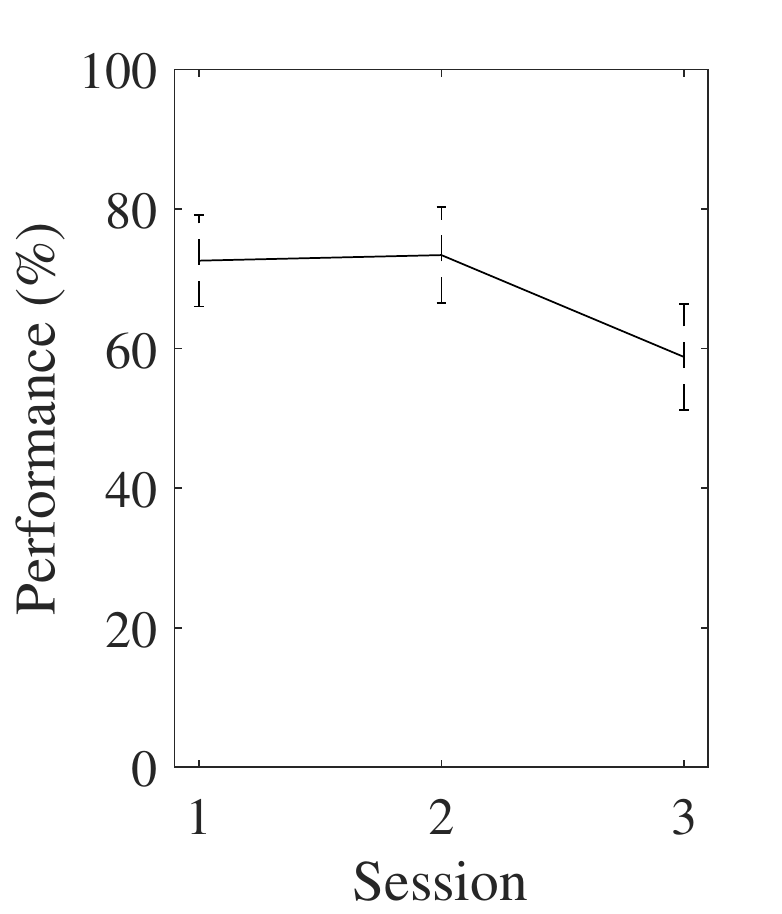}}
           \subfigure[$S_{4}$, Cued manipulation.]{\includegraphics[width=0.3\textwidth]{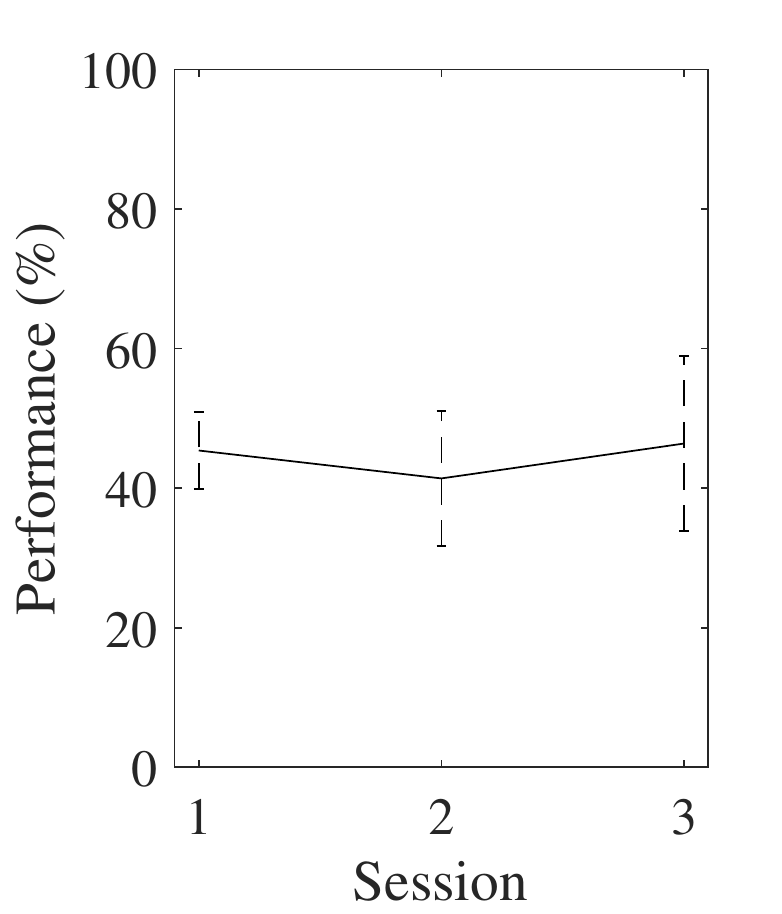}}
           \subfigure[$S_{4}$, Uncued manipulation.]{\includegraphics[width=0.3\textwidth]{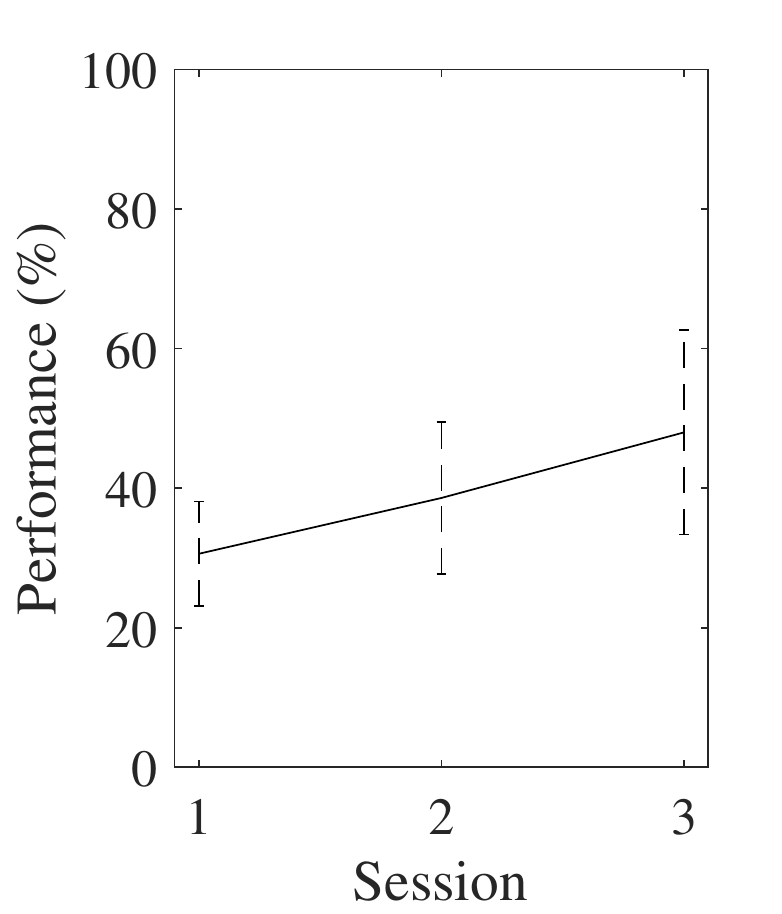}}
	\caption{Performance for Users~$S_{3}$ (top) and $S_{4}$ (bottom) in semi-autonomous goal-selection BCI during training,  cued, and  uncued  manipulation trials (left, middle and right columns respectively). Bars indicate one standard deviation.}
\label{fig:S3S4_per}
\end{figure}

Subject~$S_{3}$ started the training sessions at $56 \%$ of accuracy, $58\%$ at the second session, and reached $78\%$ at the third session. Performance for the cued manipulation trials, started at $40 \%$, increasing to $56\%$ at the second session, and decreasing to $49 \%$ by the third. During uncued manipulation trials, performance started at $60\%$  accuracy by the first session, $53\%$ at the second, and $67\%$ by the third. User~$S_{4}$ obtained performance values of $73\%$ at the first session, $72\%$ at the second, and decreased to $60\%$ at the third. During cued manipulation trials, the subject obtained performance values of  $45\%$ for the first session, $41$ for the second, and $46$ for the third. For uncued manipulation trials, user performance started at $30\%$, and increased to $38\%$ and $48\%$ by the second and third sessions, respectively. Results for goal-selection BCI performance are shown in Figure~\ref{fig:S3S4_per}  for Users~$S_{3}$ and $S_{4}$, respectively. 

\subsection{P300 analysis}
The results for the two-way ANOVA tests are presented in Table \ref{tab:anova2}. The results for the P$300$ latency two-way ANOVA showed statistical significance for  Subjects $S_{1}$ ($p=0.0147$) and $S_{2}$ ($p=0.0001$) in the trial factor, but no significance was observed for channel and interaction factors. Users~$S_{3}$ and $S_{4}$ showed no satistical significance for any of the analyzed factors. For the P$300$ amplitude two-way ANOVA, users showed smaller $p$-values in trial when compared to channel and interaction. However, our tests did not show statistical significance for any factor or interaction. 

The results for the one-way ANOVA tests are shown in Table \ref{tab:anova1}. The results for P$300$ latency one-way ANOVA showed statistical significance for User~$S_{1}$ at channel O1 ($p=0.0476$), and for User~$S_{4}$ at channel O2 ($p=0.0242$). Regarding Users~$S_{2}$ and $S_{3}$, $p$-values were not significant at any channel. For the P$300$ amplitude one-way ANOVA, User~$S_{4}$ showed statistical significance at channel Pz ($p=0.0019$). The tests for $S_{1}$, $S_{2}$, and $S_{3}$ revealed no statistical significance at none of the three analyzed channels.

The results of the performed statistical tests allowed to observe differences between analyzing latency and amplitude. Among all tests, 
greater changes were found in latency rather than in amplitude. Based on these results,  an evaluation and comparison on amplitude and latency values was performed. These values were considered as those corresponding to the channel with the lowest $p$-value on the one-way latency ANOVA results. The selected channels were O1 for $S_{1}$, Pz for $S_{2}$ and $S_{3}$, and O2 for $S_{4}$. 

\begin{figure}
	\centering
	\subfigure[Subject $S_{1}$, Session $1$]{\includegraphics[width=0.3\textwidth]{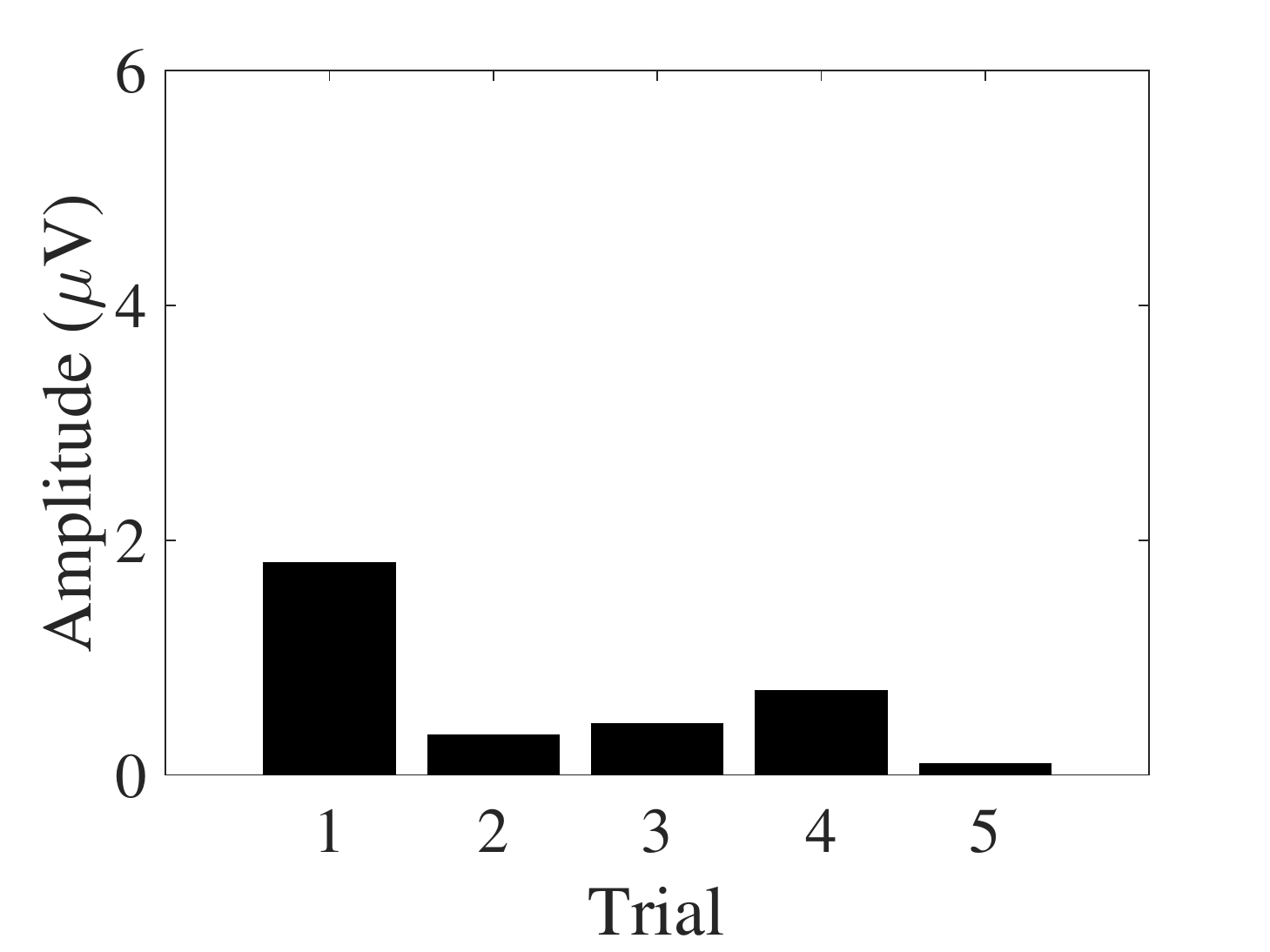}}
          \subfigure[Subject $S_{1}$, Session $2$]{\includegraphics[width=0.3\textwidth]{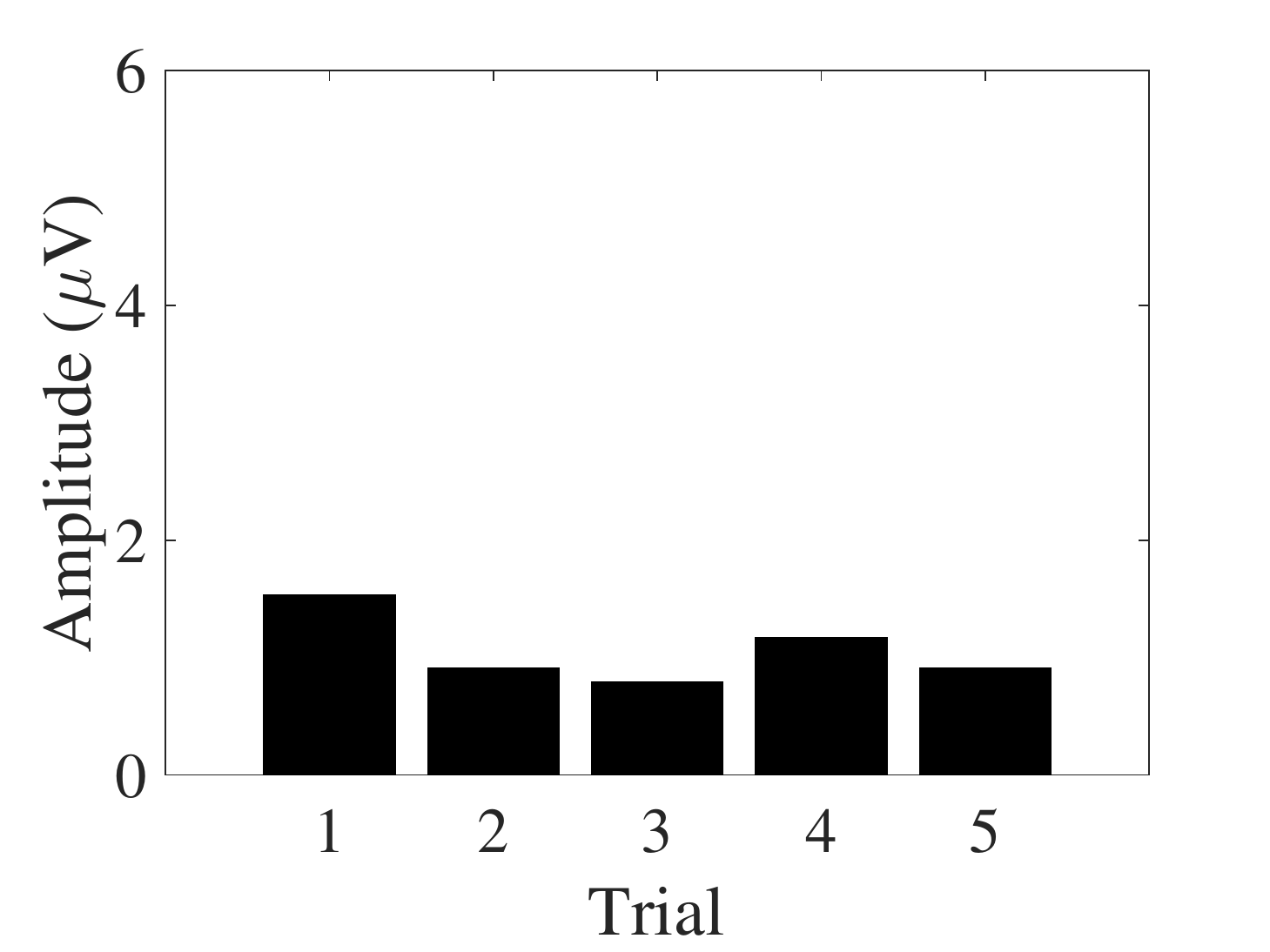}}
	\subfigure[Subject $S_{1}$, Session $3$]{\includegraphics[width=0.3\textwidth]{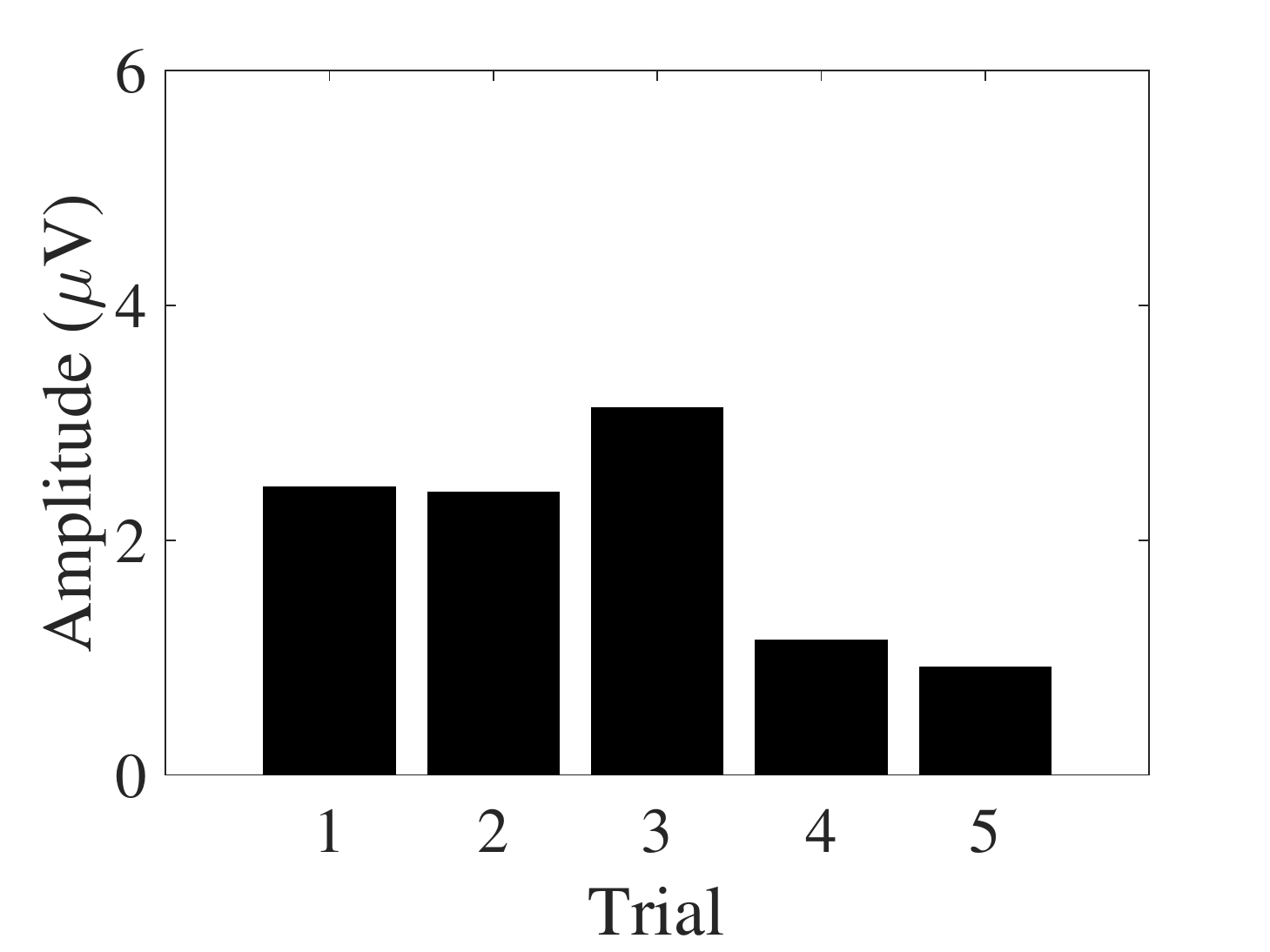}}
	 \\
	\subfigure[Subject $S_{2}$, Session $1$]{\includegraphics[width=0.3\textwidth]{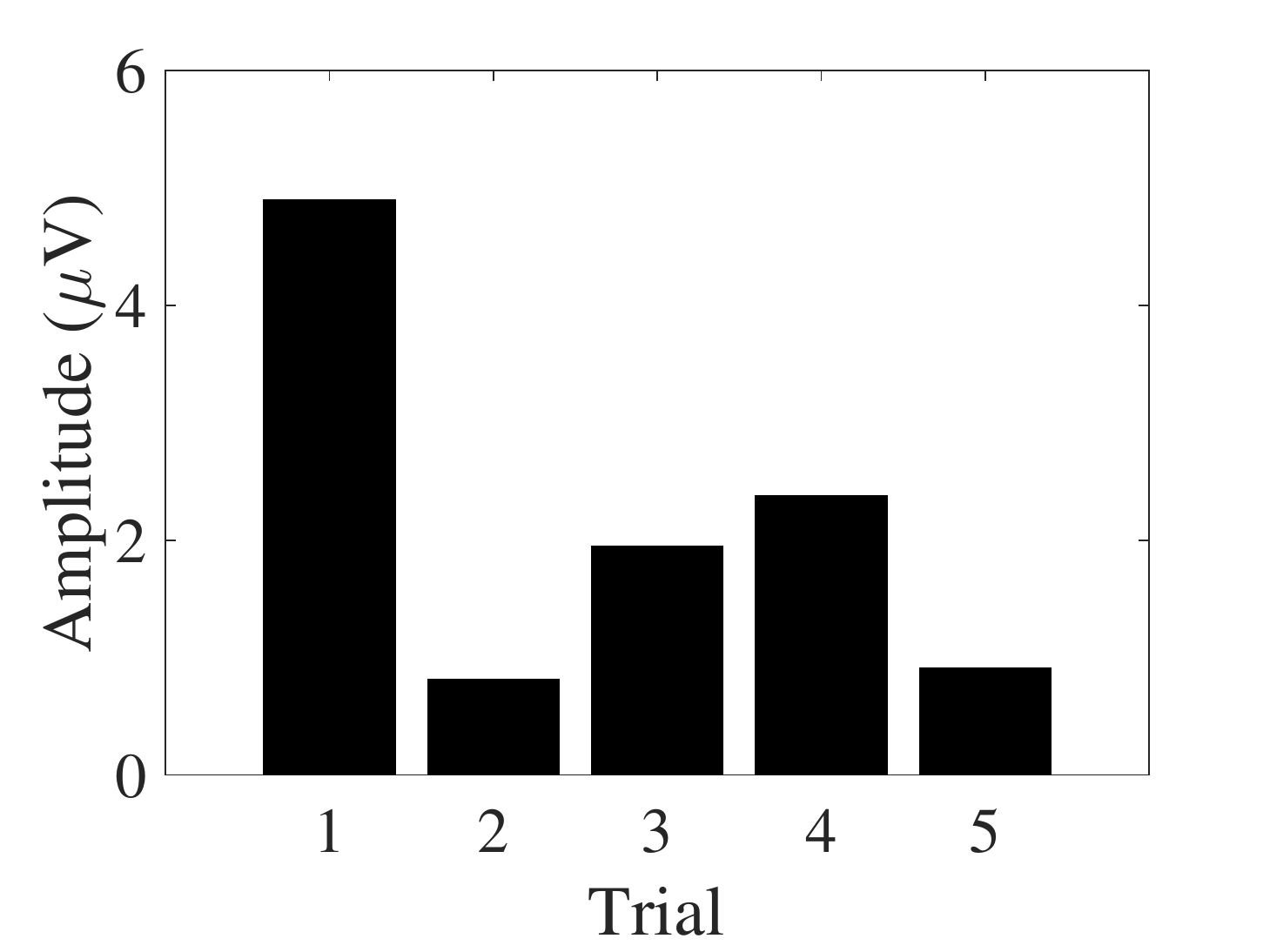}}
	\subfigure[Subject $S_{2}$, Session $2$]{\includegraphics[width=0.3\textwidth]{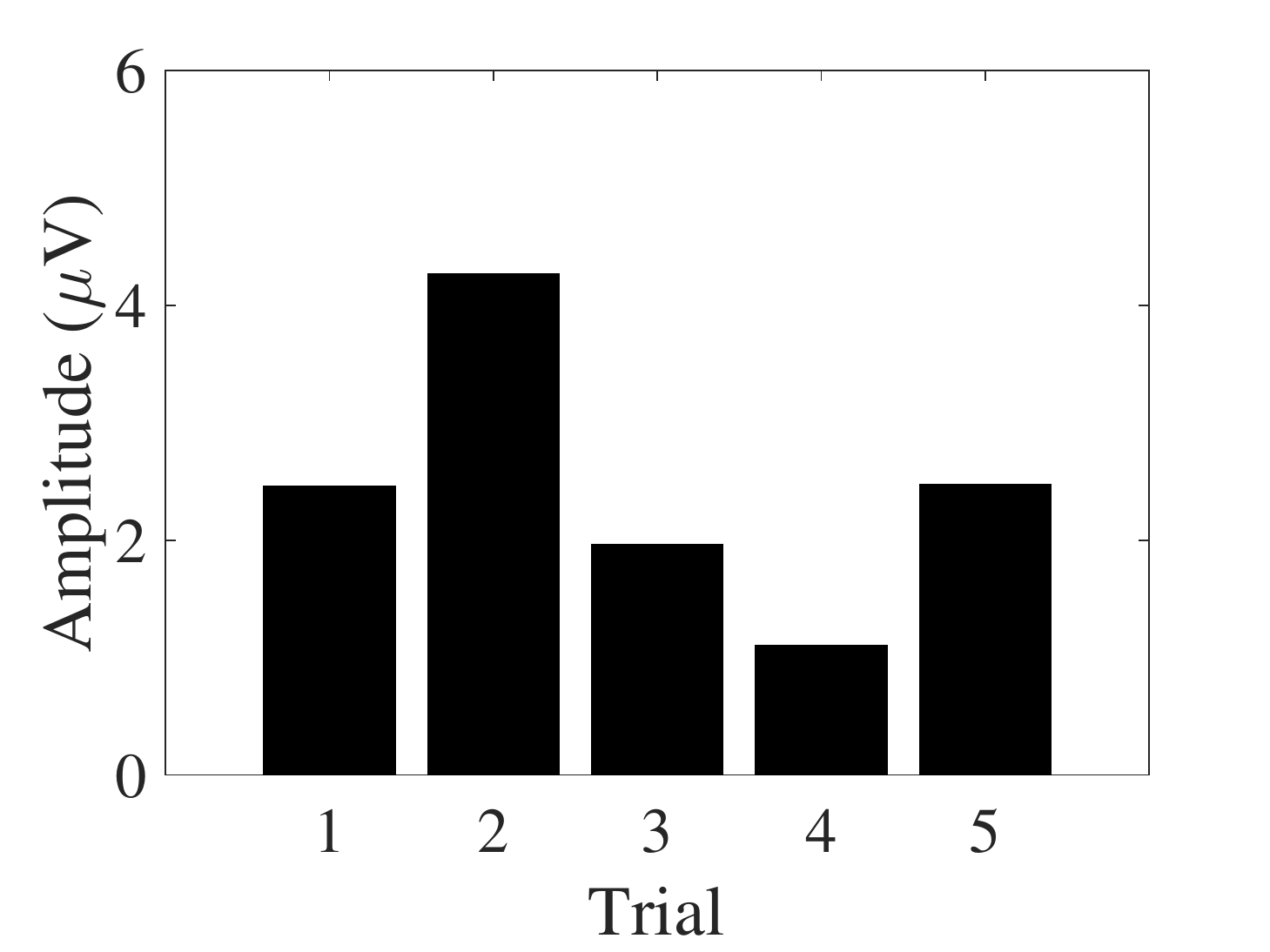}}
	\subfigure[Subject $S_{2}$, Session $3$]{\includegraphics[width=0.3\textwidth]{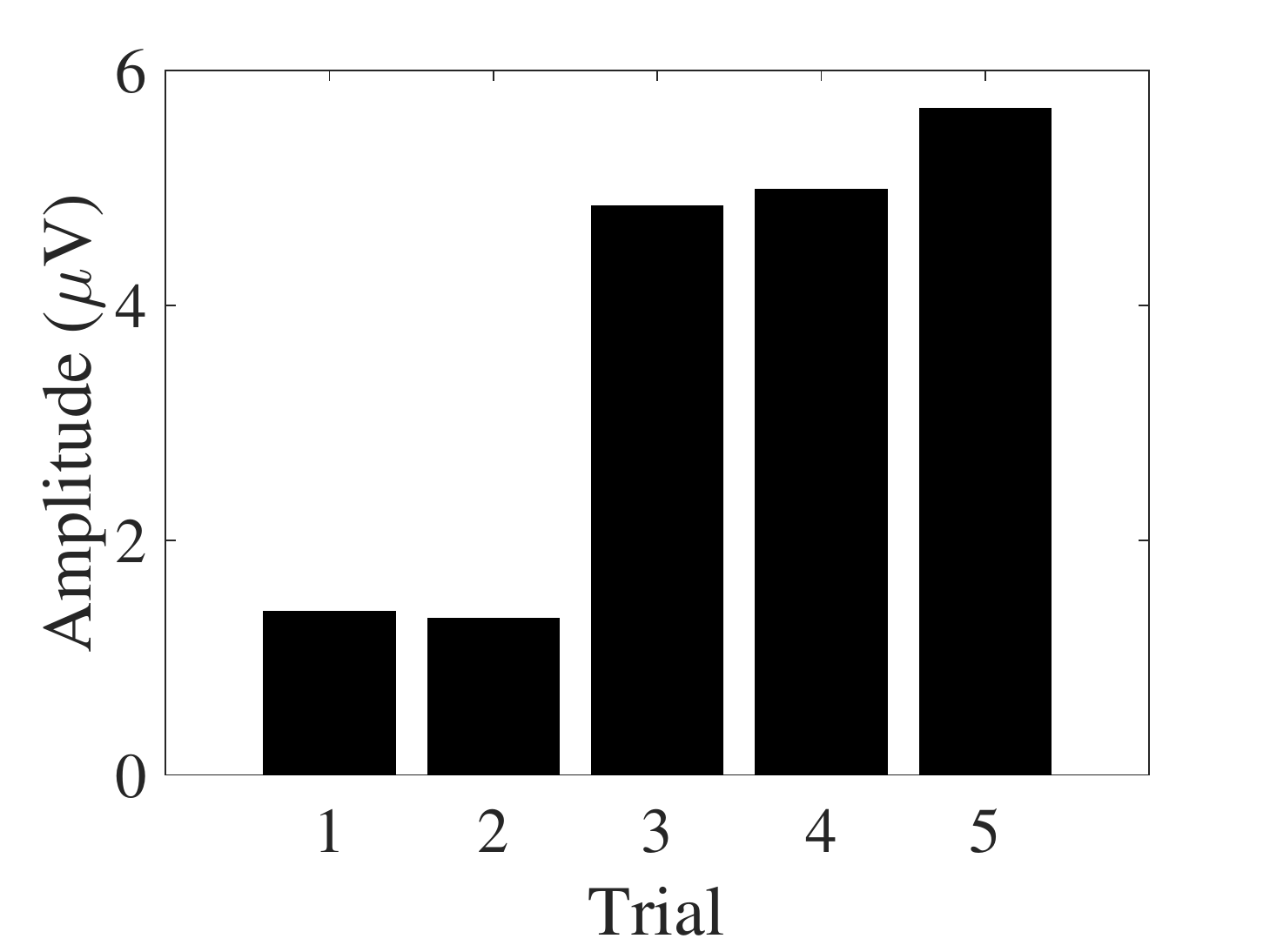}}
		\\
	\subfigure[Subject $S_{3}$, Session $1$]{\includegraphics[width=0.3\textwidth]{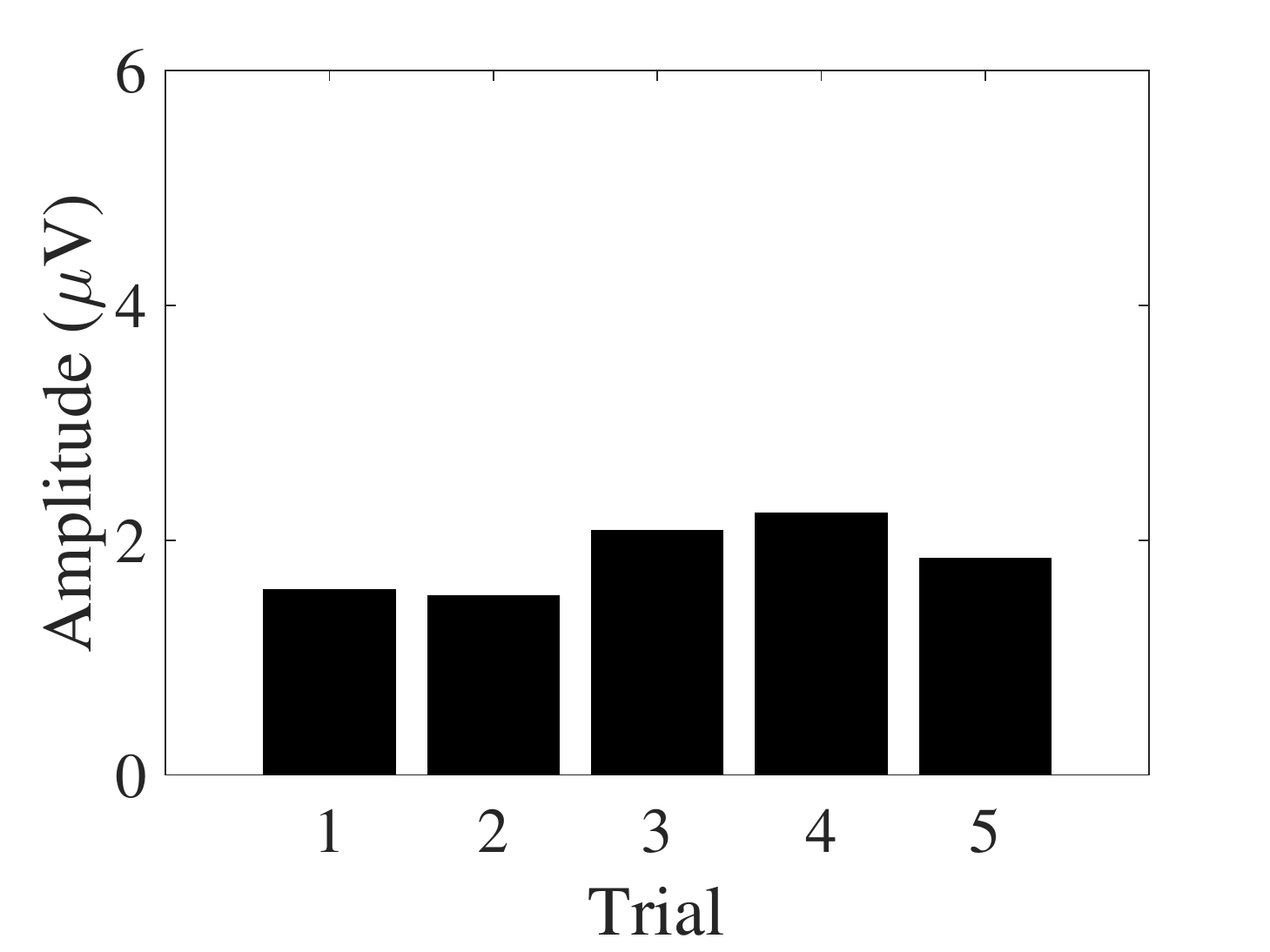}}
	\subfigure[Subject $S_{3}$, Session $2$]{\includegraphics[width=0.3\textwidth]{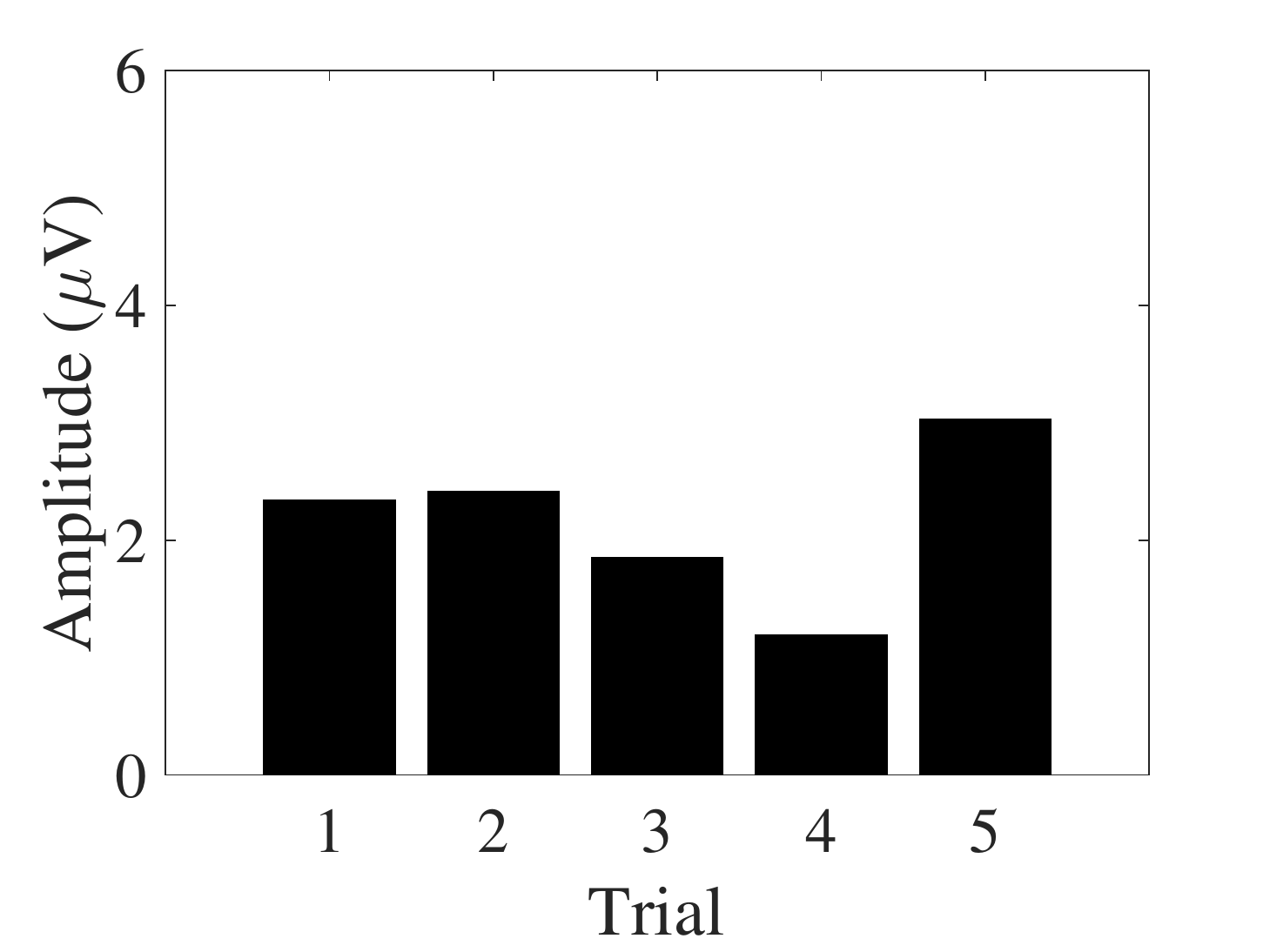}}
	\subfigure[Subject $S_{3}$, Session $3$]{\includegraphics[width=0.3\textwidth]{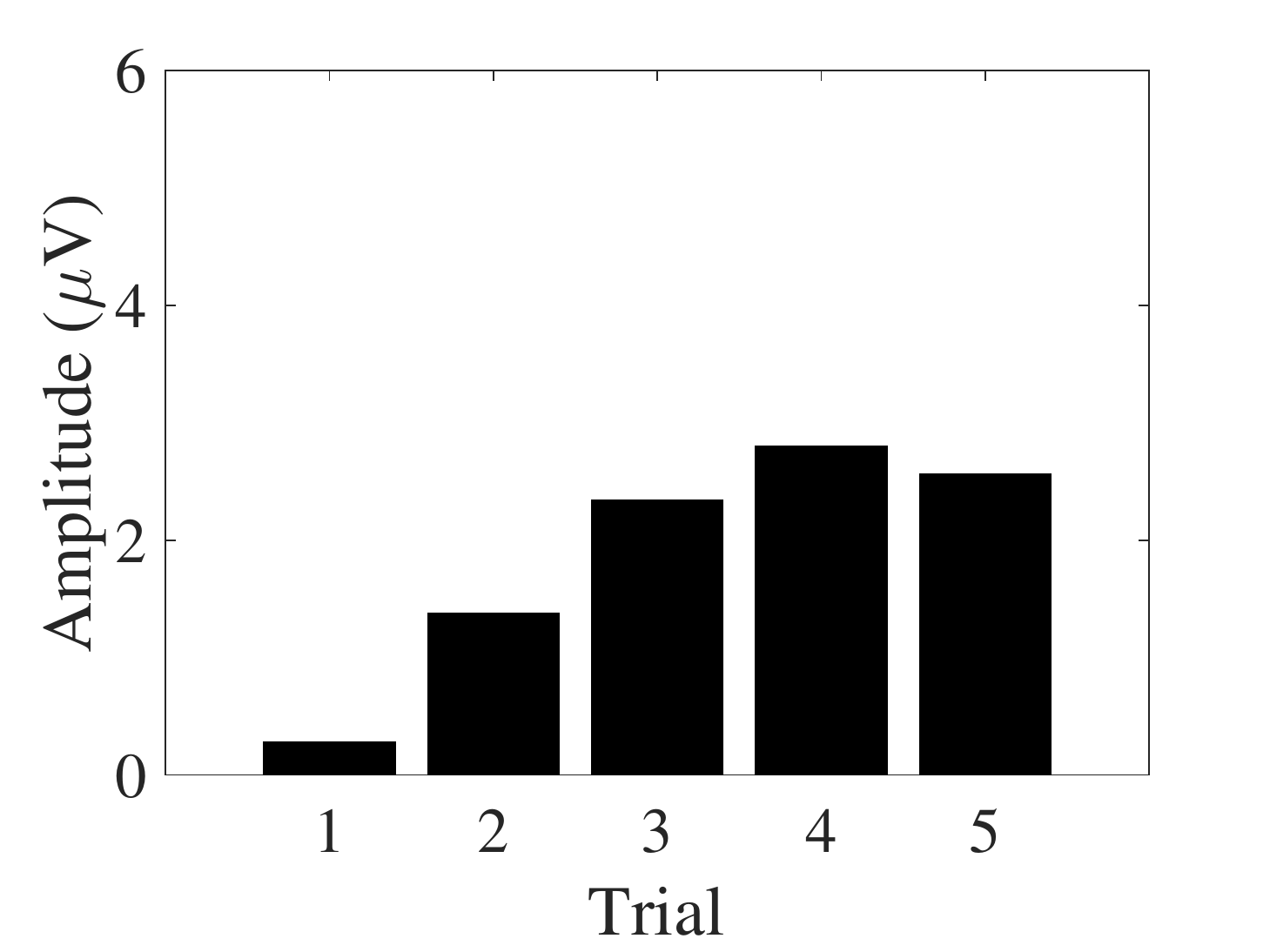}}
		\\
	\subfigure[Subject $S_{4}$, Session $1$]{\includegraphics[width=0.3\textwidth]{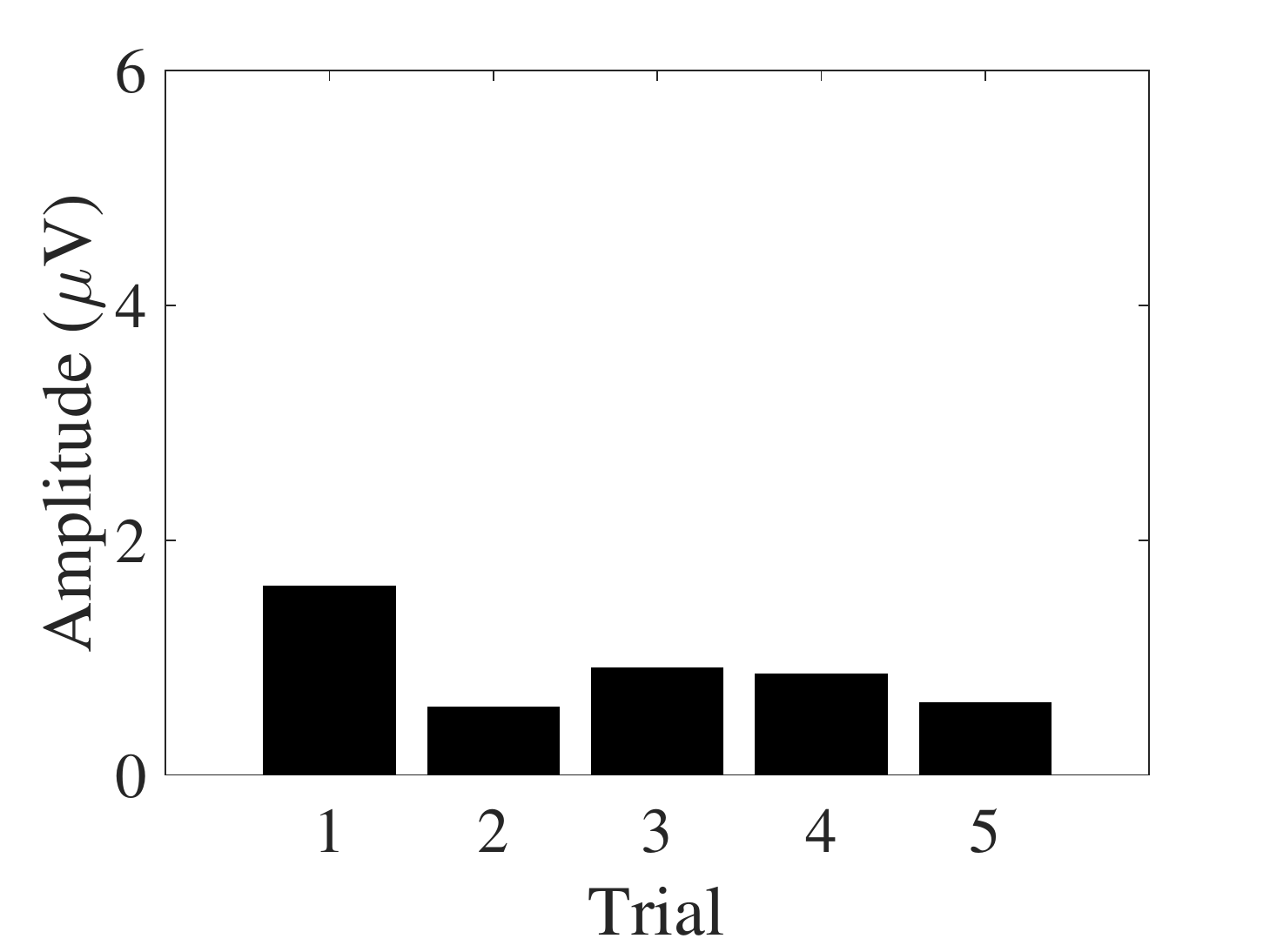}}
	\subfigure[Subject $S_{4}$, Session $2$]{\includegraphics[width=0.3\textwidth]{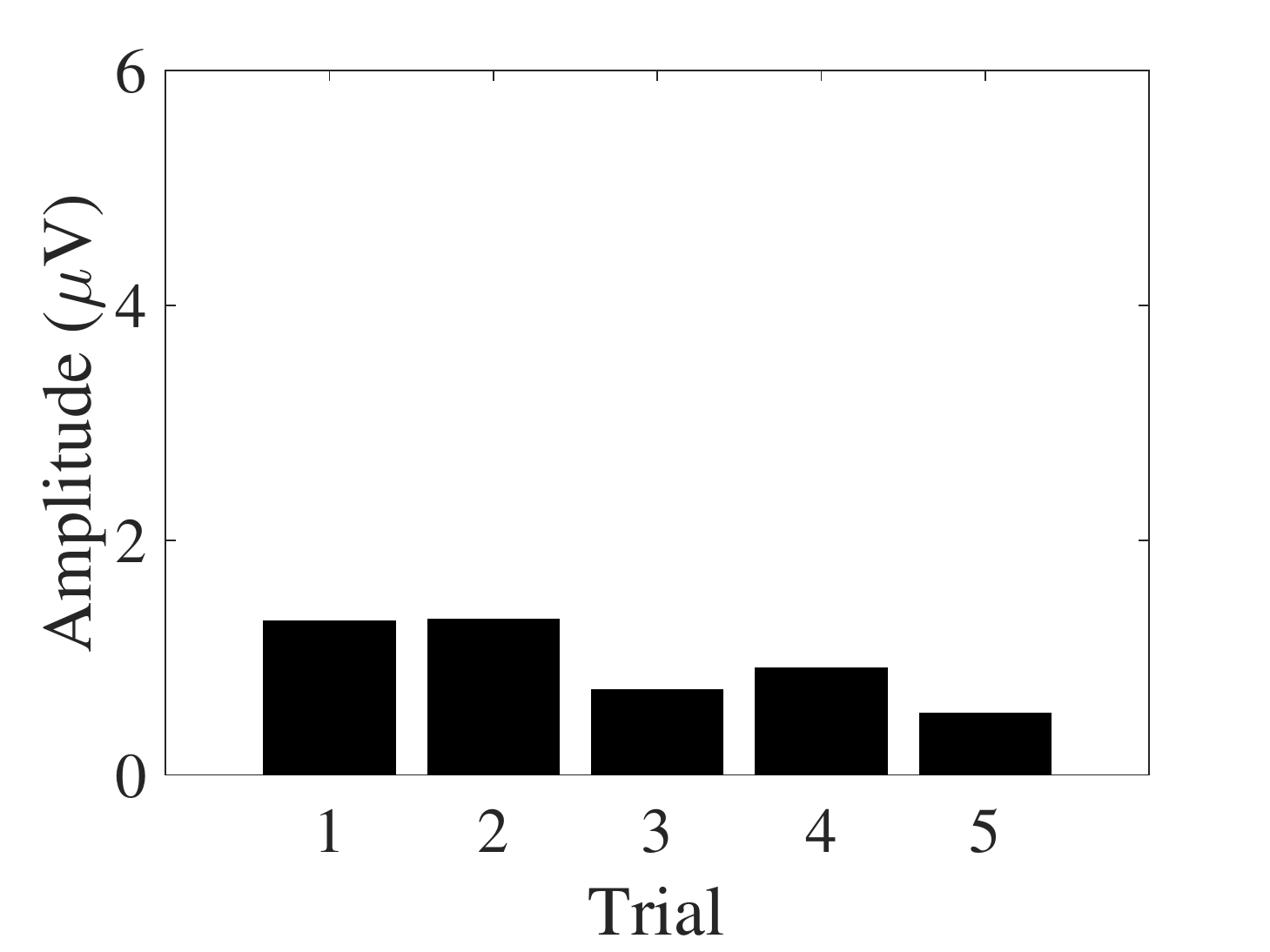}}
	\subfigure[Subject $S_{4}$, Session $3$]{\includegraphics[width=0.3\textwidth]{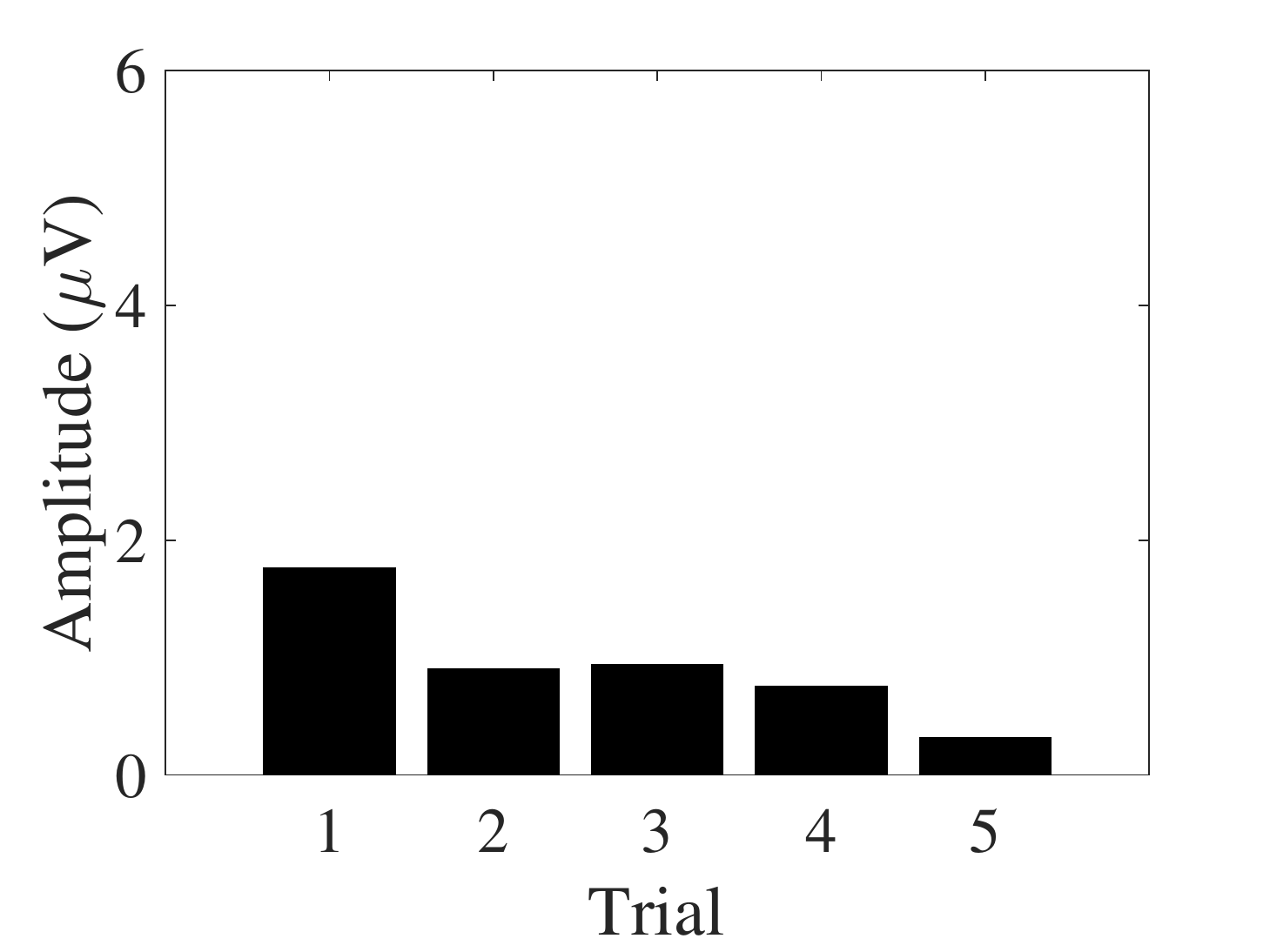}}
	\caption{Amplitude of P300 waveform during all trials and experiments for all subjects during uncued manipulation trials.}
\label{fig:p300amp}
\end{figure}

Amplitude values calculated for all  uncued manipulation trials are shown in Figure~\ref{fig:p300amp} for each session and user. Users~$S_{1}$ and $S_{4}$ showed a similar behavior: a decreasing P$300$amplitude trend in all sessions. In this case, the amplitude observed at the first trial was higher than that of the last one. $S_{2}$ showed a decreasing trend as well for the first and second sessions, yet the opposite was observed during the third session. $S_{3}$ presented an increasing amplitude trend for all sessions. Here, the amplitude obtained at the last trial was higher than the one at the first trial. 

\begin{figure}
	\centering
	\subfigure[Subject $S_{1}$, Session $1$]{\includegraphics[width=0.3\textwidth]{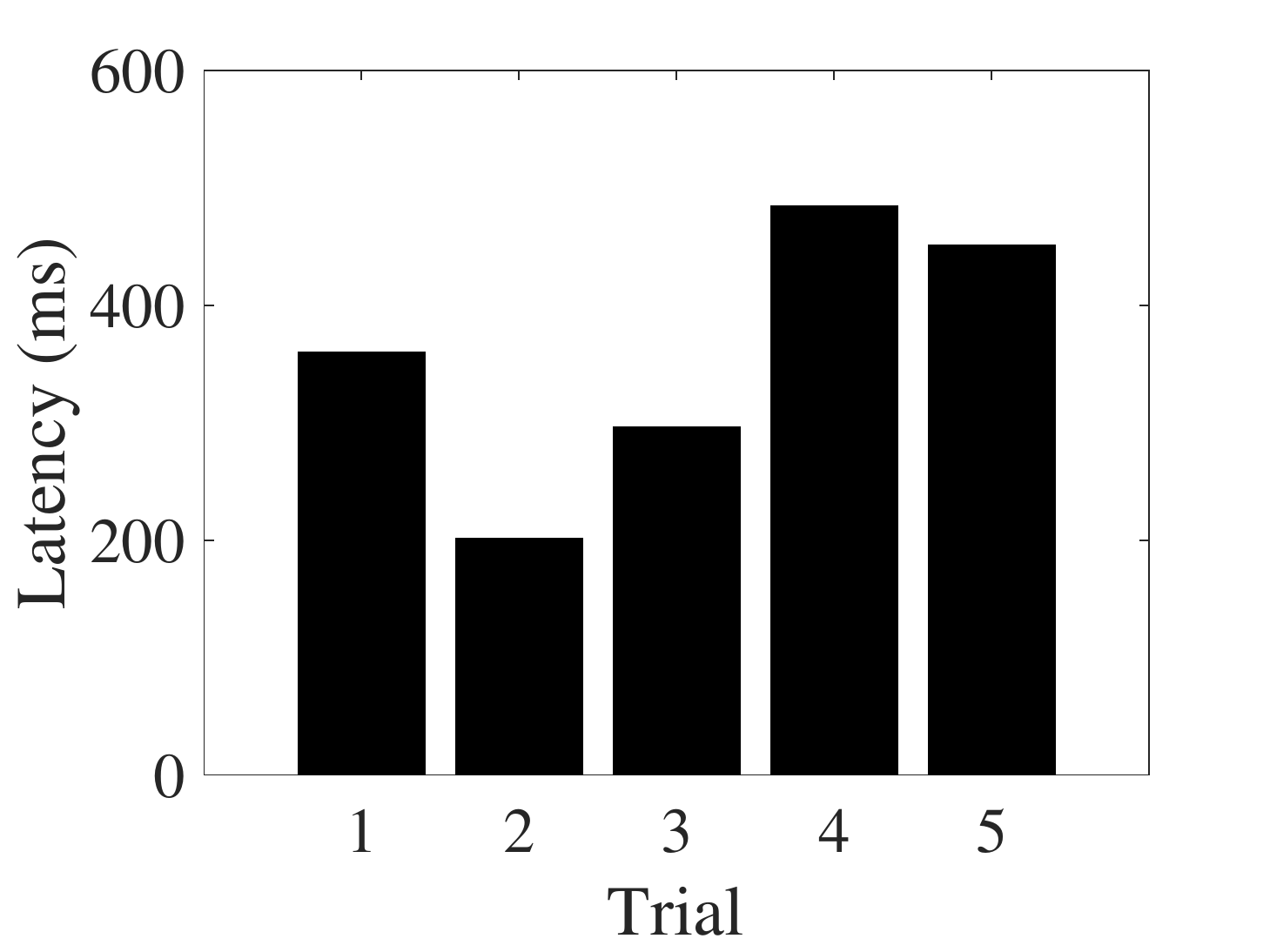}}
           \subfigure[Subject $S_{1}$, Session $2$]{\includegraphics[width=0.3\textwidth]{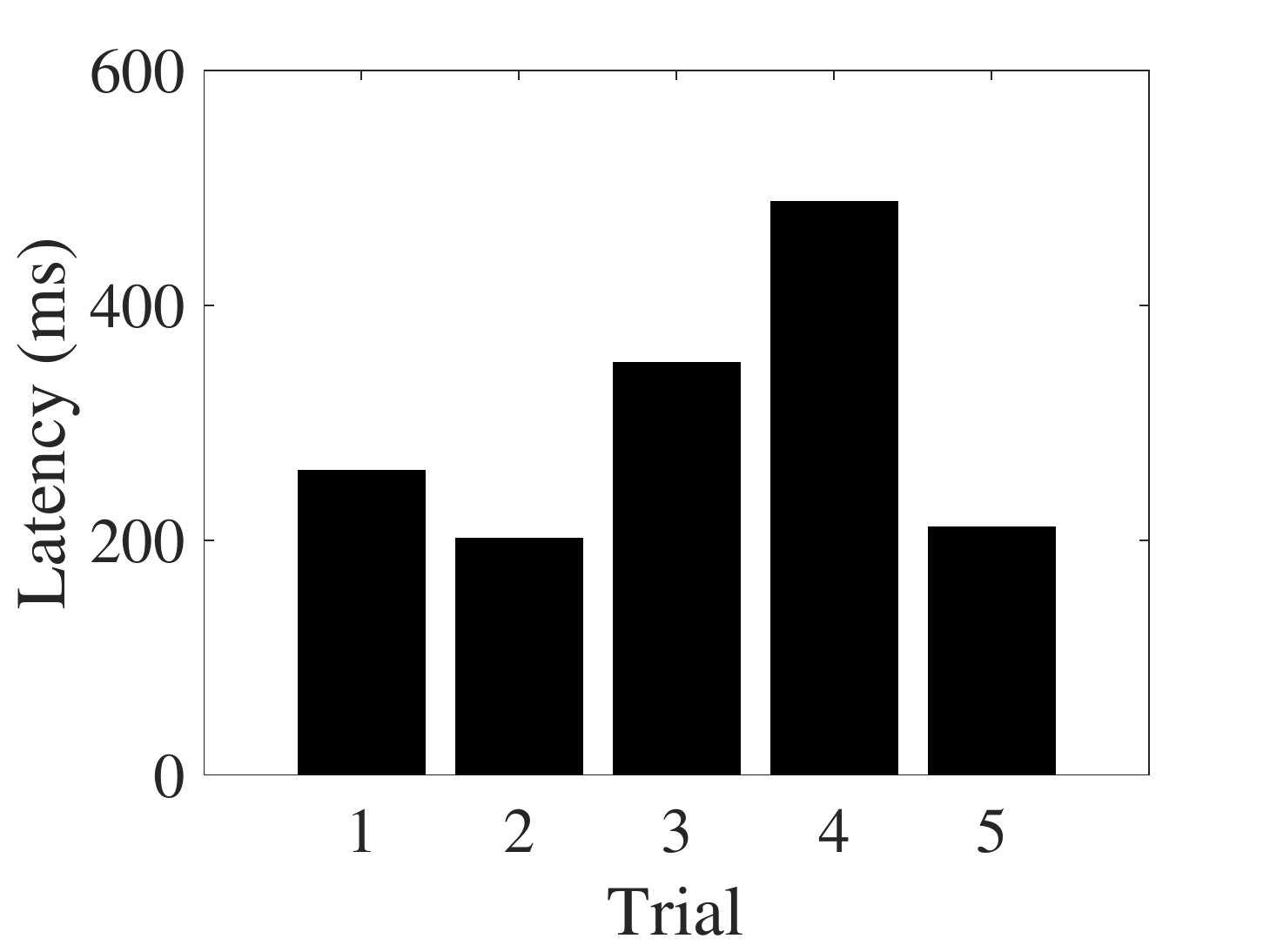}}
	\subfigure[Subject $S_{1}$, Session $3$]{\includegraphics[width=0.3\textwidth]{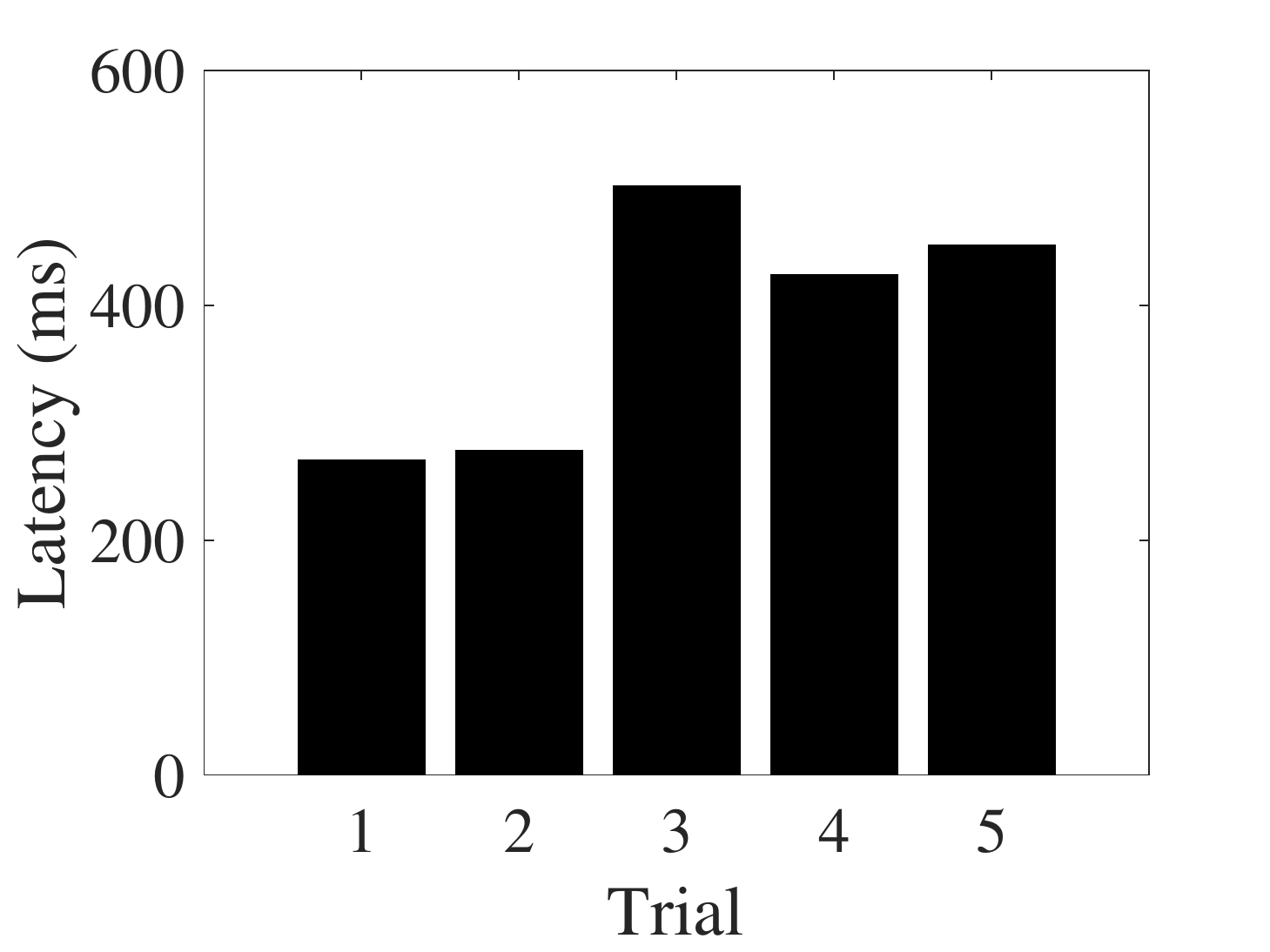}}
	 \\ 
	\subfigure[Subject $S_{2}$, Session $1$]{\includegraphics[width=0.3\textwidth]{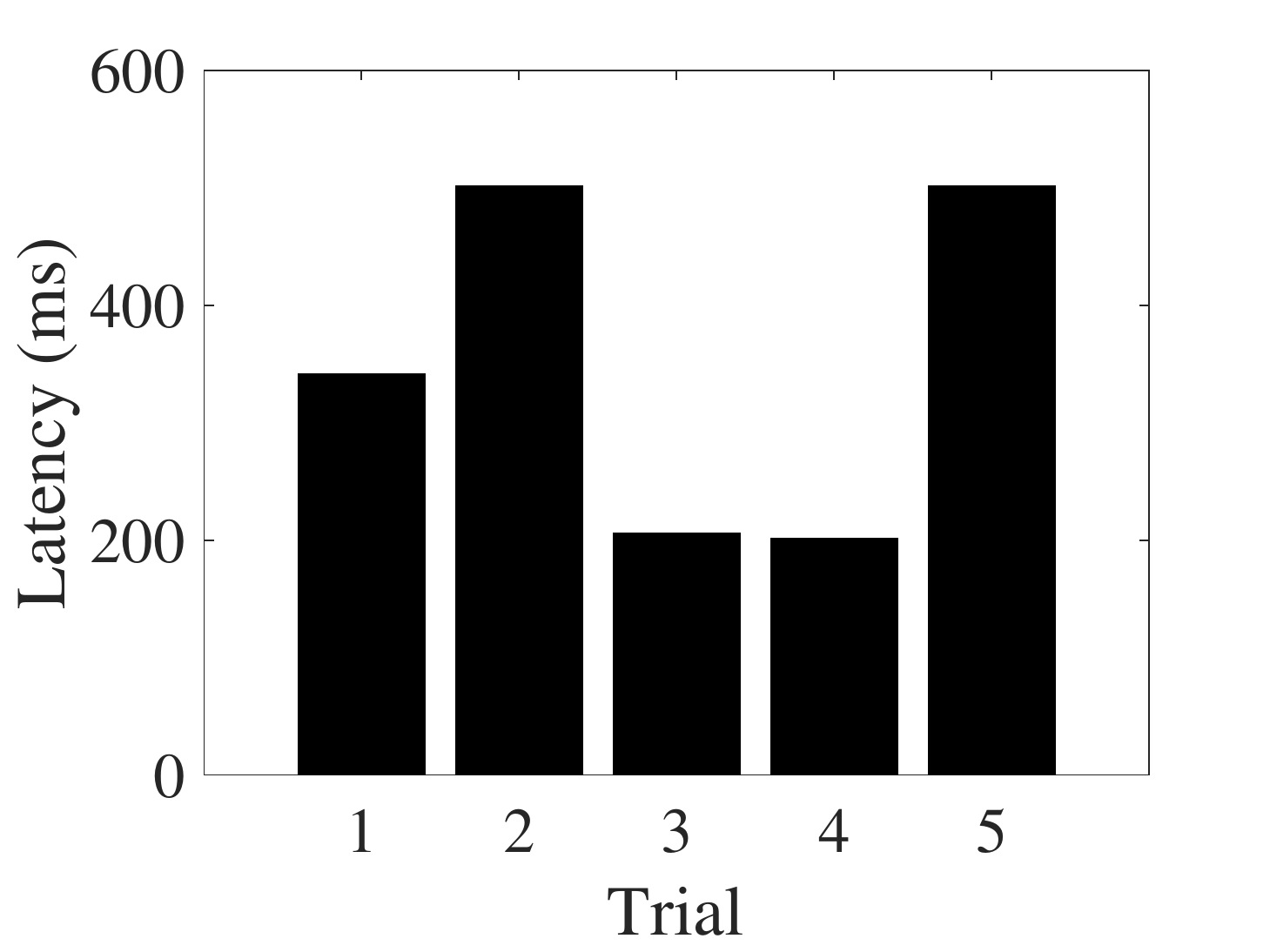}}
           \subfigure[Subject $S_{2}$, Session $2$]{\includegraphics[width=0.3\textwidth]{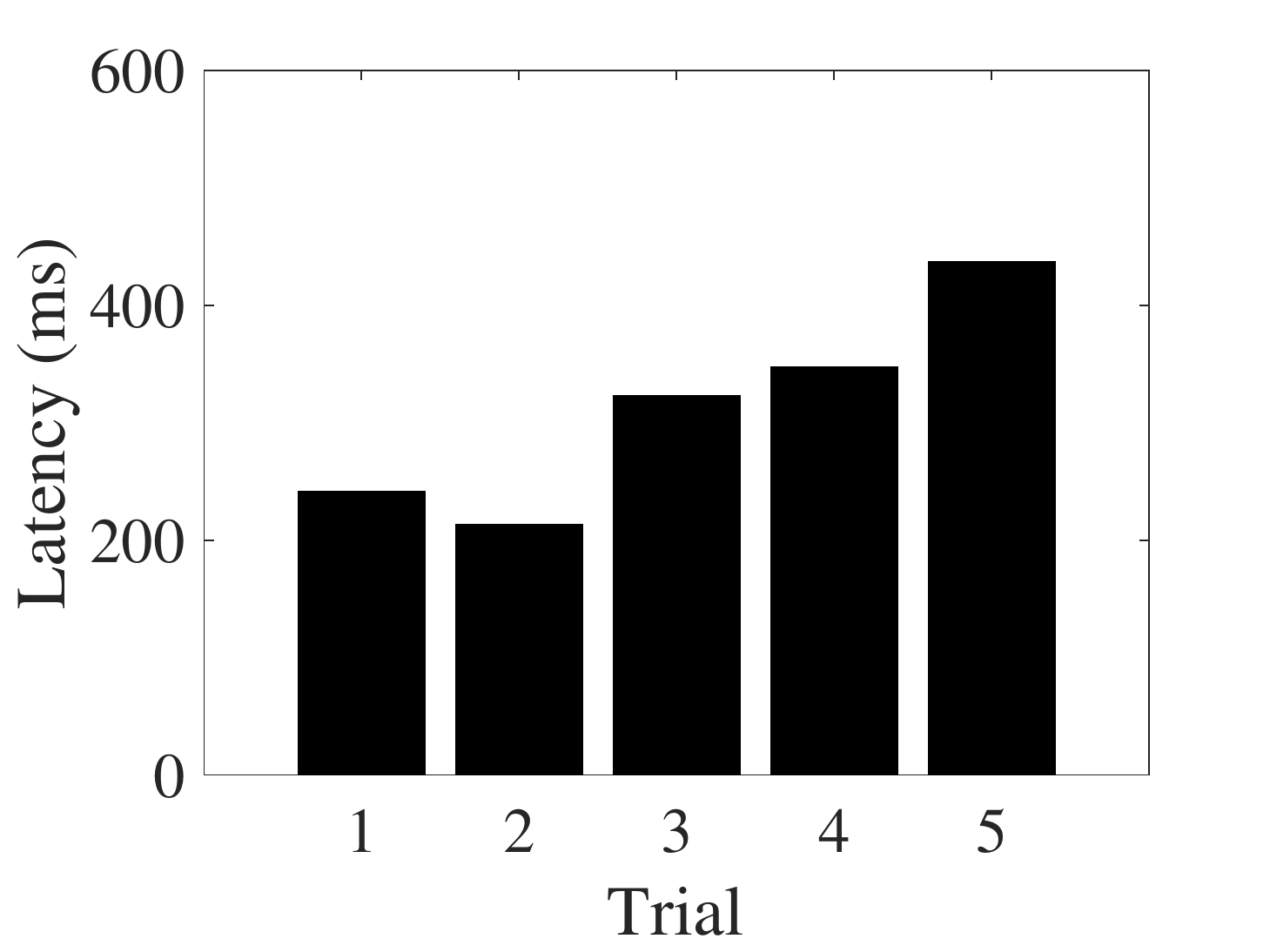}}
	\subfigure[Subject $S_{2}$, Session $3$]{\includegraphics[width=0.3\textwidth]{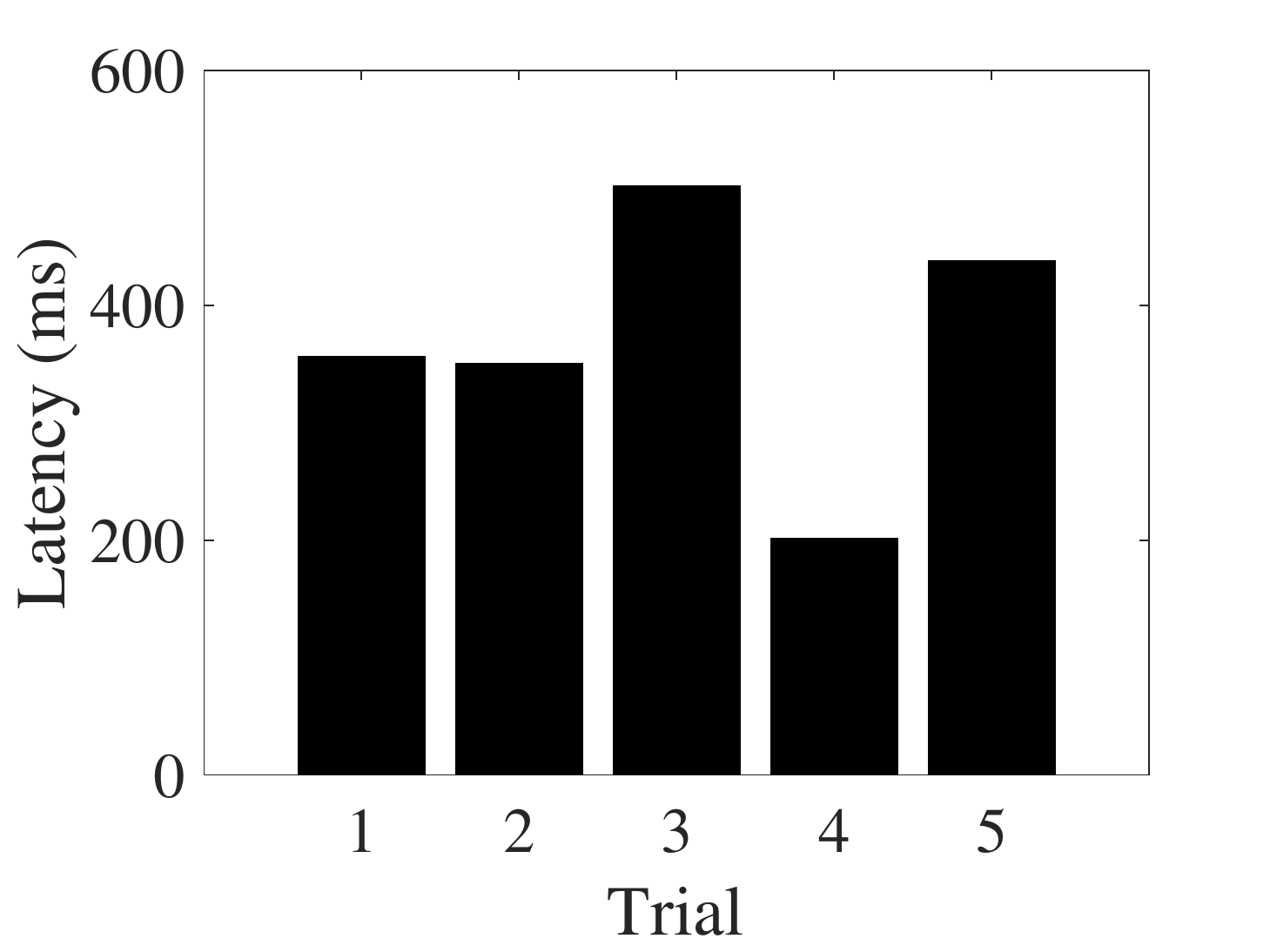}}
	 \\
	\subfigure[Subject $S_{3}$, Session $1$]{\includegraphics[width=0.3\textwidth]{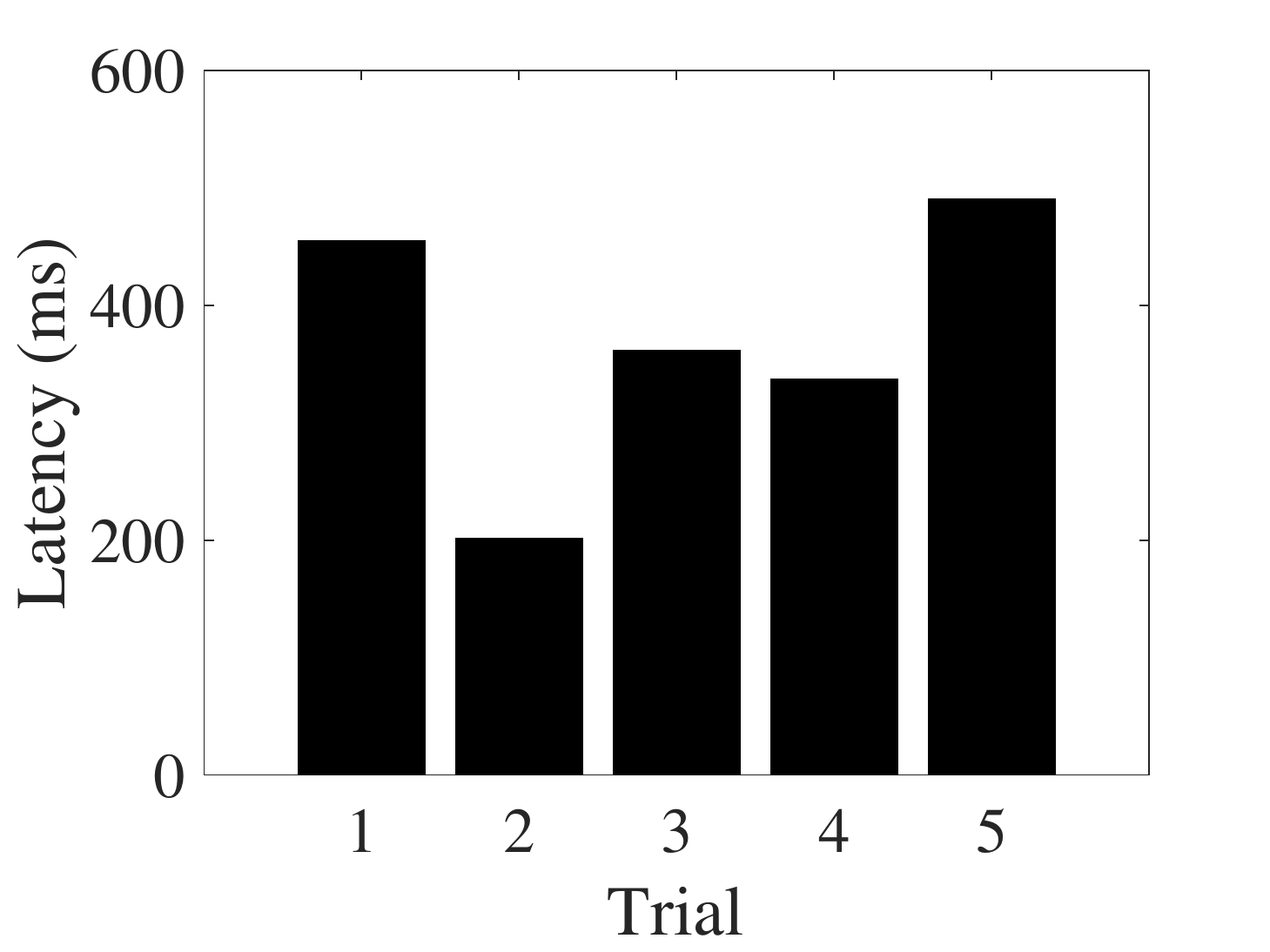}}
           \subfigure[Subject $S_{3}$, Session $2$]{\includegraphics[width=0.3\textwidth]{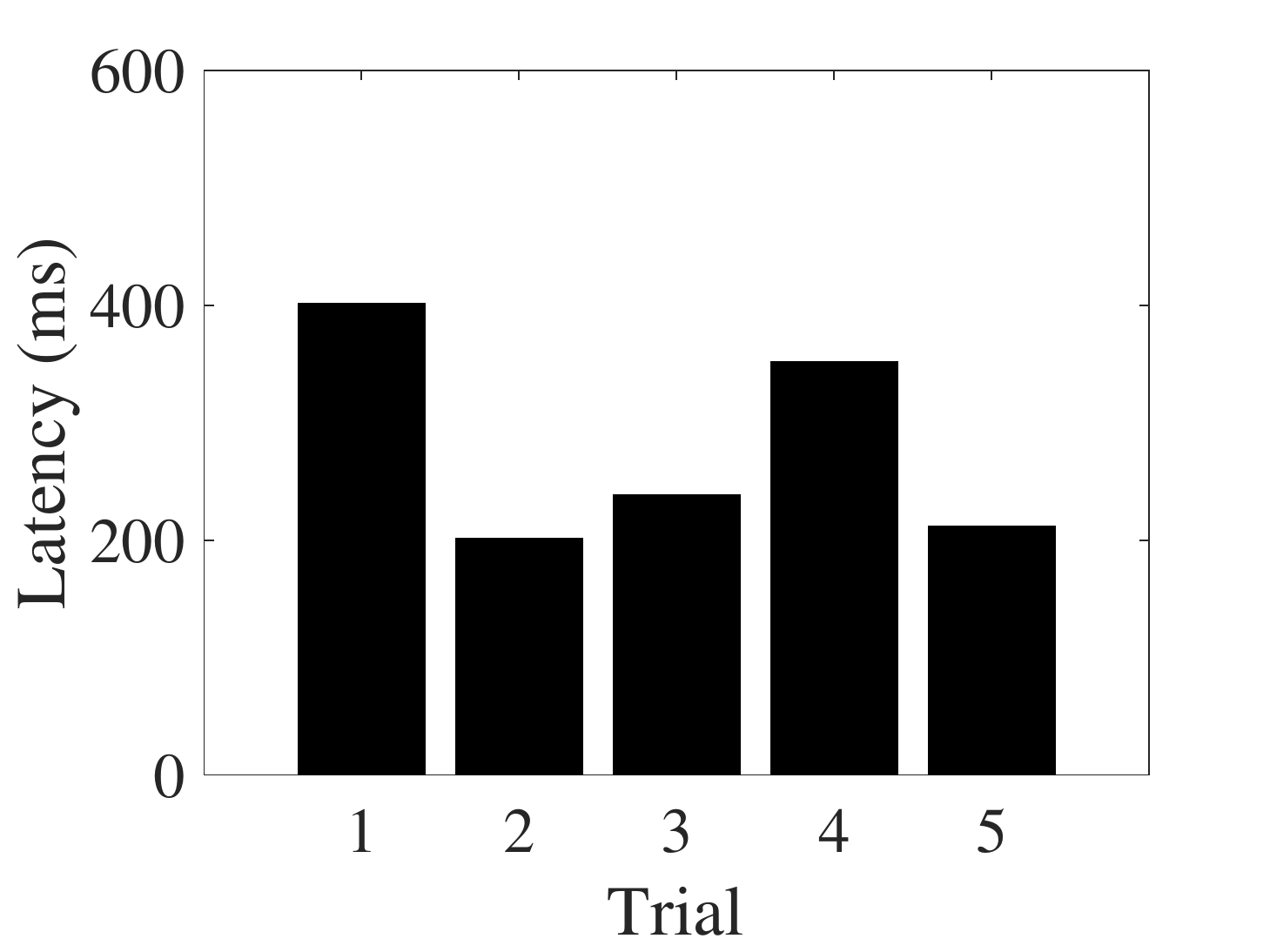}}
	\subfigure[Subject $S_{3}$, Session $3$]{\includegraphics[width=0.3\textwidth]{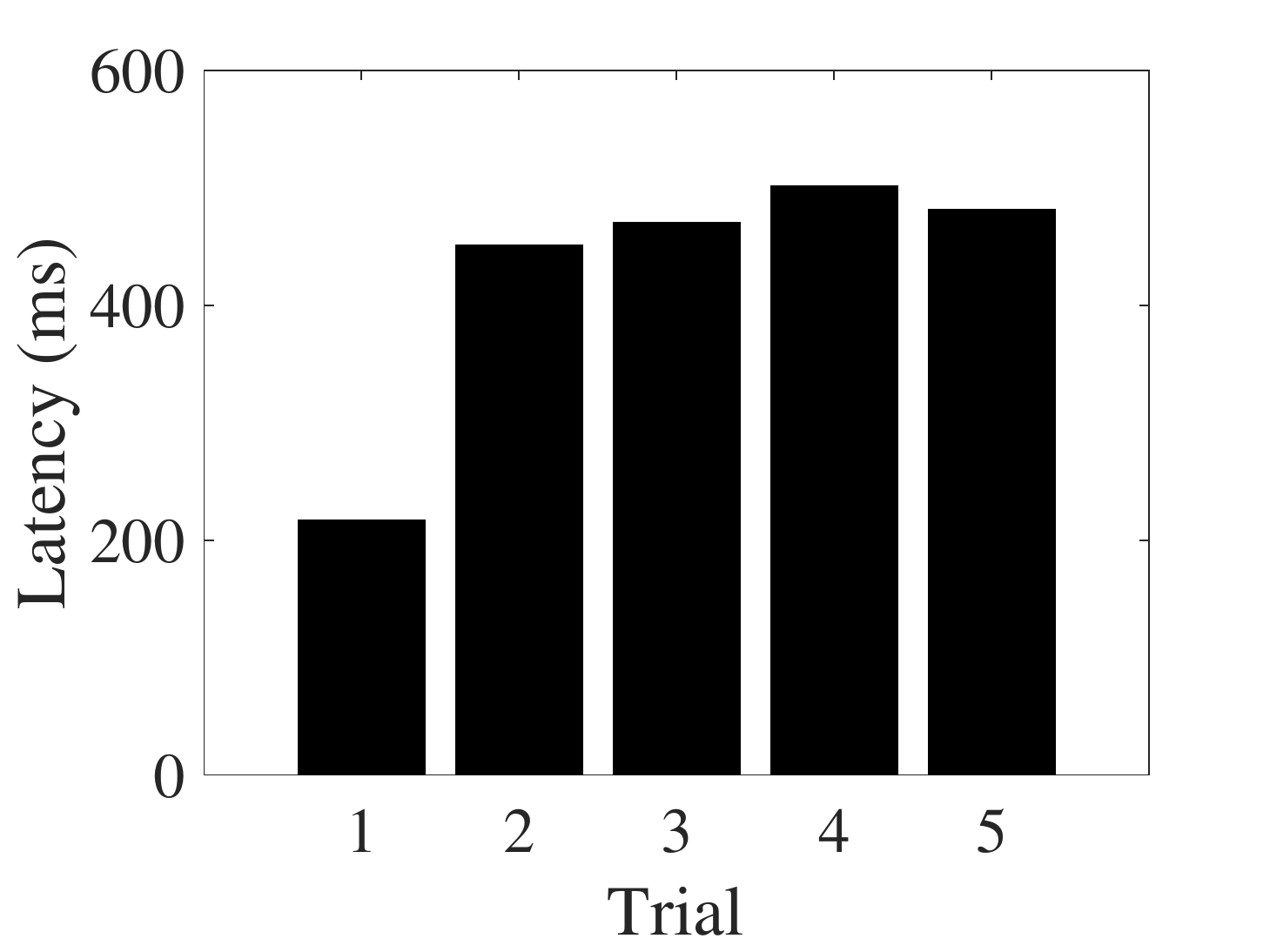}}
	 \\
          \subfigure[Subject $S_{4}$,  Session $1$]{\includegraphics[width=0.3\textwidth]{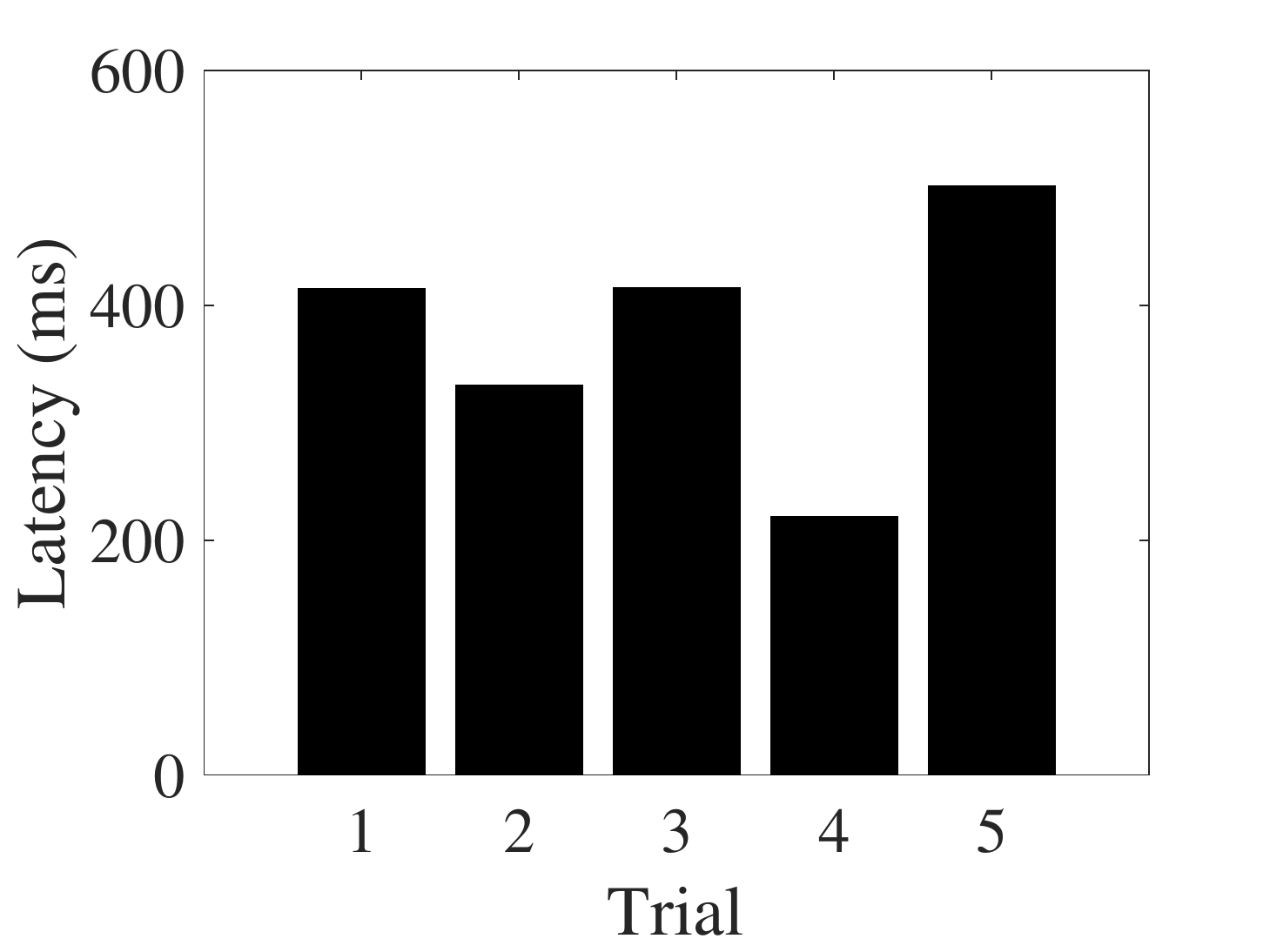}}
           \subfigure[Subject $S_{4}$, Session $2$]{\includegraphics[width=0.3\textwidth]{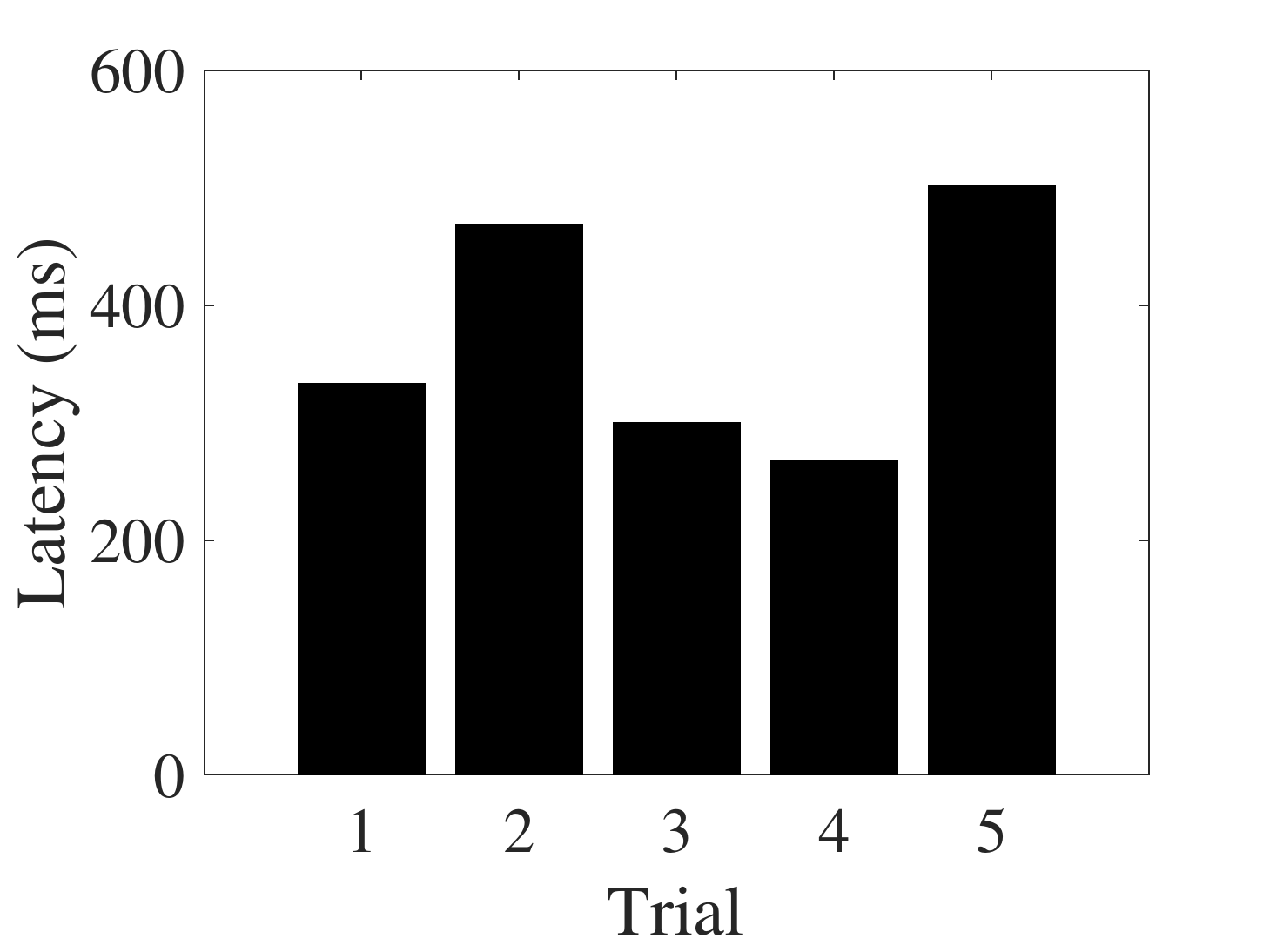}}
	\subfigure[Subject $S_{4}$, Session $3$]{\includegraphics[width=0.3\textwidth]{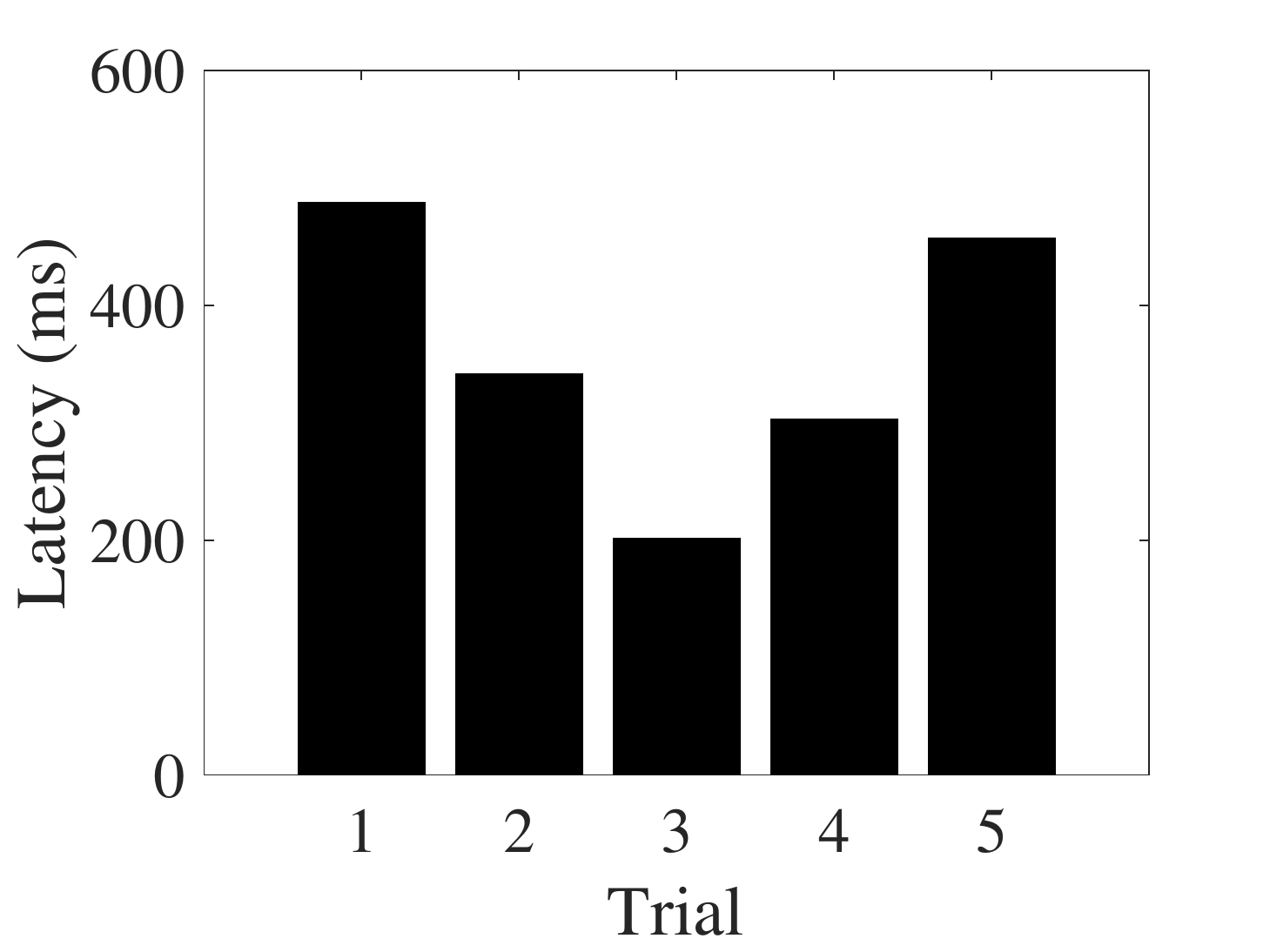}}
	\caption{Latency of P300 waveform during all trials and experiments for all subjects during uncued manipulation trials.}
\label{fig:p300lat}
\end{figure}

Latency values can be observed in Figure~\ref{fig:p300lat} for all users and sessions. Subjects~$S_{1}$ and $S_{3}$ displayed an increasing P$300$ latency trend during the first and third sessions. A decreasing trend was observed during the second session for these users. User~$S_{2}$ presented an increasing latency trend for all sessions. User~$S_{4}$ showed an increase in latency during the first and second sessions, and a decrease at the third.
 
\section{Discussion}
The implementation and integration of the CGA model and the AV algorithm allowed to successfully design a MI-based semi-autonomous BCI for manipulation tasks. When compared against a low-level system, both BCIs were similar in terms of training protocol and control commands, however the complexity of the executed tasks was different.  The semi-autonomous goal-selection BCI was superior in task complexity when compared to the process-control BCI, even though both systems used the same control commands as input. While the process-control BCI might be used to perform  more general tasks, it demands a continuous awareness state from the user. Its output are discrete low-level commands which in the long run might lead the user to a  mental fatigue condition.  Although the semi-autonomous BCI is goal specific, it requires user attention only during short time periods, making it theoretically less fatiguing. The semi-autonomous goal-selection BCI works, in essence, in a more natural way to the user than the process-control BCI. This is because when performing reaching tasks, people think on the main goal and the cerebellum process the necessary information to successfully achieve it, rather than executing several discrete low-level movements~\cite{Attwell02}.

The selected features for the general classifiers of the users were mainly frontal, central, and parietal electrodes in the $\mu$ ($8$-$13$~Hz) and $\beta$ ($13$-$30$~Hz) brain rhythms, which are known to be physiologically involved in the imaginary movement process. The selected channels for the classifiers are consistent with reports of central activity as a reflection of motor cortex contralateral desynchronization during imaginary movement~\cite{McFarland00}, and fronto-parietal activation related to control of spatial attention and motor planning during reaching tasks~\cite{Praamstra05,Naranjo07}. 

 Even though all users underwent the same training protocol, differences among them were observed. Across training sessions, $S_{1}$ and $S_{2}$ maintained a relatively constant performance, while $S_{3}$ showed a more notorious improvement. $S_{4}$ displayed a relatively high performance at the first and second session, but it decreased at the third.  During cued manipulation trials, all users obtained low performance levels and none of them showed a significant improvement across sessions.  $S_{1}$ obtained below chance level ($33\%$) performance during all sessions. Performance of users $S_{2}$, $S_{3}$ and $S_{4}$ was in general above chance level, but always remained below $60\%$. During uncued manipulation trials, users $S_{1}$ and $S_{2}$ presented the lowest performance values, close-to and below chance level. This indicates that these users were faced with difficulty while controlling the process-control BCI. Performance of $S_{3}$ and $S_{4}$ during uncued manipulation trials was higher (around $40$- $60\%$) when compared to $S_{1}$ and $S_{2}$.  Mean performance values across trials of users $S_{3}$ and $S_{4}$ failed to reach the $70\%$ considered as the theoretical threshold  for practical MI-BCI use~\cite{Guger03}. However,  their performance was evidently higher than the one obtained by users performing on the process-control BCI. This  might suggest that the designed semi-autonomous goal-selection BCI was easier to manipulate than process-control BCI. Future research will address classification optimization to increase system accuracy and ease of use.

As shown in Table~\ref{tab:tab4},  selected channels and frequencies for feature extraction showed changes across sessions for all users.  This might suggest that the used channel/frequency selection method is sensitive to intra- and inter-subjects brain variability. After training trials, a classifier with fixed parameters was selected per subject and used in all BCI trials. Yet, constant adaptation of the classifier parameters is required for optimal operation. 
Hence, an optimized feature selection algorithm should be implemented to address this issue and increase efficiency in our proposed semi-autonomous BCI. Such optimization was out of the scope of our work, but reports on how optimized correlation-based feature selection methods are used in MI-BCIs can be found in~\cite{Jin19,Feng19}. 

Another efficient approach for feature selection is the partial directed coherence (PDC) analysis, which could help to identify relevant channels and features. Recently, a PDC-based analysis was proposed in~\cite{Gaxiola17} to identify relevant features for MI tasks, and efficient classifiers were built based on this procedure.  Even more recently, a review on EEG classification algorithms highlights Riemannian geometry-based classifiers as promising, as well as adaptive classification algorithms~\cite{Lotte18}. 
A simple implementation of an adaptive classifier for MI tasks was described in~\cite{Shenoy06}, which showed an encouraging increase on classification accuracy. More novel classifiers based on Riemannian geometry have shown good results on classifying brain-related MI tasks~\cite{Guan19}.



In regards to our selection of the P$300$ component to evaluate mental fatigue, such component is not exclusively presented during non-frequent stimulus, rather its amplitude is enhanced, which makes it a suitable control command for BCIs. P$300$ amplitude is larger during non-frequent stimuli, and it is typically used/analyzed based on this argument. However, it has been demonstrated that P$300$ responses can be observed to both frequent and non-frequent stimuli~\cite{Donchin73, Verleger16}.  In fact, under a reaction-time regime, P$300$ is elicited on both predictable and unpredictable stimuli. Task demands increase in this scenario, as users must decide when to respond in a fast and correct manner. This leads to an enhancement of P$300$, independently of stimulus predictability~\cite{Donchin73}. In our study, users were instructed to perform MI commands after stimulus presentation of the word “Imagine”, and P$300$ components were analyzed immediately after stimulus onset. Although stimulus presentation during uncued manipulation trials could be considered as predictable, P$300$ analysis holds validity, as it was executed under a reaction-time regime. 

Under those conditions, the results of the two-way ANOVA and one-way ANOVA tests showed  statistically significant changes of P$300$ latency for users $S_{1}$, $S_{2}$ and $S_{4}$.  Except for $S_{4}$, the tests revealed no statistical significance for P$300$ amplitude. When comparing the amplitude and latency values from Figures~\ref{fig:p300amp} and \ref{fig:p300lat}, a general trend was found among users: a decrease of amplitude and an increase of latency. These trends in P$300$ features were presented along trial repetition, that is, after continuous manipulation of the BCI.  These changes in amplitude and latency might be related to the generation of mental fatigue, as they are presented after a continuous execution of manipulation tasks through the BCI. It has been shown that a decrease in P$300$ ampitude and an increase in latency reflect decreased cognitive processing and lower attention levels~\cite{Cheng11}. Similar results have been found on a P$300$-BCI evaluated under different levels of mental workload and fatigue~\cite{Kathner14}. When comparing subjects performing on the same BCI type, the user with the lowest performance exhibited lower amplitude and  higher latency values than the user with the highest performance (although it was more evident for amplitude values). This was observed when comparing both $S_{1}$-$S_{2}$ and $S_{3}$-$S_{4}$. Subject $S_{3}$ displayed an interesting behaviour: an increasing amplitude trend, as well as being the only subject which did not show statistical significance on any P$300$ test. At the same time, it was the subject with the highest performance values on uncued manipulation trials. A possible explanation to this particular case is that after performing manipulation trials on the BCI, mental fatigue affected differently user $S_{3}$ than the rest of the users.  This difference in mental fatigue generation was reflected as non-significant changes in P$300$ parameters during the tests, as well as higher performance values. 

\section{Conclusions}
Two BCI systems, a process-control and semi-autonomous goal-selection were implemented and compared in terms of performance and mental fatigue.  The process-control BCI allowed users to perform three-dimensional movements on a robotic arm to reach for a target. The semi-autonomous BCI allowed the user to execute manipulation tasks, using the same robotic arm, which include reaching, picking and placing movements successfully. The increase of task complexity represented by the semi-autonomous BCI was achieved without compromising the simplicity of the control procedure, as both BCIs were controlled through MI commands. Users performing on semi-autonomous BCI obtained higher performance values when compared to users performing on low-level BCI. The difference in task complexity also represented a difference in the mental fatigue experienced by the users on different systems. A P$300$ amplitude decrease and latency increase was found as users performed continuous BCI trials, which is consistent to reports of mental fatigue detection on EEG. 

We also present strong evidence of the advantages of semi-autonomous BCI in terms of performance and mental fatigue. It is also important to address the potential use of the P$300$ waveform as an indicator of mental fatigue during BCI testing, training, and evaluation.  Techniques to further reduce mental fatigue while using BCI systems might provide an increase in BCI patients acceptance rate, as well as a possible path to tackle BCI illiteracy. It is of great importance that the user finds the system as non-fatiguing and easy to use in order to provide a more comfortable and efficient assistance. This also facilitates the user in the process of learning how to control the BCI, which can be used together with different strategies to further personalize the system (see, e.g., a previous work by our group in how to select a feedback modality that better enhances the volunteer's capacity to operate a BCI system~\cite{angulo2015link}).                                                                                             

The development of more advanced semi-autonomous BCI systems which provide information about the environment during specific tasks will allow to further enhance performance and usability.  Semi-autonomous BCIs offer users the possibility to perform more complex tasks in a simple, less fatiguing way.  In our system, the integration of the AV and CGA algorithms provided a real-time calculation of the robot's inverse model, offering flexibility to implement more complex object manipulation tasks in a dynamic environment. The use of a higher DOF robotic arm, as well as the implementation of object recognition techniques might improve the complexity of the manipulation tasks to be performed, while using the same MI commands to control the BCI, ensuring control simplicity to the users.  



\newpage

\renewcommand{\listfigurename}{Captions of Figures}
\listoffigures
\renewcommand{\listtablename}{Captions of Tables}
\listoftables
\newpage
\clearpage

\graphicspath{{./}}
\DeclareGraphicsExtensions{{.pdf}}

\newpage
\clearpage

\newpage
\clearpage

\newpage
\clearpage

\newpage
\clearpage

\newpage
\clearpage

\newpage
\clearpage

\newpage
\clearpage

\newpage
\clearpage

\newpage
\clearpage

\newpage
\clearpage

\newpage
\clearpage

\newpage
\clearpage

\newpage
\clearpage

\newpage
\clearpage

\newpage
\clearpage

\newpage
\clearpage

\begin{table}
  \caption{Representations of the conformal geometric entities}
  \label{tab:tab1}
 \begin{center} 
  \begin{tabular}{ccc}
    \toprule
    Entity &Standard&Dual\\
    \midrule
    Point & $P=x+\frac{1}{2}x^{2}e_{\infty}+e_{0}$ &   \\
    Point pair & $Pp=s_{1}\wedge s_{2}\wedge s_{3}$ &  $Pp^{*}=x_{1}\wedge x_{2}$ \\
    Line & $l=\pi_{1}\wedge \pi_{2}$ &  $l^{*}=x_{1}\wedge x_{2}\wedge e_{\infty}$ \\
    Circle & $c=s_{1}\wedge s_{2}$ & $c^{*}=x_{1}\wedge x_{2}\wedge x_{3}$  \\
    Sphere & $s=P-\frac{1}{2}r^{2}e_{\infty}$ & $s^{*}=x_{1}\wedge x_{2}\wedge x_{3}\wedge x_{4}$  \\
  \bottomrule
\end{tabular}
\end{center}
\end{table}

\newpage
\clearpage

\begin{table}
  \caption{Parameters for joint angles calculation}
  \label{tab:tab2}
\begin{center}
  \begin{tabular}{ccc}
    \toprule
    $k$ & $\alpha_{k}$ & $\beta_{k}$ \\
    \midrule
    0& $e_{2}$ & $(\pi_{e}^{*} \wedge e_{\infty}) \cdot e_{0}$ \\
    2& $(L_{12}\cdot e_{0})\cdot e_{\infty}$ &  $(L_{23}\cdot e_{0})\cdot e_{\infty}$  \\
    3& $(L_{23}\cdot e_{0})\cdot e_{\infty}$  &  $(L_{3e}\cdot e_{0})\cdot e_{\infty}$ \\
  \bottomrule
\end{tabular}
\end{center}
\end{table}

\newpage
\clearpage

\begin{table}
  \caption{Features: EEG channels and frequency range (in Hz), and mean accuracy of LDA classifiers for all subjects and training sessions}
  \label{tab:tab4}
  \begin{tabular}{cccc}
    \toprule
    \textbf{Subject} & \textbf{Session} & \textbf{Features}& \textbf{Accuracy}\\
    \midrule
  &   $1$ &  C$4$ ($11$-$15$), \  P$3$ ($11$-$15$), \  C$3$ ($7$-$11$), \ Fp$2$ ($13$-$17$) &    $65$\% \\

 $S_{1}$  & $2$ & C$3$ ($7$-$13$), \  P$3$ ($9$-$13$), \ P$4$ ($9$-$13$), \ P$4$ ($13$-$17$), \ Cz ($9$-$13$)  & $64$\% \\

 &  $3$ & C$4$ ($9$-$15$), \  C$3$ ($13$-$17$), \  P$3$ ($23$-$27$), \ P$4$ ($25$-$29$) &  $63$\% \\  \hline

 &   $1$ & C$4$ ($9$-$13$), \ P$4$ ($11$-$15$), \ F$4$ ($17$-$21$), \  F$4$ ($11$-$15$), \ Cz ($11$-$15$) &  $65$\% \\

$S_{2}$ & $2$ & C$4$ ($15$-$19$), \ C$3$ ($19$-$23$), \  P$4$ ($21$-$25$), \ Cz ($19$-$23$), \ FP$1$ ($25$-$29$) &  $62$\% \\

&  $3$ & C$4$ ($19$-$23$), \ C$4$ ($17$-$21$), \   Cz ($19$-$23$), \ P$3$ ($21$-$25$), \ F$3$ ($19$-$23$) &  $60$\% \\ \hline

 & $1$ & F$4$ ($15$-$21$), \ F$3$ ($9$-$11$), \ P$4$ ($9$-$13$), \  F$4$ ($15$-$21$)  & $56$\%  \\

$S_{3}$ & $2$ & C$3$ ($11$-$15$), \ C$4$ ($7$-$11$), \  P$3$ ($15$-$19$), \  P$4$ ($11$-$15$) &   $61$\%  \\

&  $3$ &  C$3$ ($9$-$15$), \ C$4$ ($11$-$15$), \  P$3$ ($11$-$15$) &   $78$\%   \\ \hline

 & $1$ & C$4$ ($9$-$13$), \ C$3$ ($9$-$13$), \ P$4$ ($17$-$21$), \  F$4$ ($19$-$23$), \ F$3$ ($19$-$23$)  &  $73$\% \\

$S_{4}$ & $2$ & F$3$ ($11$-$15$), \ C$4$ ($7$-$13$), \   P$4$ ($17$-$21$), \  P$3$ ($17$-$21$), \ F$4$ ($19$-$23$) &  $72$\%  \\

&  $3$ &  C$4$ ($7$-$11$), \ C$3$ ($13$-$17$), \ F$4$ ($17$-$21$), \  C$4$ ($11$-$15$)   &  $60$\%   \\

  \bottomrule
\end{tabular}
\end{table}

\newpage
\clearpage

\begin{table}
  \caption{Two-way ANOVA results for P300 latency and amplitude}
  \label{tab:anova2}
  \begin{tabular}{ccccccc}
    \toprule
    & \multicolumn{3}{c}{Latency}     & \multicolumn{3}{c}{Amplitude} \\ 
    \midrule
Subject & Trial& Channel& Interaction& Trial& Channel& Interaction\\ \hline
S1  & \begin{tabular}[c]{@{}c@{}}$F=3.69$\\ $\mathbf{p=0.0147}$\end{tabular}
      & \begin{tabular}[c]{@{}c@{}}$F=0.25$\\ $p=0.782 $\end{tabular}  
      & \begin{tabular}[c]{@{}c@{}}$F=0.69$\\ $p=0.6994$\end{tabular}     
      & \begin{tabular}[c]{@{}c@{}}$F=2.45$\\ $p=0.0676$\end{tabular}
      & \begin{tabular}[c]{@{}c@{}}$F=3.08$\\ $p=0.0609$\end{tabular}
      & \begin{tabular}[c]{@{}c@{}}$F=0.42$\\ $p=0.8969$\end{tabular}       
			\\ \hline
S2  & \begin{tabular}[c]{@{}c@{}}$F=9.33$\\ $\mathbf{p=0.0001}$\end{tabular} 
     & \begin{tabular}[c]{@{}c@{}}$F=0.13$\\ $p=0.8816$\end{tabular}        
     & \begin{tabular}[c]{@{}c@{}}$F=0.1 $\\ $p=0.999 $\end{tabular}             
     & \begin{tabular}[c]{@{}c@{}}$F=1.26$\\ $p=0.3074$\end{tabular}                          
     & \begin{tabular}[c]{@{}c@{}}$F=0.48$\\ $p=0.6217$\end{tabular}       
     & \begin{tabular}[c]{@{}c@{}}$F=0.14$\\ $p=0.9970$\end{tabular}           
		  \\ \hline
S3  & \begin{tabular}[c]{@{}c@{}}$F=0.03$\\ $p=0.9983$\end{tabular} 
      & \begin{tabular}[c]{@{}c@{}}$F=0.59$\\ $p=0.5604$\end{tabular}       
      & \begin{tabular}[c]{@{}c@{}}$F=0.44$\\ $p=0.8891$\end{tabular}   
      & \begin{tabular}[c]{@{}c@{}}$F=0.83$\\ $p=0.5191$\end{tabular}         
      & \begin{tabular}[c]{@{}c@{}}$F=0.01$\\ $p=0.9924$\end{tabular}        
      & \begin{tabular}[c]{@{}c@{}}$F=0   $\\ $p=1     $\end{tabular}            
		  \\ \hline
S4  & \begin{tabular}[c]{@{}c@{}}$F=1.05$\\ $p=0.4003$\end{tabular} 
      & \begin{tabular}[c]{@{}c@{}}$F=1.5 $\\ $p=0.24  $\end{tabular}    
      & \begin{tabular}[c]{@{}c@{}}$F=0.3 $\\ $p=0.9589$\end{tabular}            
      & \begin{tabular}[c]{@{}c@{}}$F=1.81$\\ $p=0.1534$\end{tabular}          
      & \begin{tabular}[c]{@{}c@{}}$F=0.17$\\ $p=0.8405$\end{tabular}      
      & \begin{tabular}[c]{@{}c@{}}$F=0.13$\\ $p=0.9972$\end{tabular}    \\        
 
 \bottomrule
\end{tabular}
\end{table}

\newpage
\clearpage

\begin{table}
\caption{One-way ANOVA results for P300 latency and amplitude on channels O1, O2 and Pz.}
\begin{center}
\begin{tabular}{ccccccc}
\toprule
& \multicolumn{3}{c}{Latency}     & \multicolumn{3}{c}{Amplitude} \\ 
\midrule 
Subject & O1 & O2 & Pz & O1 & O2 & Pz \\ \hline
S1  & \begin{tabular}[c]{@{}c@{}}$F=3.54$\\ $\mathbf{p=0.0476}$\end{tabular}
      & \begin{tabular}[c]{@{}c@{}}$F=0.76$\\ $p=0.5767$\end{tabular}  
      & \begin{tabular}[c]{@{}c@{}}$F=0.55$\\ $p=0.7055$\end{tabular}     
      & \begin{tabular}[c]{@{}c@{}}$F=1.79$\\ $p=0.2066$\end{tabular}
      & \begin{tabular}[c]{@{}c@{}}$F=1.61$\\ $p=0.246$\end{tabular}
      & \begin{tabular}[c]{@{}c@{}}$F=0.82$\\ $p=0.5421$\end{tabular}       
			\\ \hline
S2  & \begin{tabular}[c]{@{}c@{}}$F=2.1 $\\ $p=0.1554$\end{tabular} 
    & \begin{tabular}[c]{@{}c@{}}$F=2.13$\\ $p=0.1518$\end{tabular}        
    & \begin{tabular}[c]{@{}c@{}}$F=2.5 $\\ $p=0.1091$\end{tabular}             
    & \begin{tabular}[c]{@{}c@{}}$F=0.3 $\\ $p=0.8705$\end{tabular}                          
    & \begin{tabular}[c]{@{}c@{}}$F=0.23$\\ $p=0.9171$\end{tabular}       
    & \begin{tabular}[c]{@{}c@{}}$F=0.1 $\\ $p=0.9806$\end{tabular}           
		  \\ \hline
S3  & \begin{tabular}[c]{@{}c@{}}$F=0.37$\\ $p=0.8238$\end{tabular} 
      & \begin{tabular}[c]{@{}c@{}}$F=0.12$\\ $p=0.9731$\end{tabular}       
      & \begin{tabular}[c]{@{}c@{}}$F=0.54$\\ $p=0.7089$\end{tabular}   
      & \begin{tabular}[c]{@{}c@{}}$F=1.31$\\ $p=0.3322$\end{tabular}         
      & \begin{tabular}[c]{@{}c@{}}$F=0.13$\\ $p=0.9687$\end{tabular}        
      & \begin{tabular}[c]{@{}c@{}}$F=0.41$\\ $p=0.8043$\end{tabular}            
		  \\ \hline
S4  & \begin{tabular}[c]{@{}c@{}}$F=1.68$\\ $p=0.2311$\end{tabular} 
      & \begin{tabular}[c]{@{}c@{}}$F=4.52$\\ $\mathbf{p=0.0242}$\end{tabular}    
      & \begin{tabular}[c]{@{}c@{}}$F=2.45$\\ $p=0.1138$\end{tabular}            
      & \begin{tabular}[c]{@{}c@{}}$F=2.99$\\ $p=0.0728$\end{tabular}          
      & \begin{tabular}[c]{@{}c@{}}$F=2.27$\\ $p=0.1336$\end{tabular}      
      & \begin{tabular}[c]{@{}c@{}}$F=9.5 $\\ $\mathbf{p=0.0019}$\end{tabular}     \\ 
\bottomrule
\end{tabular}
\label{tab:anova1}
\end{center}
\end{table}

\end{document}